\documentclass[twocolumn]{aastex631}

\newcommand{\scm}{~s~cm$^{2}$}
\newcommand{\scmpx}{~s$^{-1}$~cm$^{-2}$~px$^{-1}$}
\newcommand{\ergcmdegs}{~erg~cm$^{-2}$~deg$^{-2}$~s$^{-1}$}

\newcounter{daggerfootnote}

\usepackage{amsmath}
\usepackage{multirow}
\usepackage[abs]{overpic}
\usepackage{breakurl}
\usepackage{hyperref}
\usepackage{textcomp}
\usepackage{fancyhdr}
\usepackage{graphicx}
\usepackage{amssymb}
\usepackage[toc,page]{appendix}
\usepackage{floatrow}
\usepackage{nicefrac}
\usepackage{varwidth}
\usepackage{scrextend}
\usepackage{xcolor}
\usepackage{colortbl}
\usepackage{tablefootnote}
\usepackage{array}
\usepackage[]{graphicx}

\newcommand{\SAUNAS}{\texttt{SAUNAS}}

\newcommand\asr{\ref@jnl{Adv. Space Res.}}

\newcommand{\s}{$\sim$}
\newcommand{\n}{$-$}
\newcommand{\Chandra}{\emph{Chandra}}
\newcommand{\Hubble}{\emph{HST}}

\newcommand{\ciao}{\texttt{CIAO}}
\newcommand{\LIRA}{\texttt{LIRA}}
\newcommand{\vorbin}{\texttt{vorbin}}

\definecolor{navyblue}{rgb}{0.0, 0.0, 0.5}

\newcommand{\escmarc}{s$^{-1}$ cm$^{-2}$ arcsec$^{-2}$}
\usepackage{makecell}

\usepackage[symbol]{footmisc}
\renewcommand*{\thefootnote}{\arabic{footnote}}
\begin{document}

\title{\textbf{SAUNAS} I: Searching for Low Surface Brightness X-ray Emission with Chandra/ACIS}

\author[0000-0003-3249-4431]{Alejandro S. Borlaff}
\affiliation{NASA Ames Research Center, Moffett Field, CA 94035, USA}
\affiliation{Bay Area Environmental Research Institute, Moffett Field, California 94035, USA}

\author{Pamela M. Marcum}
\affiliation{NASA Ames Research Center, Moffett Field, CA 94035, USA}

\author{Mehmet Alpaslan}
\affiliation{ConstructConnect, Rookwood Exchange, 3825 Edwards Rd 800, Cincinnati, OH 45209}

\author[0000-0002-8341-342X]{Pasquale Temi}
\affiliation{NASA Ames Research Center, Moffett Field, CA 94035, USA}

\author[0000-0002-1598-5995]{Nushkia Chamba}
\affiliation{NASA Ames Research Center, Moffett Field, CA 94035, USA}

\author[0000-0003-0346-6722]{Drew S. Chojnowski}
\affiliation{NASA Ames Research Center, Moffett Field, CA 94035, USA}

\author{Michael N. Fanelli}
\affiliation{NASA Ames Research Center, Moffett Field, CA 94035, USA}

\author[0000-0002-6610-2048]{Anton M. Koekemoer}
\affiliation{Space Telescope Science Institute, 3700 San Martin Dr., Baltimore, MD 21218, USA}

\author[0000-0003-1250-8314]{Seppo Laine}
\affiliation{IPAC, Mail Code 314-6, Caltech, 1200 E. California Blvd., Pasadena, CA 91125, USA}

\author[0000-0001-5357-6538]{Enrique Lopez-Rodriguez}
\affiliation{Kavli Institute for Particle Astrophysics \& Cosmology (KIPAC), Stanford University, Stanford, CA 94305, USA}

\author[0000-0002-0905-7375]{Aneta Siemiginowska}
\affiliation{Harvard Smithsonian Center for Astrophysics, 60 Garden St, Cambridge, MA 02138, USA}

\begin{abstract}
We present \SAUNAS\ (Selective Amplification of Ultra Noisy Astronomical Signal), a pipeline designed for detecting diffuse X-ray emission in the data obtained with the Advanced CCD Imaging Spectrometer (ACIS) of the Chandra X-ray Observatory. \SAUNAS\ queries the available observations in the \Chandra\ archive, performs photometric calibration, PSF (point spread function) modeling and deconvolution, point-source removal, adaptive smoothing, and background correction. This pipeline builds on existing and well-tested software including \ciao, \vorbin, and \LIRA. We characterize the performance of \SAUNAS\ through several quality performance tests, and demonstrate the broad applications and capabilities of \SAUNAS\ using two galaxies already known to show X-ray emitting structures. \SAUNAS\ successfully detects the 30~kpc X-ray super-wind of NGC\,3079 using \Chandra/ACIS datasets, matching the spatial distribution detected with more sensitive XMM-Newton observations. The analysis performed by \SAUNAS\ reveals an extended low surface brightness source in the field of UGC\,5101 in the 0.3--1.0~keV and 1.0--2.0~keV bands. This source is potentially a background galaxy cluster or a hot gas plume associated with UGC\,5101. \SAUNAS\ demonstrates its ability to recover previously undetected structures in archival data, expanding exploration into the low surface brightness X-ray universe with \Chandra/ACIS. 

\end{abstract}



\section{Introduction} 
\label{sec:intro}

The Advanced CCD Imaging Spectrometer on the \Chandra\ X-ray Observatory \citep[][hearafter \Chandra/ACIS]{weisskopf+2000inproceedings_2} provides an effective balance between angular resolution and sensitivity for the study of diffuse galactic hot gas emission, with its field of view (FOV)  up to $16.9\times16.9$ arcmin$^2$ and 0.492 arcsec of spatial resolution. Stacking multiple observations made over \Chandra's 25$+$ year mission is one of the keys to obtaining the deepest observations of the universe in X-ray. However, in most cases, the position of the target on the detector changes within observations, introducing serious challenges to acquiring a meaningful combined image. The PSF (point spread function) broadens and becomes more ellipse-shaped with increasing off-axis angle\footnote{Understanding the \Chandra\ PSF \url{https://cxc.cfa.harvard.edu/ciao/PSFs/psf_central.html}}\footnote{\Chandra/CIAO PSF presentation from 233rd AAS meeting: \url{https://cxc.harvard.edu/ciao/workshop/nov14/02-Jerius.pdf}}, necessitating an elaborate deconvolution scheme and hampering the ability to exploit the full capabilities of the archive. Consequently, \Chandra\ observations are under-explored to date in studies advancing the low X-ray surface brightness (SB) domain. 

Future studies of low X-ray SB emission (\s$10^{-8}$ to $10^{-11}$~s$^{-1}$~cm$^{-2}$~arcsec$^{-2}$ and beyond) enabled by data processed to enhance detection of low-count regions could advance progress in several currently open questions relevant to galaxy evolution, including the origins of \emph{diffuse} soft X-ray emission in galaxies and feedback involvement \citep{kelly+2021mnras502_2934, henley+2010apj723_935}. 

Lambda Cold Dark Matter (LCDM) cosmology predicts filaments of diffuse gas from the cosmic web to  accrete during their infall onto proto-galactic dark matter (DM) halos \citep{white+1978mnras183_341, white+1991apj379_52, benson+2010mnras402_2321} where gas is heated to approximately the halo virial temperature ($T>10^{6}$ K). 
This plasma, further shaped by energy injection from active galactic nuclei \citep[AGN,][]{diehl+2008apj680_897}, supernovae (SN) and stellar winds \citep{hopkins+2012mnras421_3488}, is detected as diffuse soft X-ray band emission around galaxies \citep{mulchaey2000araa38_289, osullivan+2001mnras328_461, sato+2000apj537_73,aguerri+2017mnras468_364}.  The origins and evolution of hot gas halos are important open questions in astrophysics, as halos are both the aftermath and active players of gas feedback processes, which modulate the star formation efficiency in galaxies \citep{rees+1977mnras179_541, silk1977apj211_638,  binney1977apj215_483, white+1978mnras183_341, white+1991apj379_52}.  The largely-unexplored realm of extreme diffuse gas emission, likely associated with large departures from equilibrium \citep{strickland+2004apj151_193}, is likely to preserve a unique historical record of these events. Such emission is also likely to be disregarded in studies using standard pipelines that are not optimized for preservation of statistically significant but low SB detections.

This project is the first in a series that will study the hot gas halos around galaxies using X-ray observations from the \Chandra\ X-ray observatory. The first step is to test the pipeline to reduce the \Chandra/ACIS data products, named \SAUNAS\ (Selective Amplification of Ultra Noisy Astronomical Signal).
This paper describes the \SAUNAS\ pipeline processing of data from the \Chandra\ Data Archive\footnote{\Chandra\ Data Archive: \url{https://cxc.harvard.edu/cda/}} and benchmarks it to previous works. In particular, we focus on the comparison of results between our analyses and those from other investigations for two well-detected X-ray sources characterized in the literature: NGC\,3079 and UGC\,5101. The latter has complex and extended X-ray emission, previously unexplored and only revealed by the current work. \par 

This paper is organized as follows. The \SAUNAS\ pipeline is described in Sec.\,\ref{sec:methods}. The selection of published results for \SAUNAS\ performance comparison is discussed in Sec.\,\ref{subsec:sample_selection}. The benchmark analysis is presented in Secs.\,\ref{subsec:NGC3079}, and \ref{subsec:UGC5101}. The discussion and conclusions are presented in Secs.\,\ref{sec:discussion} and \,\ref{sec:conclusions}, respectively. We assume a concordance cosmology \citep[$\Omega_{\mathrm{M}} = 0.3$, $\Omega_{\mathrm{\Lambda}}=0.7, H_{0} =70 $ km s$^{-1}$ Mpc$^{-1}$, see][]{spergel+2007apj170_377}. All magnitudes are in the AB system \citep{oke1971apj170_193} unless otherwise noted.

\section{Methodology} 
\label{sec:methods}

\subsection{Observational challenges}
\label{subsec:Objectives}
From an observational perspective, measuring diffuse X-ray halo properties in galaxies involves at least four technical challenges:
\begin{enumerate} 
\item Detection: The outskirts of X-ray halos are extremely faint ($\lesssim10^{-8}$ -- $10^{-11}$~s$^{-1}$~cm$^{-2}$~arcsec$^{-2}$). Separating the faint emission associated with sources from that of the X-ray background \citep{anderson+2011apj737_22}  within such low count regimes is an extraordinarily challenging task. Statistical methods that assume a normal (Gaussian) distribution may not produce accurate results.

\item Deblending: AGNs and XRBs are typically unresolved point sources that may contribute to the same X-ray bands where the hot gas halos are expected to emit (from $\sim 0.3-0.5$ to $1.2-2$~keV). While in principle the detection of hot gas halos in nearby galaxies may not require very high spatial resolution observations or spectral capabilities, the separation of such emission from that of point sources does require them. High spatial resolution observations reduce systematic contamination in low surface brightness regimes.

\item Point spread function (PSF) contamination: The distribution of diffuse emission is easily confused with the scattered, extended emission of the unresolved bright cores that contaminate the outskirts of the target through the extended wings of the PSF of the detector \citep{sandin2014aap567_97, sandin2015aap577_106}. Most studies do not correct for this type of scattering effect, although a few works, such as \citet{anderson+2013apj762_106}, have explored the combined stacked hot gas halo emission of 2165 galaxies observed with ROSAT (0.5--2.0~keV), convolving the combined surface brightness profiles by the PSF model to take into account the dispersion of light.


\item Reproducibility \& Accessibility: The methodologies for calibration, detection, and characterization of X-ray emission have substantial differences between studies. Due to the Poissonian nature of the X-ray emission, most studies employ different types of adaptive smoothing in their analysis. These software methods tend to be custom-made and infrequently made publicly available. Likewise, the final data products (final science frames) are seldom offered to the community.

\end{enumerate} 

The \SAUNAS\ methodology presented in the current paper attempts to address most of these points by 1) correcting the PSF in the images, 2) separating the emission of point sources from that of diffuse extended ones, and 3) providing a quantitative metric to determine if a detection is real or not. These two points implemented in \SAUNAS\ are the major difference with other existing codes for detection of extended X-ray emission, such as \texttt{vtpdetect} \citep{ebeling+1993pre47_704} or \texttt{EXSdetect} \citep{liu+2013aap549_143}, as they do not attempt to deconvolve the observations using dedicated PSF models or to separate diffuse emission from point sources.
\subsection{SAUNAS pipeline} \label{subsec:methods_saunas}

\begin{figure*}[t!]
 \begin{center}
\includegraphics[trim={0 0 0 0}, clip, width=\textwidth]{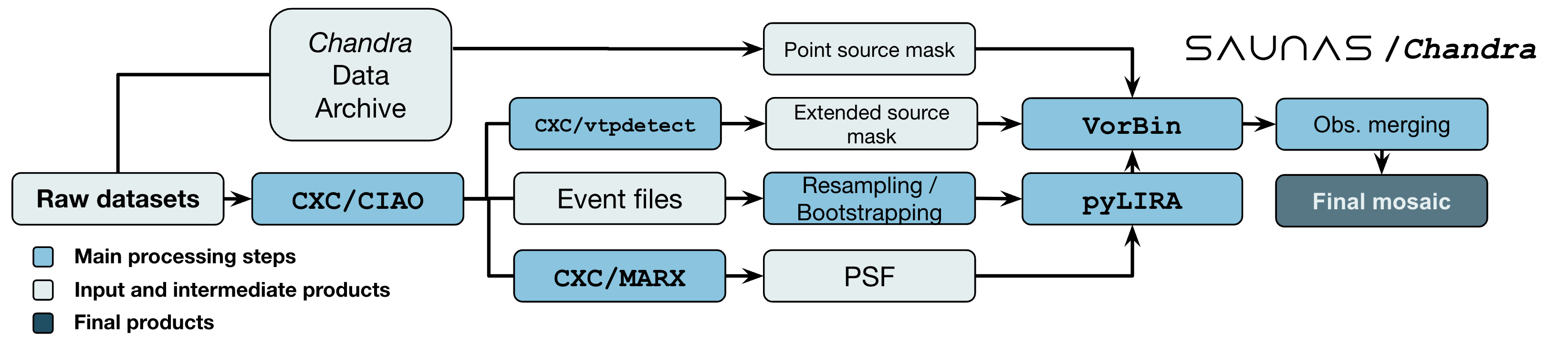}
\caption{\texttt{SAUNAS} pipeline flowchart. From left to right: \SAUNAS\ pre-calibrates the \Chandra\ observations by first using \Chandra\ X-ray Center (CXC)/\texttt{CIAO}, which generates the event files, extended source masks, and the PSFs. The events in each individual visit are first resampled via bootstrapping and then deconvolved using \LIRA. Voronoi binning is applied to each deconvolved observation, and merged into a single flux map after sky background correction.} 
\label{fig:saunas_flowchart}
\end{center}
\end{figure*}

\begin{figure*}[t!]
 \begin{center}
\begin{overpic}[trim={35 20 35 35}, clip, width=0.495\textwidth]{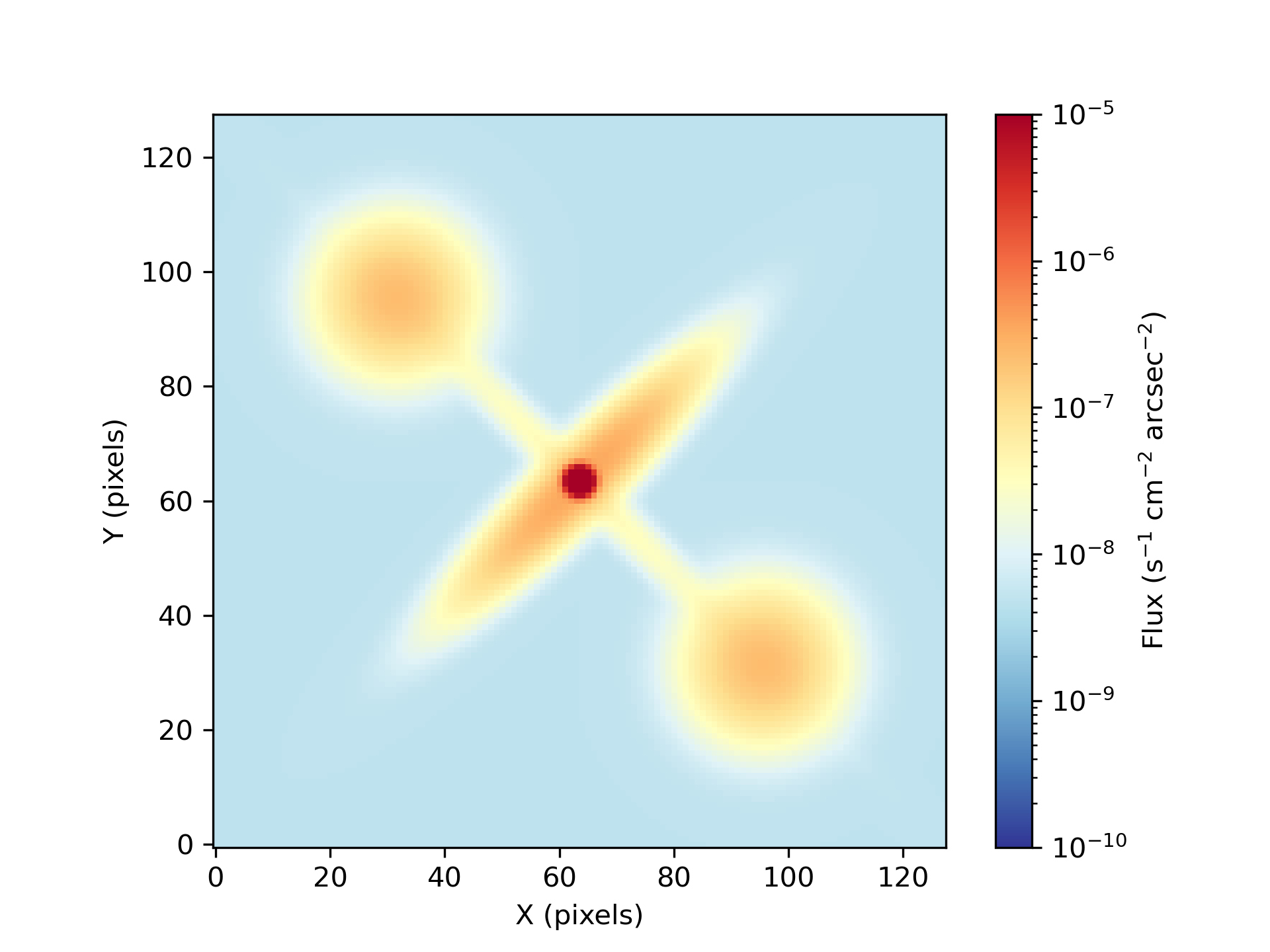}
\put(126,173){\color{black} \colorbox{white}{\textsf{Simulated source}}}
\end{overpic}
\begin{overpic}[trim={35 20 35 35}, clip, width=0.495\textwidth]{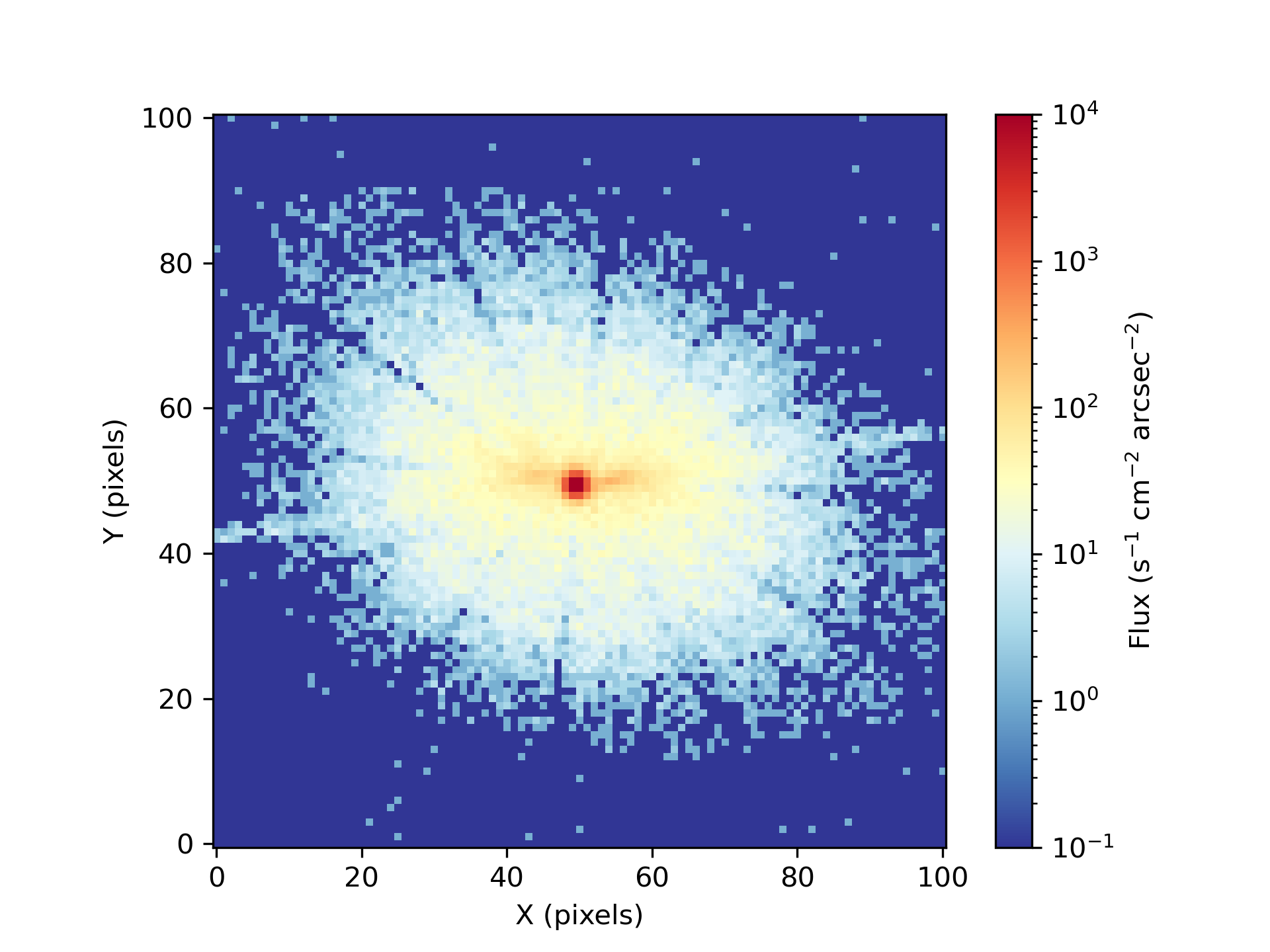}
\put(175.5,173){\color{black} \colorbox{white}{\textsf{PSF}}}
\end{overpic}

\begin{overpic}[trim={35 20 35 35}, clip, width=0.495\textwidth]{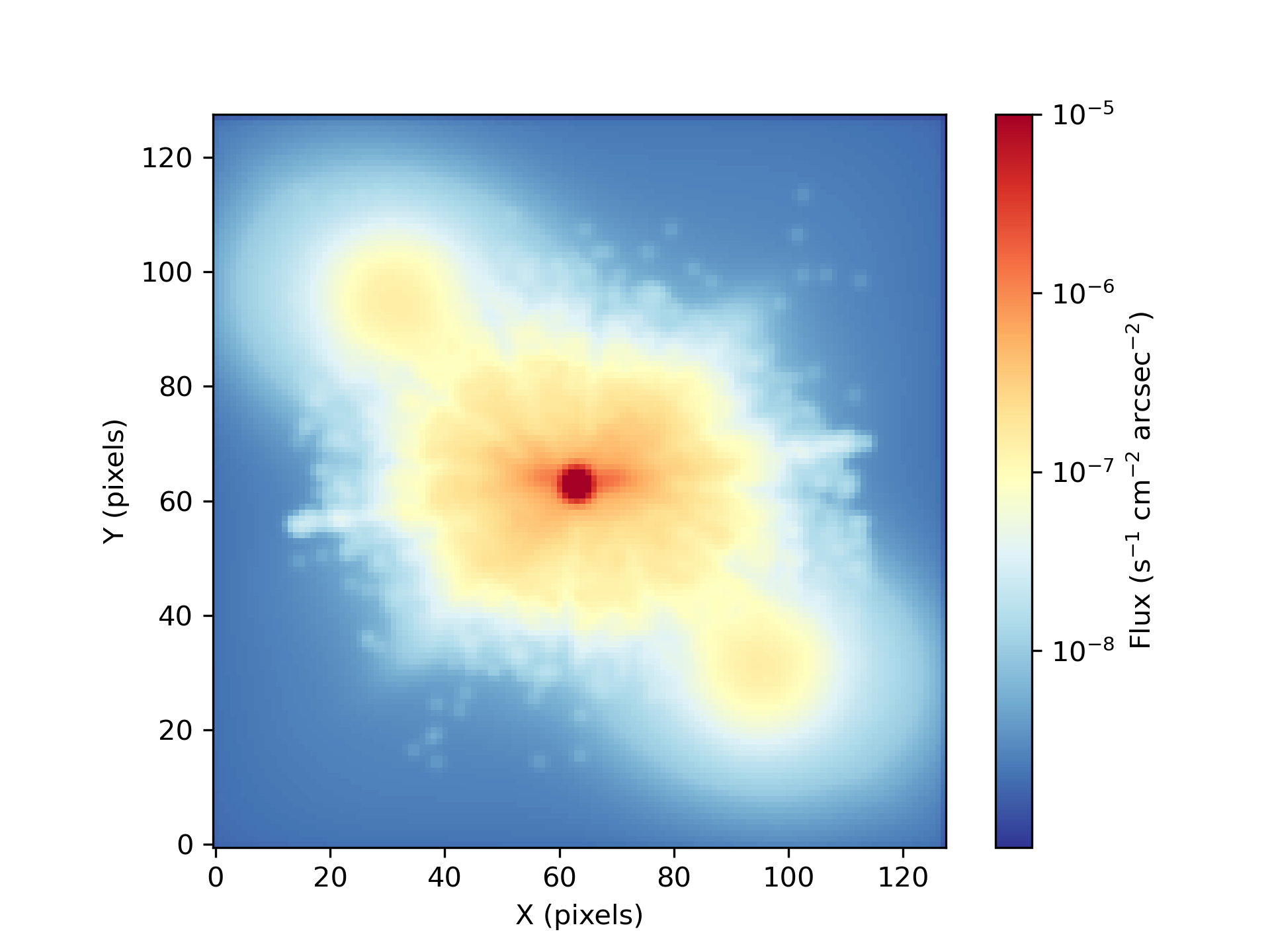}
\put(107,173){\color{black} \colorbox{white}{\textsf{PSF convolved source}}}
\end{overpic}
\begin{overpic}[trim={35 20 35 35}, clip, width=0.495\textwidth]{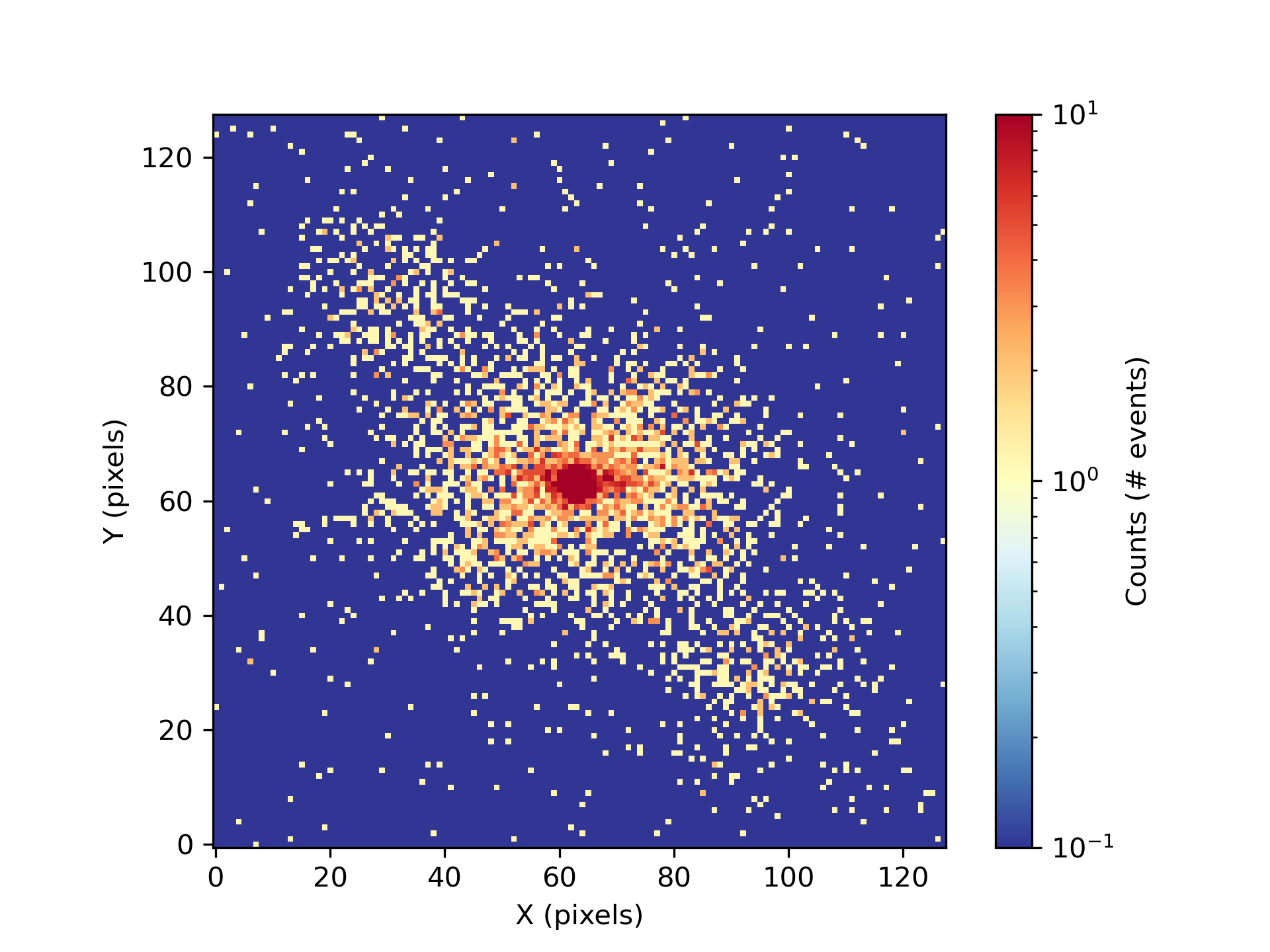}
\put(106.5,173){\color{black} \colorbox{white}{\textsf{Simulated observation}}}
\end{overpic}

\begin{overpic}[trim={35 0 35 35}, clip, width=0.495\textwidth]{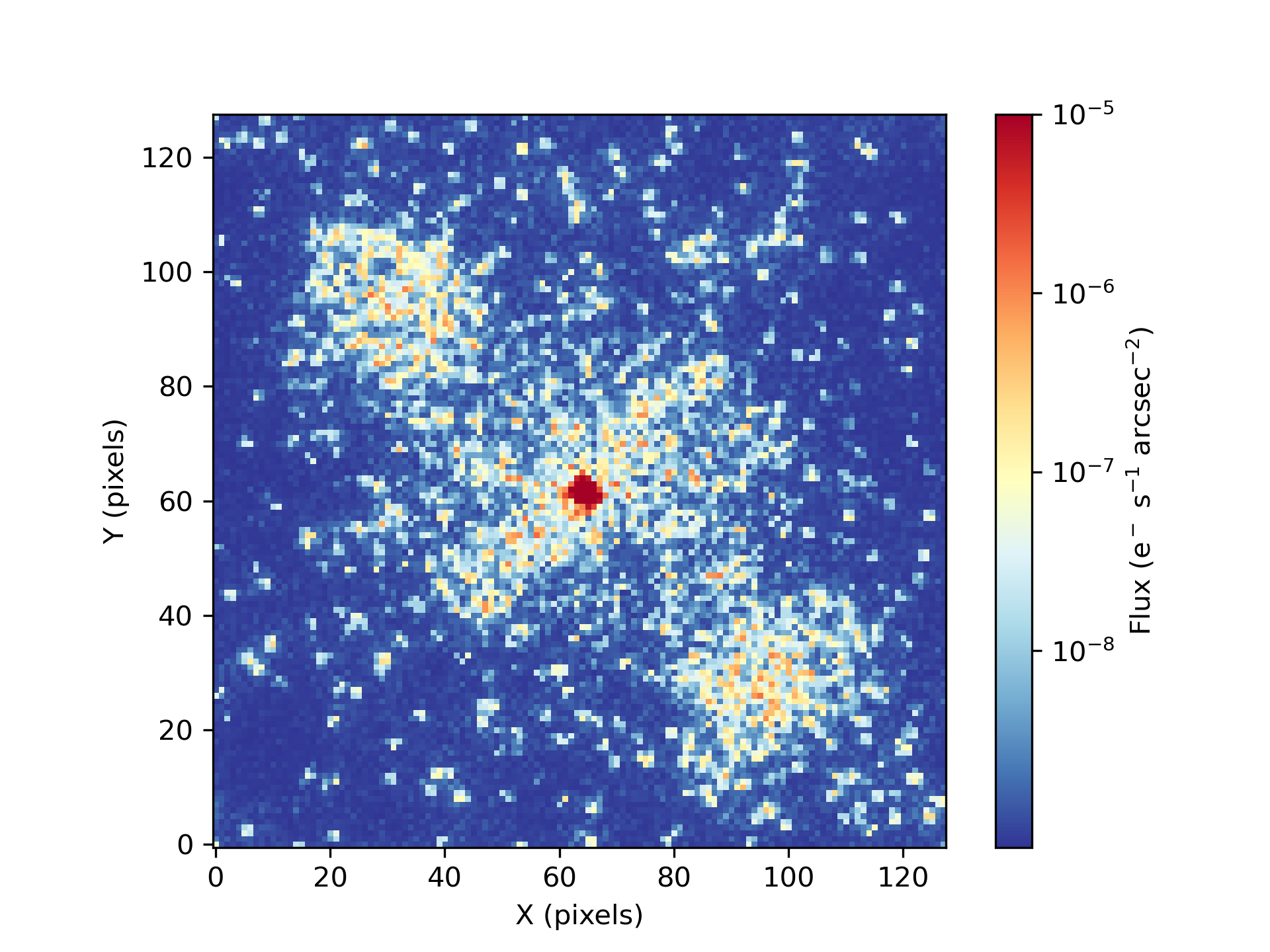}
\put(122.5,188){\color{black} \colorbox{white}{\textsf{LIRA deconvolved}}}
\end{overpic}
\begin{overpic}[trim={35 0 35 35}, clip, width=0.495\textwidth]{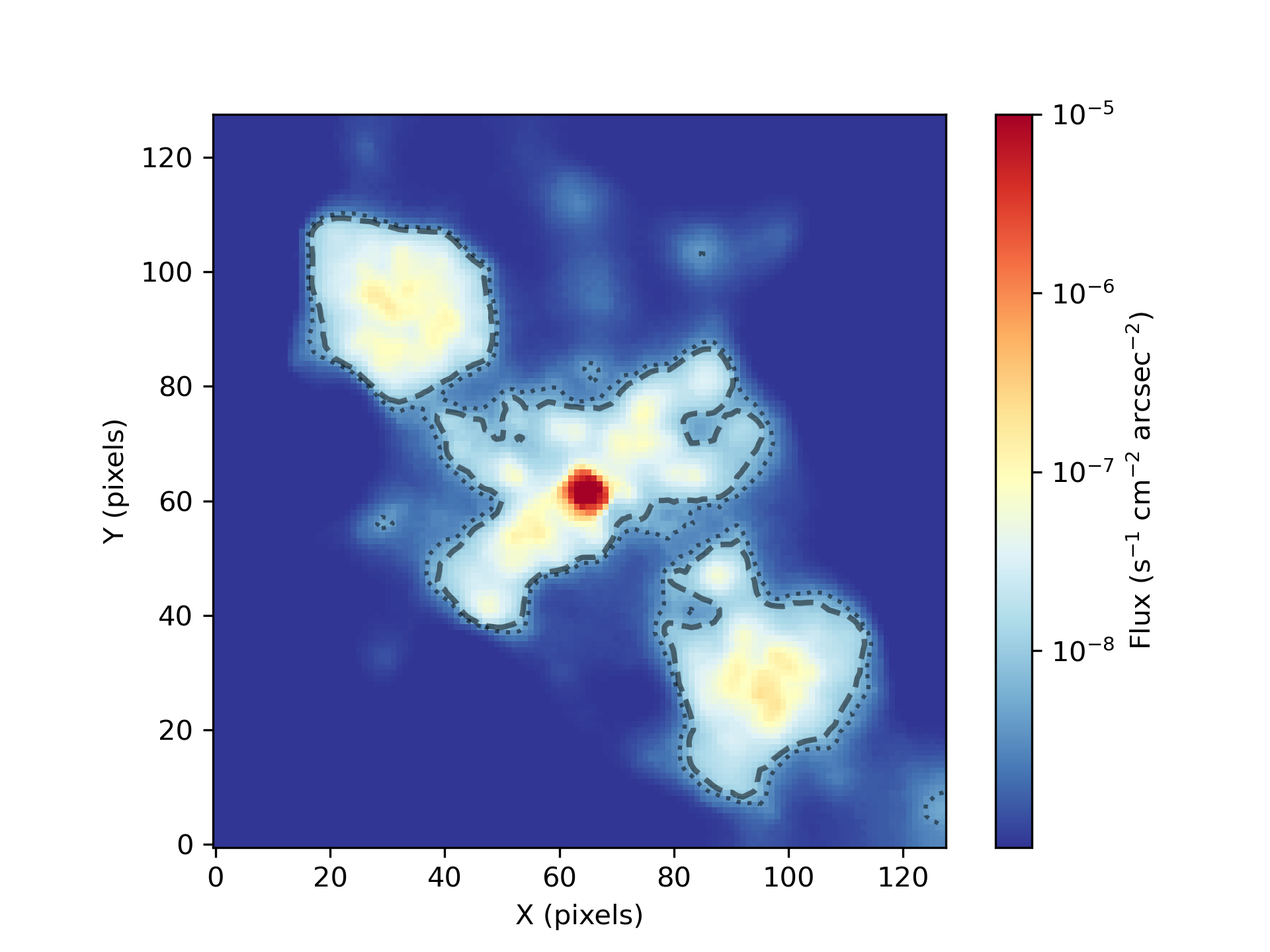}
\put(103,188){\color{black} \colorbox{white}{\textsf{Final smoothed mosaic}}}
\end{overpic}
\vspace{-0.5cm}
\caption{\SAUNAS\ analysis test on a synthetic dataset. \emph{Top left:} Underlying distribution of the simulated test source. \emph{Top right:} Point spread function (PSF) of the simulated observation. \emph{Central left:} Simulated underlying distribution of the test source convolved by the PSF. \emph{Central right:} Simulated observed events based on the PSF-convolved distribution. \emph{Bottom left:} LIRA PSF deconvolved average posterior image. \emph{Bottom right:} Adaptively smoothed final mosaic. Dashed contours represent the $3\sigma$ and dotted contours the $2\sigma$ detection level of X-ray emission. The equivalent exposure time for this test is $\tau_{\rm exp} = 5\times10^6$\scm.}
\label{fig:test_model}
\end{center}
\end{figure*}

\SAUNAS\ generates two main products: a) PSF-deconvolved X-ray adaptively smoothed surface brightness maps and b) signal-to-noise ratio (SNR) detection maps. The X-ray adaptively smoothed surface brightness maps provide the flux and luminosity of the hot gas X-ray halos, while the SNR detection maps provide the probability that the flux associated with each region on those maps is statistically higher than the local X-ray background noise. \par 

\SAUNAS\ creates these products in four major steps (see Fig.\,\ref{fig:saunas_flowchart}): 1) pre-processing of the archival \Chandra\ X-ray observations using the \emph{Chandra Interactive Analysis of Observations}\footnote{\texttt{CIAO}: Chandra Interactive Analysis of Observations \url{https://cxc.cfa.harvard.edu/ciao/}} software, \citep[][\ciao, hereafter, see Sec.\,\ref{subsubsec:methods_saunas_ciao}]{fruscione+2006inproceedings_62701V}, 2) statistical resampling of the X-ray detection events by bootstrapping\footnote{Bootstrapping: \url{https://link.springer.com/chapter/10.1007/978-1-4612-4380-9_41}}, 3) PSF deconvolution of the event maps using the Bayesian Markov Chain Monte Carlo (MCMC) \LIRA\ tool (\emph{Low-counts Image Reconstruction and Analysis}, \citet{donath+2022inproceedings_108}; see  Sec.\,\ref{subsubsec:methods_saunas_LIRA})\footnote{\LIRA: Low-counts Image Reconstruction and Analysis - \url{https://pypi.org/project/pylira/}}, and 4) adaptive smoothing using \texttt{VorBin} (see Sec.\,\ref{subsubsec:methods_saunas_voronoi})\footnote{\texttt{VorBin}: Adaptive Voronoi Binning of Two Dimensional Data - \url{https://pypi.org/project/vorbin/}}. \SAUNAS\ requires a few user-input parameters, including the location of the target ($\alpha$, $\delta$), field-of-view (FOV), and energy band. The main steps of the pipeline are described in the following subsections.

\subsubsection{CIAO pre-processing} \label{subsubsec:methods_saunas_ciao}

First, the data is pre-processed using \ciao\ in the following way: 

\begin{enumerate}
    \item All available \Chandra/ACIS observations containing the user-supplied sky coordinates are identified using \texttt{find\_chandra\_obsid}. The datasets and their best available calibration files are automatically downloaded using (\texttt{download\_chandra\_obsid} and \texttt{download\_obsid\_caldb}).

    \item The raw observations are reprocessed using \texttt{chandra\_repro} (v4.16). To avoid over-subtraction of both the source and background counts necessary for the statistical analysis, the particle background cleaning subprocess is set  (\texttt{check\_vf\_pha}) to ``no". See the main \ciao\ manual\footnote{ACIS VFAINT Background Cleaning: \url{https://cxc.harvard.edu/ciao/why/aciscleanvf.html}} for more information on this step.   

    \item All the available ACIS datasets are merged into a single events file (\texttt{merge\_obs}). This product serves as the phase~1 (first pass) observation file and is used to identify emission regions and to determine the source spectra needed for PSF construction. 

    \item The phase~1 merged observation file is used to define the angular extent of detected emission sufficient for basic spectral characterization. The spectral information is used in the step following this one. The  \texttt{VorBin} \citep{cappellari+2003mnras342_345} library generates a map of Voronoi bins, from which and a surface brightness profile is constructed. The preliminary detection radius ($R_{\rm lim, 0}$), defined as the radial limit having a surface brightness equal to $10\%$ of the surface brightness at the central coordinates, is computed. If $R_{\rm lim, 0}$ is undefined due to a low central surface brightness, the presence of detectable emission is unlikely. For such cases, $R_{\rm lim, 0}$ is arbitrarily set to 1/4 of the FOV defined by the user. 
    The events inside this detection radius are used to construct a spectrum employed in the next step to define the deconvolution kernel (e.g., PSF) appropriate for this target. The choice of a $10\%$ limit is an optimal compromise based on the analysis of \Chandra/ACIS observations: including as much emission as possible from the source enhances the spectra used to generate the PSF. However, including a region too large reduces computational efficiency. Note that the spectrum derived in this step serves the sole purpose of informing PSF construction and is not intended for physical characterization of the gas.    

    \item \ciao 's task \texttt{simulate\_psf}, in combination with the spectral information provided by the previous step, is used to generate a PSF representative of each observing visit to the target. The PSF modeling is dependent on the spectra of both the source and the background region, as well as the target position within the detector (off-axis angle). The latter is unique to each visit. The preliminary detection radius defines both the circular ($R<R_{\rm lim, 0}$) and annular ($R_{\rm lim, 0}<R<2\,R_{\rm lim, 0}$) apertures used to measure the source and background spectra, respectively (\texttt{specextract}). The \texttt{aspectblur}\footnote{\texttt{Aspectblur} in \ciao: \url{https://cxc.cfa.harvard.edu/ciao/why/aspectblur.html}} is set to 0.25, and the number of iterations to 1000 per dataset.

    \item Finally, the individual event files and PSFs corresponding to each visit are cropped to a cutout, with the preferred energy range selected.  
\end{enumerate}

The outputs from the pre-processing procedure with \ciao\, described above, are: 1) the detected event maps (named \texttt{obsid\_Elow-upper\_flt\_evt.fits}, where \texttt{low} and \texttt{upper} refer to the energy range limits and \texttt{obsid} is the observation ID identification in the \Chandra\ archive), 2) the exposure time maps (\texttt{obsid\_Elow-upper\_flt\_expmap.fits}), 3) the flux maps (\texttt{obsid\_Elow-upper\_flt\_flux.fits}), and 4) the PSF (\texttt{obsid\_Elow-upper\_flt\_psf.fits}). This set of intermediate files is used in the remaining steps of the \SAUNAS\ pipeline to generate the final maps.

\subsubsection{X-ray event resampling: bootstrapping}
\label{subsubsec:methods_saunas_bootstrapping}

The X-ray sky background is a very low count regime. \citet{bartalucci+2014aap566_25} obtained a flux of $10.2^{+0.5}_{-0.4}\times10^{-13}$\ergcmdegs\ for the 1--2~keV band and 3.8$\pm{0.2}\times10^{-12}$\ergcmdegs\ for the 2--8~keV band. This flux is equivalent to $\sim$ 0.03--0.003 photons~arcsec$^{-2}$ (1--2~keV) and $\sim$ 0.01--0.001~photons~arcsec$^{-2}$ for a typical $t=10^4$--$10^5$~s exposure \citep{she+2017apj835_223}. As a consequence, the signal from spurious groups of a few counts can dominate the shape of the Voronoi bins used for adaptive smoothing in each simulation if appropriate statistical methods are not implemented.  \par 

To enhance the robustness of the adaptively smoothed mosaics and to reduce contamination from non-significant signal in the background, the X-ray events are re-sampled via replacement (\emph{bootstrapping}) as an additional (and user-optional) step before deconvolution. Bootstrapping is especially well-suited for inferring the uncertainties associated with an estimand -- such as the median surface brightness in cases for which the Gaussian standard deviation regime does not apply or parametric solutions are too complicated or otherwise unknown. Bootstrapping effectively reduces the leverage that single events or very low count sources may have in the background of the final mosaics by accounting for the photon-noise uncertainties in the PSF deconvolution and Voronoi binning steps through a non-parametric approach, allowing for a better assessment of the uncertainties in the final simulations. \par 

In our application, bootstrapping generates $N$$\sim100$ (hereafter, $N_{\rm{boot}}$) new X-ray event samples from the observed sample, preserving size (flux) and permitting events to be repeated. While the number of bootstrapping simulations is set to 100 by default as a compromise between precision and computational resources,  $N_{\rm{boot}}$ can be defined by the user in \SAUNAS. Each resampled list of events is translated into an image, which is fed into the next step, PSF deconvolution (Sec.\,\ref{subsubsec:methods_saunas_LIRA}).

\subsubsection{LIRA PSF deconvolution}
\label{subsubsec:methods_saunas_LIRA}

The \texttt{LIRA} \citep{connors+2011inproceedings_463,donath+2022inproceedings_108} package deconvolves the emission from sources in X-ray data. Through the use of \LIRA, \SAUNAS\ removes the contamination from active galactic nuclei (AGNs) and X-ray binary stars (XRBs) which can be significantly extended and easily confused with a diffuse halo if the PSF is not accurately corrected. \LIRA\ uses a Bayesian framework to obtain the best-fit PSF convolved model to the observations, allowing the user to evaluate the probability that a detection is statistically significant. \LIRA\ was designed to provide robust statistics in the low-count Poissonian regimes representative of faint extended halos, the primary science focus of our project.

As detailed in Sec.\,\ref{subsubsec:methods_saunas_ciao}, the PSF models are generated specifically for each target, taking into account their location in the detector and their spectral energy distributions, on a per-visit basis. \SAUNAS\ deconvolves data from individual visits, using these PSF models as input into \LIRA. 
Discrete hard-band emission is produced primarily by point sources, including AGNs \citep{fabbiano+1989apj347_127, fabbiano2019inbook_7}, young stellar objects, and mass transfer onto the compact stellar object within XRB pairs \citep{wang2012inproceedings_183}. Because these point sources contaminate the soft band emission, they are excised from the data. They are identified using the \Chandra\ Source Catalog \citep{evans+2010apj189_37}, and then removed from the event file by deleting events that lay within the cataloged positional uncertainty ellipse of the source.  

The \texttt{Python} implementation of \LIRA\ is used to deconvolve the X-ray event files, thus minimizing the effects of the off-axis dependency associated with \Chandra's PSF, such that data from different visits can be combined in a later stage. \LIRA\ accepts five input arrays: a) counts (number of events), b) flux (in \scmpx), c) exposure (\scm), d) PSF, e) a first approximation to the background (counts). The first four inputs are generated by the \ciao\ pipeline (Sec.\,\ref{subsubsec:methods_saunas_ciao}), while the initial baseline background is set to one. The number of LIRA simulations is set to 1000 (\texttt{n\_iter\_max}), in addition to 100 initial burn-in simulations (\texttt{num\_burn\_in}). To speed up the process\footnote{Even in parallel processing mode, PSF deconvolution takes the largest fraction of time of the \SAUNAS\ pipeline. As a reference, in an Apple M1 Max 2021 laptop (32~Gb of RAM, 10 cores), the computation of a $1024\times1024$ mosaic typically takes two hours, with $\sim$90\% of the time spent in deconvolution.}, \SAUNAS\ splits the LIRA simulations in parallel processing blocks (defined by the number of bootstrapping simulations), to be combined after the deconvolution process has finished. While 1000 \LIRA\ simulations are run on each of the N$\sim$100 bootstrapping-resampled images described in Section\,\ref{subsubsec:methods_saunas_bootstrapping}, only the last \LIRA\ realizations (those produced after the deconvolution process has stabilized) for each resampled image are used (hereafter, $N_{\rm{stable}}$), which typically is equal to $\sim$100. 
To save computational resources, $N_{\rm{stable}}$ is adapted based on the number of bootstrapping simulations so that the deconvolved dataset consists of a maximum of $N=N_{\rm{boot}}~\times N_{\rm{stable}} =~1000$ deconvolved images (posterior samples).

\subsubsection{Adaptive Voronoi smoothing} \label{subsubsec:methods_saunas_voronoi}

The deconvolved datacubes, hereafter referred to as "Bootstrapping-\LIRA" realizations, serve as a proxy of the probability density distribution of the true X-ray emission on a pixel-per-pixel basis, at the \Chandra/ACIS spatial resolution (a minimum of 0.492" px$^{-1}$, depending on the binning set by the user). To facilitate the detection of extended, low surface brightness structures such as hot gas halos -- with apparent sizes substantially larger than the spatial resolution limit for the galaxies -- the use of spatial binning enhances the detection of regions with very low signal-to-noise ratio. 

Voronoi binning \citep[\texttt{VorBin},][] {cappellari+2003mnras342_345} is applied to each of the $N$ posterior samples in the deconvolved datacube. This process generates $N$ Voronoi tesselation maps, each one differing from the other because they were calculated from the Bootstrapping-\LIRA\ realizations. This dataset is a Voronoi map datacube representing the probability density distribution of the surface brightness of the target. 

A consequence of this binning approach is the loss of spatial resolution in the faintest regions of the image (halos, background) compared to the brightest regions (i.e., the galactic cores). This loss is caused by the fact that the Voronoi technique varies the bin size in order to achieve a fixed signal-to-noise ratio in the resulting map. As we are primarily interested in mapping the large scale halo structures, this loss in spatial resolution does not significantly impact our science goals.

A surface brightness map is created by calculating the median across one of the axes of the Voronoi datacube. To prevent background emission from contaminating the final image, the scalar background level is determined individually for each realization of the Bootstrapping-\LIRA\ datacube. All sources, both resolved and unresolved, must be meticulously masked prior to measuring the background level, to prevent systematically over-subtracting the background in the final mosaics. The source masking and background correction process are conducted iteratively:

\begin{enumerate}
    \item After the \LIRA\ deconvolution process and before the Voronoi binning is performed, point sources from the Chandra Source Catalog (CSC 2.0;\footnote{\Chandra\ Source Catalog Release 2.0: \url{https://cxc.harvard.edu/csc2/}} likely X-ray binaries, SNe, AGNs) that lay in the image footprint are removed from the associated event file. Point source removal prevents the associated emission from impacting the adaptive Voronoi maps and resulting in diffuse contamination that could be confused with a gas halo component. 

    \item A secondary mask is generated using \ciao's routine \texttt{vtpdetect}\footnote{\texttt{CIAO/vtpdetect:} \url{https://cxc.cfa.harvard.edu/ciao/ahelp/vtpdetect.html}}. This mask identifies the regions with detectable extended X-ray emission that are removed from the maps before measuring the background level. A mask is generated for each CCD of each visit through independent analysis. The masks are then combined into a single master extended source mask.

    \item If a source was detected in the preliminary surface brightness profile generated a part of the \ciao\ pre-processing step (see Sec.\,\ref{subsubsec:methods_saunas_ciao}, step 4), then those pixels with $R<R_{\rm lim, 0}$ are also masked before the background assessment. 
 
    \item After removing all the masked pixels using the masks from the three previous steps, the first approximation of the background level ($B_0$) is made by measuring the median value of the unmasked sigma clipped ($\sigma=3$) pixels. The background value is then subtracted from the voronoi binned maps.
    
\end{enumerate}

Once the individual observations have been background corrected, all the flux maps are combined using mean-weighting by the respective exposure times. Finally, a refined background value ($B_1$) is calculated using the combined observations by repeating the process described above. The noise level is then estimated from the background distribution as the ratio between the median background level and the lower limit of the $1\sigma$ error bar (equivalent to the $15.8\%$ percentile). The final background-subtracted, PSF-corrected, and Voronoi binned surface brightness maps are derived by using a median of the background-corrected bootstrapping-\LIRA\ realizations. The final mosaics and the noise level are used to generate three different frames to be stored in the final products: 1) an average adaptive X-ray surface brightness map, 2) a noise level map, and 3) an SNR map.

\subsection{Quality tests} \label{subsec:methods_test}

\begin{figure*}[t!]
\begin{center}
\resizebox{\linewidth}{!}{
\begin{overpic}[trim={80 0 100 40}, clip, height=7cm]{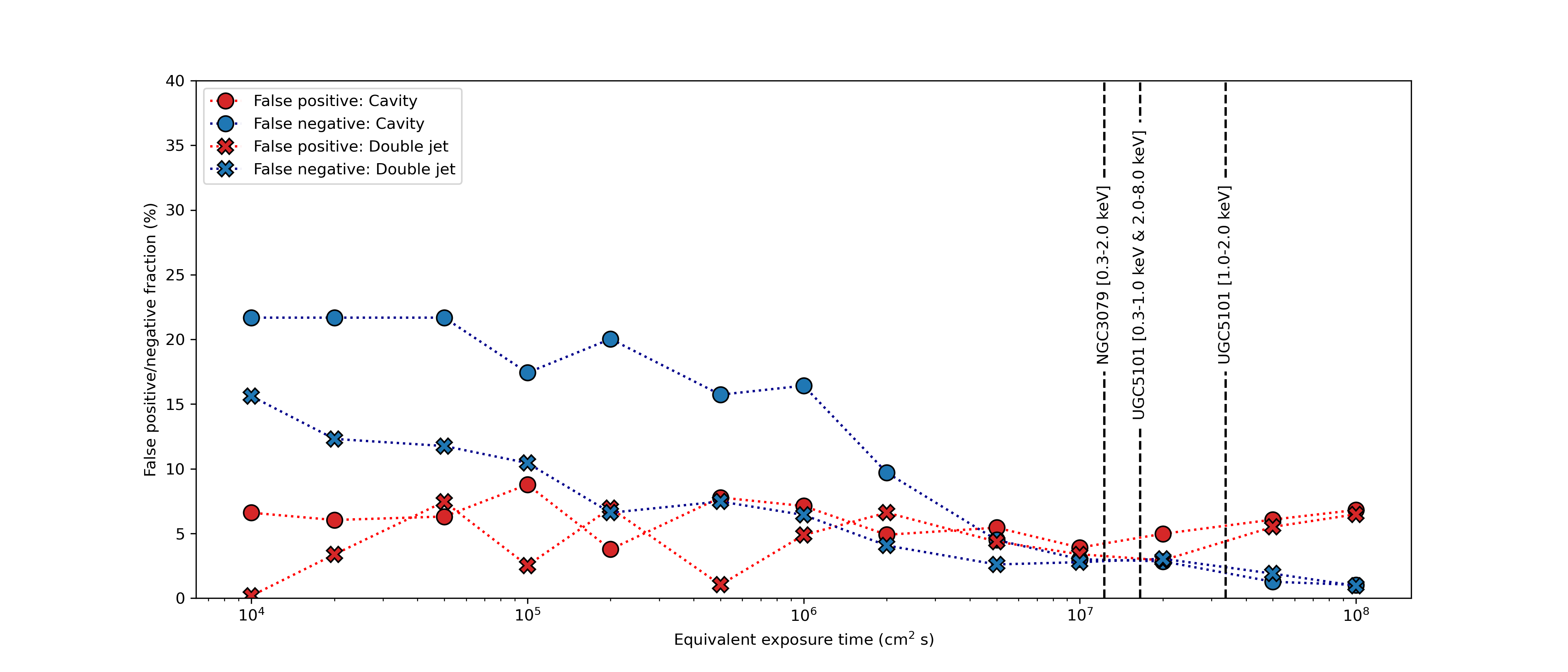}
\end{overpic}
}
\caption{Fraction of false positives and false negatives in the \SAUNAS\ detection maps derived from two truth models as a function of the equivalent exposure time (cm$^2$ s). Blue symbols and lines represent the fraction of false negatives, while red represents the fraction of false positive detections in the mock maps. Cross symbols correspond to the double jet model and filled circles represent the cavity model (see Table \ref{tab:test_model_properties}). Vertical dashed lines indicate the median equivalent exposure times for the analyzed \emph{real} observations in their respective bands.} 
\label{fig:FPFN_test}
\end{center}
\end{figure*}

This section presents the results from a series of quality tests designed to  evaluate specific aspects of the output mosaics generated with \SAUNAS:

\begin{enumerate}
    \item Identify the fraction of false positives and false negative detections (Sec.\,\ref{subsubsec:quality_FPFN}). 
    \item Estimate the flux conservation of the deconvolution / Voronoi binning process (Sec.\,\ref{subsubsec:quality_Flux})
    \item Quantify the quality of \SAUNAS\ performance as compared to that of other methods (\texttt{arestore}, Sec.\,\ref{subsubsec:quality_PSFdeco}). 
\end{enumerate}

\subsubsection{False positive / False negative ratio} \label{subsubsec:quality_FPFN}

For quality assessment, \SAUNAS\ is tested using two different models varying the exposure time to reduce the photon flux and the detectability conditions:
\begin{enumerate}
    \item A model of an idealized edge-on galaxy with two lobes emerging from a jet (double jet model). 

    \item A shell-like structure with a central bright source (cavity model). 
\end{enumerate}

The models are created as combinations of Gaussian 2D probability distributions (\texttt{astropy.convolution.Gaussian2DKernel}) with different ellipticities and rotations as described in Table \ref{tab:test_model_properties}. Following PSF convolution, a synthetic observed events map is generated using a random Poisson distribution (\texttt{numpy.random.poisson}). 

The double jet model includes the emission from three sources: the galactic disk, a bright core, and the lobes. The range of surface brightnesses is  $\sim$ 10$^{-6}$--10$^{-8}$ \escmarc, excluding the considerably brighter (three to five orders of magnitude brighter) peak surface brightness of the core. Its morphology mimics the predominant structure observed in double jet radio galaxies such as Centaurus\,A \citep{hardcastle+2007apj670_81}. 

The other test simulation, a cavity model, contains a hollow shell with a central bright source. This model provides an important pipeline test 
for the reconstruction of cavities found in the intergalactic medium. The detection of cavity rims seen in projection against the diffuse emission from the hot intracluster and/or intergalactic medium is challenging. These large bubbles potentially provide a useful record of interactions between AGNs and the intergalactic medium, in which the expansion of the associated radio lobes excavate the surrounding medium \citep{pandge+2021mnras504_1644}. Our test model is designed to be particularly challenging: an X-ray cavity with a dominant central source representing an AGN 
\citep{blanton+2001apj558_15, dunn+2010mnras404_180}. The surface brightness background level of both models is fixed at $5\times10^{-9}$ \escmarc, and the equivalent exposure time is assumed to be flat an varying from $\tau_{\rm exp} = 10^8$\ to $\tau_{\rm exp} = 10^4$\scm. For reference, $\tau_{\rm exp} = 5\times10^5$\scm, equals $\sim$10~ks at $0.3-1.0$~keV band\footnote{\Chandra\ variation of effective area with energy \url{https://cxc.cfa.harvard.edu/proposer/POG/html/INTRO.html}}.

\begin{deluxetable*}{cccccc}
\tabletypesize{\footnotesize}
\tablecolumns{7}
\tablewidth{0pt}
\vspace{-0cm}
\tablecaption{Photometric and structural properties of the synthetic test models. Columns: 1) Name, 2) Component, 3) Size, 4) Surface brightness, 5) Eccentricity, 6) Position angle. \label{tab:test_model_properties}}
\tablehead{\thead[t]{Mock model} & \thead[t]{Component} & \thead[t]{Size} & \thead[t]{$\mu$} & \colhead{$q$} & \thead[t]{PA}\\
\colhead{(1)} & \colhead{(2)} & \colhead{(3)} & \colhead{(4)} & \colhead{(5)} & \colhead{(6)} \\ 
& \colhead{} & \colhead{\makecell{($\sigma_{x}, \sigma_{y}$, pixels)}} &  \colhead{\makecell{[s$^{-1}$ cm$^{-2}$ arcsec$^{-2}$]}} & \colhead{} & \colhead{[$^{\circ}$]}} 
\vspace{0.5cm}
\startdata 
Double jet & Core & 1,1 & 1.2$\times$10$^{-4}$ & 1 & 0 \\
 & Disk & 15,3 & $\sim$ 10$^{-6}$--10$^{-8}$ & 0.2 & 135\\
 & Lobes & 7,7 & $\sim$ 10$^{-6}$--10$^{-8}$ & 1 & 0\\
 & Jet & 25,2 & 5$\times$10$^{-7}$ & 0.08 & 45\\
 & Background & -- & 5$\times$10$^{-9}$ & -- & -- \\
\hline
Cavity & Core & 1,1 & 1.2$\times$10$^{-4}$ & 1 & 0 \\
 & Shell & [30--45] & $\sim$ 10$^{-7}$ & 1 & 0\\
 & Background & -- & 5$\times$10$^{-9}$ & -- & -- \\
\enddata
\end{deluxetable*}

The synthetic data are generated using the real PSF associated with the \Chandra/ACIS datasets of NGC\,3656 \citep[Arp 155, PID:10610775, Fabiano, G.,][]{smith+2012aj143_144}. This PSF, which displays the characteristic ellipsoid pattern of off-axis ACIS observations, is selected as a worst-case scenario, given its extreme ellipticity due to its off-axis position in the detector array. The readout streak\footnote{\Chandra/ACIS PSF: \url{https://cxc.cfa.harvard.edu/ciao/PSFs/psf_central.html}} is visible as a spike departing from the center of the PSF at a position angle of -70$^{\circ}$ approximately (North $= 0^{\circ}$, positive counter-clockwise). 

The simulated observed events are passed to the \SAUNAS\ pipeline for processing, followed by a comparison between the detected ($3\sigma$) maps and  truth models. The quantitative quality test includes identification of  the fraction of pixels that were incorrectly identified as false negatives (FN) and false positives (FP).

Fig.\,\ref{fig:test_model} demonstrates the deconvolution and smoothing process for a mock galaxy with $\tau_{\rm exp} = 5\times10^6$ s, having both diffuse X-ray emission and an extended PSF. The position angle selected for the model galaxy (Table \ref{tab:test_model_properties}) is selected specifically to offer a nontrivial test for the PSF deconvolution method. By using a position angle of $45^{\circ}$, the resulting convolved image displays two elongated features with apparently similar intensity (central left panel in Fig.\,\ref{fig:test_model}, PSF convolved source): one real, and one created by the PSF. If the PSF elongated feature is removed in the final images, we can conclude that the image reconstruction was successful. 

After Poisson sampling (see Simulated observation panel in Fig.\,\ref{fig:test_model}), the resulting events map is equivalent to the processed \ciao\ event files. The events map shows broad emission for the core of the galaxy model in which the disk is indistinguishable. The two lobes are still present, but considerably blended with the emission from the inner regions. The events are then processed using \SAUNAS\ (\LIRA\ deconvolution, Bootstrapping, and \vorbin\ steps). 

The results from the PSF deconvolution (\LIRA\ deconvolved panel in Fig.\,\ref{fig:test_model}) show a removal of most of the PSF emission, recovering the signal from the disk of the galaxy and removing the PSF spike emission. However, a significant amount of noise is still visible, and the background level is difficult to estimate (lower left panel of Fig.\,\ref{fig:test_model}). 

After applying the bootstrapping and Voronoi binning methods, the resulting final corrected mosaic (final smoothed mosaic panel, Fig.\,\ref{fig:test_model}) clearly shows the signal from the X-ray lobes, the disk, and the central bright core over the background. The $2\sigma$ and $3\sigma$ contours show the detected features following the calibration procedures described in Sec.\,\ref{sec:methods}, demonstrating complete removal of the PSF streak in the final mosaics (at a 99.7$\%$ of confidence level). The original shape and orientation of the disk is recovered, with the flux correctly deconvolved into the bright core of the model galaxy. Due to its dim brightness, the jet that connects the lobes with the main disk is notably distorted in the final mosaic, but still visible at a $2\sigma$ confidence level. For this test, the fraction of pixels unrecovered by the pipeline that were part of the model sources (false negatives, FN) is $FN=3.2\%$. On the other hand, the fraction of misidentified pixels that were part of the background (false positives, FP) is $FP=4.0\%$. The maps of false positives and false negatives for this test are available in Appendix \ref{Appendix:SAUNAS_extended_psf_deconvolution_test}.

The test for the cavity model is repeated, sampling different equivalent exposure times. The results are shown in Appendix \ref{Appendix:SAUNAS_extended_psf_deconvolution_test}. Fig.\,\ref{fig:FPFN_test} presents a comparison of the false positive and false negative fraction as a function of the equivalent exposure time and model. For equivalent exposure times higher than $\tau_{\rm exp} = 10^6$\scm,  the FP and FN are lower than 5--10\%. These fractions increase towards shorter exposures as expected, showing a notable increase to 20$\%$ of false negatives (true source emission that is unrecovered by \SAUNAS) at approximately $\tau_{\rm exp} = 5\cdot10^5$\scm. The reason for this increase is the lack of detection of the dimmer outer regions in contrast with the brighter core (the lobes in the case of the double jet model, and the outer shell in the cavity model). Interestingly, the fraction of false positives does not increase substantially even at extremely low equivalent exposure times, remaining stable at $\sim10\%$ down to $\tau_{\rm exp} < 10^4$\scm. \emph{This result demonstrates that even in cases of extremely short exposure times, \SAUNAS\ is not expected to generate false positive detections, which is a critical requirement for our study.}

\subsubsection{Flux conservation} \label{subsubsec:quality_Flux}

In an ideal scenario, the total flux of the events processed by \SAUNAS\ should be equal to the total flux in the pre-processed frames by \ciao. In practice, the baseline model assumptions during the deconvolution process may affect the total flux in the resulting frames. \LIRA\ assumes a flat background model that--combined with the counts in the source--tries to fit all the events in the image. However, deviations from this ideal scenario (non-uniform background, regions with different exposure time) generate differences between the input and output flux. In order to understand the impact of flux conservation in \LIRA\ deconvolved images, we must 1) analyze the relative difference of flux before and after deconvolution, and 2) determine if the residuals of the deconvolution process generate any systematic artificial structure (i.e., photons may be preferentially lost around bright sources, generating holes in the image or erasing the dim signal from halos). 

\begin{figure}[t!]
\begin{center}
\resizebox{\linewidth}{!}{
\begin{overpic}[trim={0 20 40 0}]{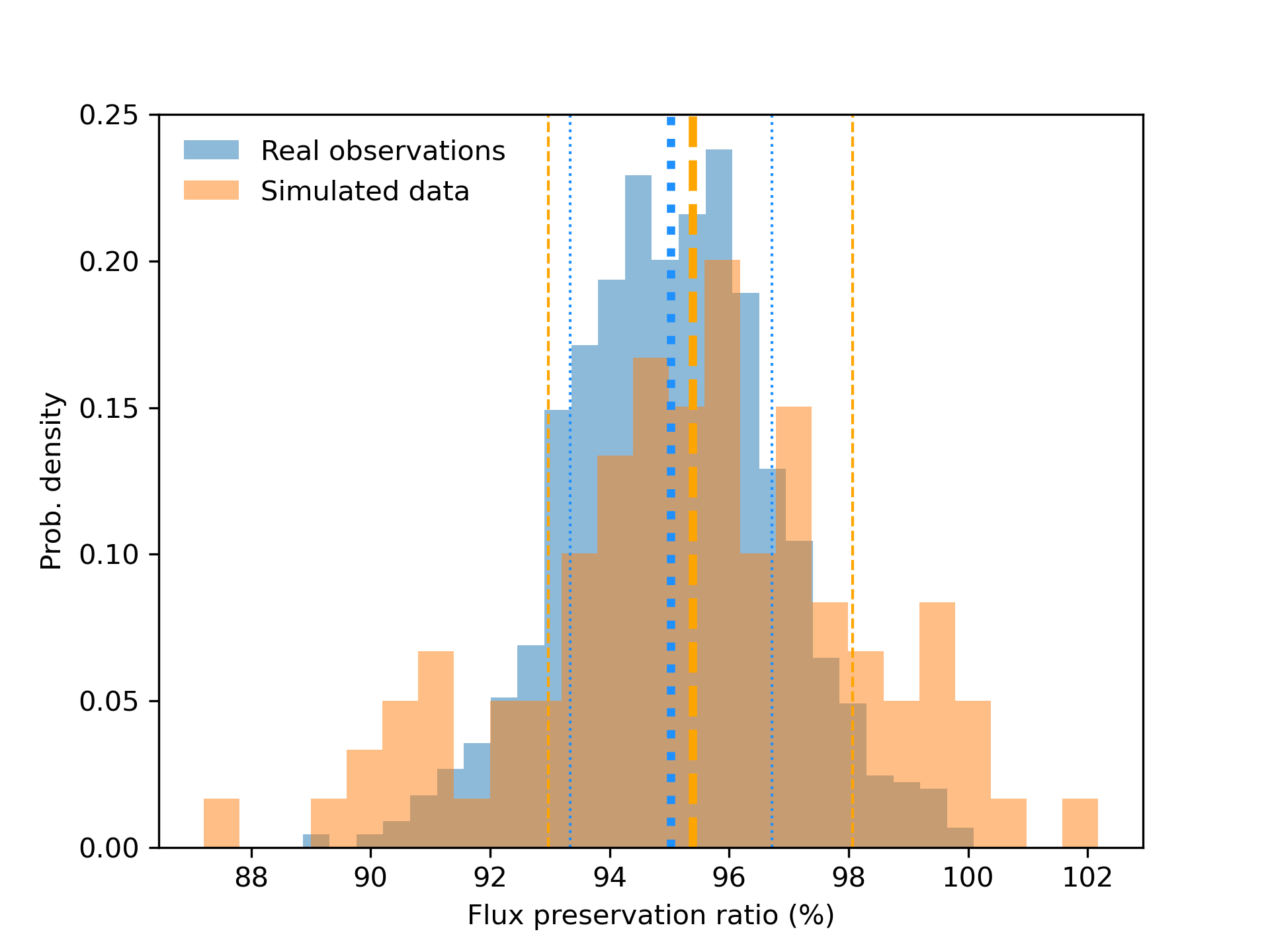}
\end{overpic}
}
\caption{Flux conservation in \SAUNAS\ frames. The histogram represents the probability distribution of the ratio between the recovered flux after \SAUNAS\ processing and the total flux of the input, pre-processed frames.} 
\label{fig:flux_conservation}
\end{center}
\end{figure}

Total flux conservation is tested by measuring the ratio between the total flux in the input frames (those obtained at the end of the \ciao\ pre-processing, see Sec.\,\ref{subsubsec:methods_saunas_ciao}) divided by the total flux in the final, \SAUNAS\ processed frames. We perform this test on real (UGC\,5101, see Sec.\,\ref{subsec:UGC5101}) and synthetic observations (Sec.\,\ref{subsubsec:quality_FPFN}). The results are shown in Fig.\,\ref{fig:flux_conservation}.

A total flux loss of $\sim5\%$ is detected in the \SAUNAS\ processed frames when compared with the pre-processed event maps by \ciao. The results are consistent in real observations (recovered flux ratio of $95.0\pm1.7$\%) and in synthetic observations ($95.4^{+2.7}_{-2.4}$\%). Using different simulations, we determined that this small flux loss is independent of the size of the FOV (in pixels), remaining stable at $\sim5\%$. For the total area of the images analyzed, a 5\% of lost flux is negligible and well within the stochastic uncertainty of typical photometry (see the error bars in the profiles described in Fig.\,\ref{fig:NGC3079_filaments}). We consider a flux conservation ratio lower than 100\% (i.e., 90\% -- 99\%) as erring on the side of caution from a statistical perspective: the bias of \LIRA\ to lose flux implies that \SAUNAS\ will not generate false positive detections of hot gas halos. 

\subsubsection{Quality PSF deconvolution test} \label{subsubsec:quality_PSFdeco}

While Sec.\,\ref{subsubsec:quality_Flux} reported on the conservation of total flux in the image as a whole, this section discusses whether \SAUNAS\ introduces unwanted artificial structures (fake halos, or oversubtracted regions) in the processed maps. For this test, two additional types of test sources are used: 1) a point source, and 2) a circular extended source. Both of these sources have been previously combined with a \Chandra/ACIS PSF. To provide context, the results of \LIRA\ are compared with those from \texttt{CIAO}/\texttt{arestore}\footnote{\texttt{arestore}: \url{https://cxc.cfa.harvard.edu/ciao/ahelp/arestore.html}}.  

The results are displayed in Fig.\,\ref{fig:example_deco_point_source} (point source) and Fig.\,\ref{fig:example_deco_circle} (circular extended source) and detailed in Appendices \ref{Appendix:SAUNAS_psf_deconvolution_test} and \ref{Appendix:SAUNAS_extended_psf_deconvolution_test}. To quantify the quality of the different deconvolution methods, radial surface brightness profiles of the truth (non convolved) model, the convolved simulated observations, and the resulting deconvolved maps are constructed. The profiles show that \texttt{arestore} tends to oversubtract the PSF, generating regions of negative flux around the simulated source. In the point source case scenario, \texttt{arestore} oversubtracts the background by more than $5\times10^{-8}$\scmpx, while \LIRA\ recovers the background level with five times less residuals. The superiority of \LIRA\ over \texttt{arestore} to recover diffuse structures is even more obvious in the extended source scenario (Fig.\,\ref{fig:example_deco_circle}):  \texttt{arestore} shows a clear oversubtraction ring-like region around the source, dipping the background level to  $10^{-7.8}$\scmpx\ as compared to the real (truth model) level of $10^{-7}$\scmpx. \LIRA\ fits the background level significantly more faithfully, at a level of $\sim 10^{-7.2}$\scmpx.

We conclude that \LIRA\ deconvolution results are better suited for the detection of diffuse X-ray emission, such as extended hot gas halos, compared to other PSF correction techniques, such as \texttt{CIAO}'s \texttt{arestore}. Despite the model limitations described in Sec.\,\ref{subsubsec:quality_Flux}, \SAUNAS\ suppresses false positive extended emission detections without over-fitting the PSF, while recovering the true morphologies of X-ray hot gas distributions. Thanks to the modularity of \SAUNAS, future updates in the \LIRA\ deconvolution software will be automatically implemented in our pipeline, improving the quality of the processed frames. 

\section{Application to real observations} \label{sec:results}

\subsection{Sample selection}\label{subsec:sample_selection}

We identified two astrophysical targets of interest for testing the pipeline:
\begin{enumerate}
    \item NGC\,3079, a highly inclined barred spiral galaxy with a prominent Fermi bubble \citep[][the primary benchmarking target, see Sec.\,\ref{subsec:NGC3079}]{hodgeskluck+2020apj903_35}.

    \item UGC\,5101, an ultra-luminous IR galaxy that is undergoing a galactic merger \citep[][the secondary benchmarking target, see Sec.\,\ref{subsec:UGC5101}]{sanders+1988apj325_74,imanishi+2001apj558_93}.


 

\end{enumerate}

The targets used to demonstrate \SAUNAS\ capabilities were selected because they were known \emph{apriori} to have extended soft X-ray emission detected by telescopes other than \Chandra\ (NGC\,3079), and the characterization of the extended emission was well-documented with a detailed methodology that could be replicated in the published research. Insisting that the data come from a different platform provides a truth model independent of systematic effects inherently associated with \Chandra. Finally, these specific targets were selected in order to test \SAUNAS\ against simple and complex emission structures associated with the different morphologies (a disk galaxy and an interacting system).

\subsection{NGC\,3079} \label{subsec:NGC3079}

\begin{figure*}
\centering

\begin{overpic}[trim={0 0 0 0}, clip, width=\textwidth]{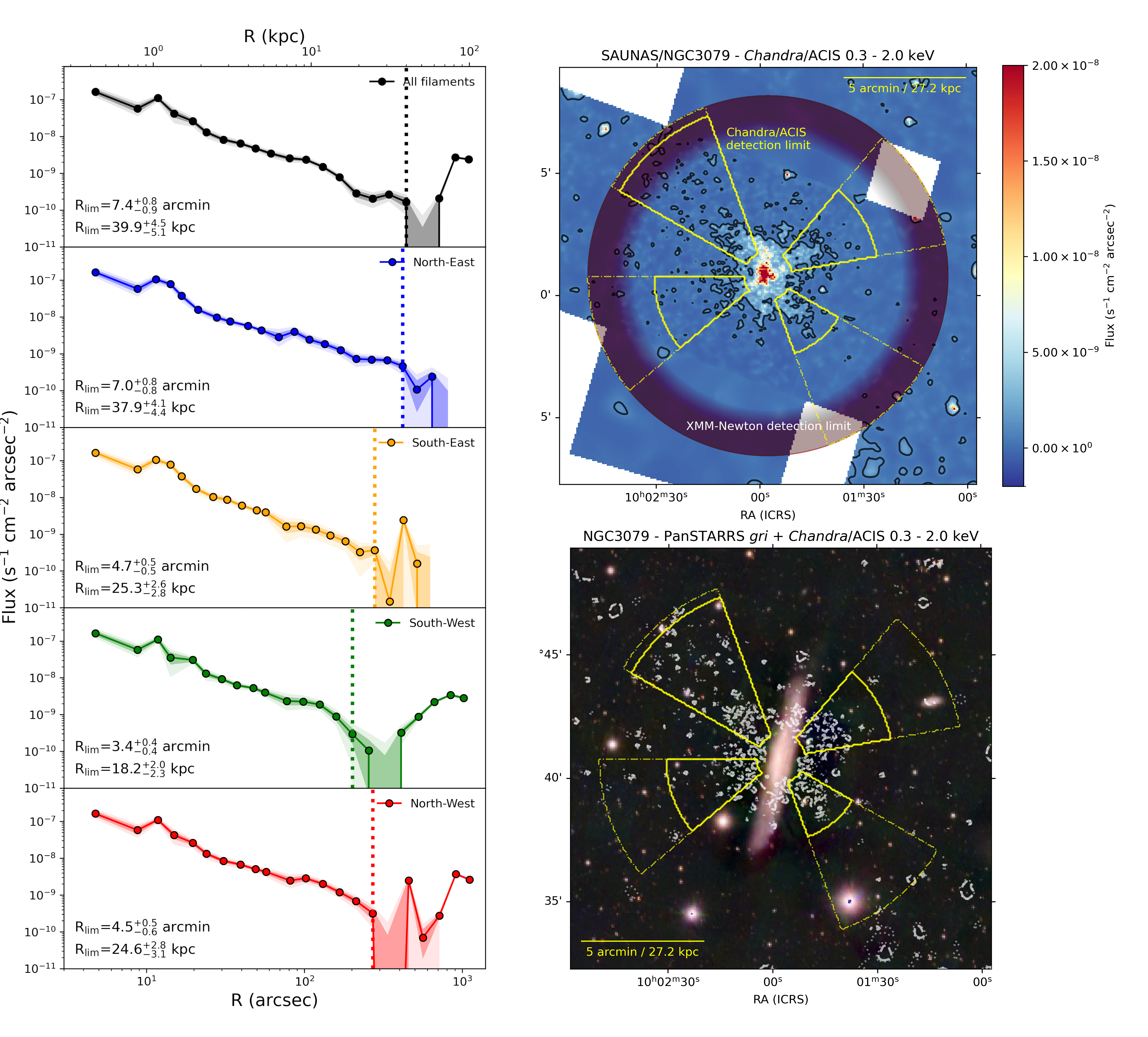}
\put(10,440){\color{black} \Large{\textsf{a)}}}
\put(230,440){\color{black} \Large{\textsf{b)}}}
\put(230,220){\color{black} \Large{\textsf{c)}}}
\end{overpic}

\caption{Extended X-ray wind cones in NGC\,3079, recovered in the \Chandra/ACIS observations using \SAUNAS. \emph{a)} Broad-band (\emph{Chandra}: 0.3--2.0~keV) surface brightness profiles of the four filaments identified by \citet{hodgeskluck+2020apj903_35} using XMM-\emph{Newton} and \emph{GALEX} observations. \emph{Top to bottom:} All filaments, north-east, south-east, south-west, and north-west. Radial detection limits are in the panels (95\% confidence level). \emph{b)} \SAUNAS\ processed image showing $2\sigma$ contours (black, shown in white in panel c)) with filament sectors in yellow. The radial detection limit indicated in panel a) for each of the four filaments  is shown as solid yellow sectors, while that of ``all filaments'' is shown as dashed yellow, following the methodology found in \citet{hodgeskluck+2020apj903_35}. The thick magenta circle in b) shows the maximum detection limit found with XMM-\emph{Newton}, compatible with our results. \emph{c)} Comparison of the optical morphology (Pan-STARRS $gri$) of NGC\,3079 with the extended X-ray emission.}
\label{fig:NGC3079_filaments}
\end{figure*}

Large-scale bipolar winds, Fermi and radio bubbles, are examples of extended structures observed around the center of the Milky Way in multi-wavelength observations, including radio (MeerKAT, S-PASS), microwave (WMAP), mid-infrared (\emph{MSX}), UV (XMM), X-rays (\Chandra, XMM-\emph{Newton}, \emph{ROSAT}) and gamma rays (\emph{Fermi}-LAT) \citep[][]{sofue1995pasj47_527, blandhawthorn+2003apj582_246, su+2010apj724_1044, finkbeiner2004apj614_186, carretti+2013nat493_66, heywood+2019nat573_235}. While the presence of these structures is well-known in our own galaxy, \citet{li+2019apj873_27} reported the first non-thermal hard X-ray detection of a Fermi bubble in an external galaxy, NGC\,3079 \citep[$\alpha=150.491^{\circ}$, $\delta=+55.680^{\circ}$, $D=18.68\pm1.32$~Mpc, 11.04~arcsec~kpc$^{-1}$][]{springob+2005apj160_149}, using \Chandra\ observations. Further works in X-ray and UV using XMM-\emph{Newton} and GALEX revealed a 30~kpc long X-ray Galactic Wind Cone in NGC\,3079 \citep[up to 60~kpc in FUV,][]{hodgeskluck+2020apj903_35}, potentially associated with material that has been shocked by Type~II supernovae. 

The length of the X-ray wind cone of NGC\,3079 ($R\sim3$ arcmin, 16.3~kpc) contrasts with that of the bubble found by \citet{li+2019apj873_27} using \Chandra\ observations ($R\lesssim0.75$~arcmin, 4.1~kpc). \citet{hodgeskluck+2020apj903_35} argued that the sensitivity of the longest \Chandra\ observations in the soft X-ray band ($E < 1$~keV) is affected by the molecular contaminant buildup on the detector window, and as a consequence, these \Chandra/ACIS observations were only used for point source identification on NGC\,3079 and subsequent masking for XMM-\emph{Newton}. \par 

Additionally, the available \Chandra\ observations were much shallower (124.2~ks, with only 26.6~ks of usable exposure time due to contamination) than those of XMM (300.6~ks). Despite Fig.\,6 in \citet{hodgeskluck+2020apj903_35} showing signs of faint extended emission in the \Chandra/ACIS datasets, the authors did not attempt to characterize it. Because ancillary X-ray observations from XMM-\emph{Newton} are available for this object, NGC\,3079 is an ideal case for benchmarking the low surface brightness recovery capabilities of the \SAUNAS\ pipeline.

To detect the X-ray galactic wind in NGC\,3079,  the same bandpass (0.3--2.0~keV) as in \citet{hodgeskluck+2020apj903_35} is used. The available \Chandra/ACIS observations of NGC\,3079 are detailed in Table \ref{tab:NGC3079_data}. Each visit was reprocessed with independent PSF deconvolution, and then the visits were combined for Voronoi binning. Observations 19307 and 20947 were processed but discarded due to the presence of very large-scale gradients and unusually high background  levels in the detectors where the main emission from NGC\,3079 is located. After processing the remaining observations (2038 and 7851) with \SAUNAS, extended emission observed by \Chandra\ is compared to the results from XMM-\emph{Newton}. The PSFs of the 2038 and 7851 observations and their unprocessed events are available in Figs.\, \ref{fig:NGC3079_psf} and \ref{fig:NGC3079_events} in Appendices \ref{Appendix:Observed_PSFs} and \ref{Appendix:Observed_events} respectively. \par 

Following the results from Fig.\,2 in \citet{hodgeskluck+2020apj903_35}, four angular cone regions display diffuse emission: north-east ($\theta=40^{\circ}$), south-east ($\theta=110^{\circ}$), south-west ($\theta=-140^{\circ}$), and north-west ($\theta=-60^{\circ}$), ($\theta$ is measured counter-clockwise, north corresponds to $0^{\circ}$, see Fig.\,\ref{fig:NGC3079_filaments}). Mimicking the methodology in the original article, an amplitude of $\pm20^{\circ}$ is set for all the cones around their central axis. Surface brightness profiles are generated from the reprocessed \Chandra\ observations, providing a direct comparison with previous results. 

The results show that the extended X-ray wind emission is detectable using \emph{Chandra} observations, up to a limit of $R_{\lim}\sim40$~kpc from the center of NGC\,3079  ($R_{\lim}=39.9^{+4.5}_{-5.1}$~kpc on average, extending up to $R_{\lim}=37.9^{+4.1}_{-4.4}$~kpc in the North-East filament) at a confidence level of 95\% (2$\sigma$). The filament in the south-west of the galaxy is shortest at $R\sim$ 16--20~kpc. Interestingly, the XMM observations reveal a slightly larger extent in the X-ray emission on the west side (40~kpc) compared to the east side (30--35~kpc) according to \citet{hodgeskluck+2020apj903_35}.\footnote{Note that the authors do not specify the details of their methodology for measuring the radial limits in their X-ray observations, but rather infer the dimensions of the X-ray filaments by visual inspection of their Fig.\,1b. In this work, we adopt a 95\% confidence level ($p=0.05$) to claim statistical significance.}. The average limiting surface brightness (95\% confidence level) is $\mu=1.66^{+0.5}_{-0.5}\times10^{-10}$~\escmarc. Limiting surface brightness reaches its lowest limit when combining all the filaments, suggesting that the observations are limited by noise and not by systematic effects (if dominated by systematic gradients, a lower SNR would result from combining all the regions).

\begin{deluxetable}{cccccc}
\tabletypesize{\footnotesize}
\tablecolumns{7}
\tablewidth{0pt}
\vspace{-0cm}
\tablecaption{\Chandra/ACIS observations available within 10~arcmins of NGC\,3079 and UGC\,5101, retrieved from the \Chandra\ Data Archive, as of February~2024 . Columns: 1) Observation ID, 2) \Chandra\ instrument, 3) total exposure time per observation, 4) observation mode, 5) average count rate, 6) exposure start date. \label{tab:NGC3079_data}}
\tablehead{
\thead[t]{Obs ID} & \thead[t]{Instrument} & \thead[t]{Exposure\\time} & \colhead{Mode} & \thead[t]{Count\\rate} & \colhead{Start date}\\
\colhead{(1)} & \colhead{(2)} & \colhead{(3)} & \colhead{(4)} & \colhead{(5)} & \colhead{(6)}\\ 
\colhead{} & \colhead{} & \colhead{[$ks$]} & \colhead{} & \colhead{[s$^{-1}$]} & \colhead{}}
\vspace{0.5cm}
\startdata
NGC\,3079 &  &  &  &  &  \\\hline
2038 & ACIS-S & 26.58 & FAINT & 10.27 & 2001-03-07 \\
7851 & ACIS-S & 5.00 & FAINT &  14.88 & 2006-12-27 \\
19307 & ACIS-S & 53.16 & FAINT &  6.14 & 2018-01-30 \\
20947 & ACIS-S & 44.48 & FAINT &  6.10 & 2018-02-01\\\hline
UGC\,5101 &  &  &  &  &  \\\hline
2033 & ACIS-S & 49.32 & FAINT & 9.53 & 2001-05-31 \\
\enddata
\end{deluxetable}





\subsection{UGC\,5101}\label{subsec:UGC5101}

UGC\,5101 \citep[$z=0.039$, $D=161.8$~Mpc, 0.784~kpc~arcsec$^{-1}$,][]{rothberg+2006aj131_185} is an irregular galaxy that is undergoing a potential major merger. This object has previously been identified as a Seyfert~1.5 \citep{sanders+1988apj325_74}, a LINER (low-ionization nuclear emission-line region) galaxy \citep{veilleux+1995apj98_171}, and a Seyfert~2 galaxy \citep{yuan+2010apj709_884}. UGC\,5101 has a very extended optical tidal tail ($\sim40$~kpc) to the west from the nucleus, with a second semicircular tidal tail that surrounds the bright core of the galaxy with a radius of 17~kpc \citep{surace+2000apj529_170}. Radio, \citep{lonsdale+2003apj592_804}, IR \citep{genzel+1998apj498_579, soifer+2000aj119_509, armus+2007apj656_148, imanishi+2001apj558_93}, and X-ray observations with \Chandra\ and XMM-\emph{Newton} \citep{ptak+2003apj592_782, gonzalezmartin+2009aap506_1107} suggest the presence of a heavily dust-obscured AGN in the nucleus of this galaxy. 

The total exposure time and other information relevant to the \Chandra/ACIS observations of UGC\,5101 are provided in Table \ref{tab:NGC3079_data}. The diffuse X-ray emission of UGC\,5101 has been previously analyzed in the literature. \citet{huo+2004apj611_208} found evidence for an inner hot gas halo of $8.7$~kpc (10.4\arcsec) and an outer halo of $14.3$~kpc (17.0\arcsec). \citet{grimes+2005apj628_187} found that 95\% of the 0.3--1.0~keV emission is enclosed in the inner 8.75~kpc galactocentric radius (10.5\arcsec).  \citet{smith+2018aj155_81, smith+2019aj158_169} analyzed the \Chandra/ACIS observations, finding that the 0.3--1.0~keV emission has a size of $24.0\arcsec\times14.2\arcsec$ ($\sim19.1\times11.3$~kpc, position angle of 90$^{\circ}$), and a total X-ray luminosity of $\log L_{X}=41.6$~erg~s$^{-1}$. 

\begin{figure*}[t!]
\begin{center}
\resizebox{\linewidth}{!}{
\begin{overpic}[trim={20 25 85 38}, clip, height=7cm]{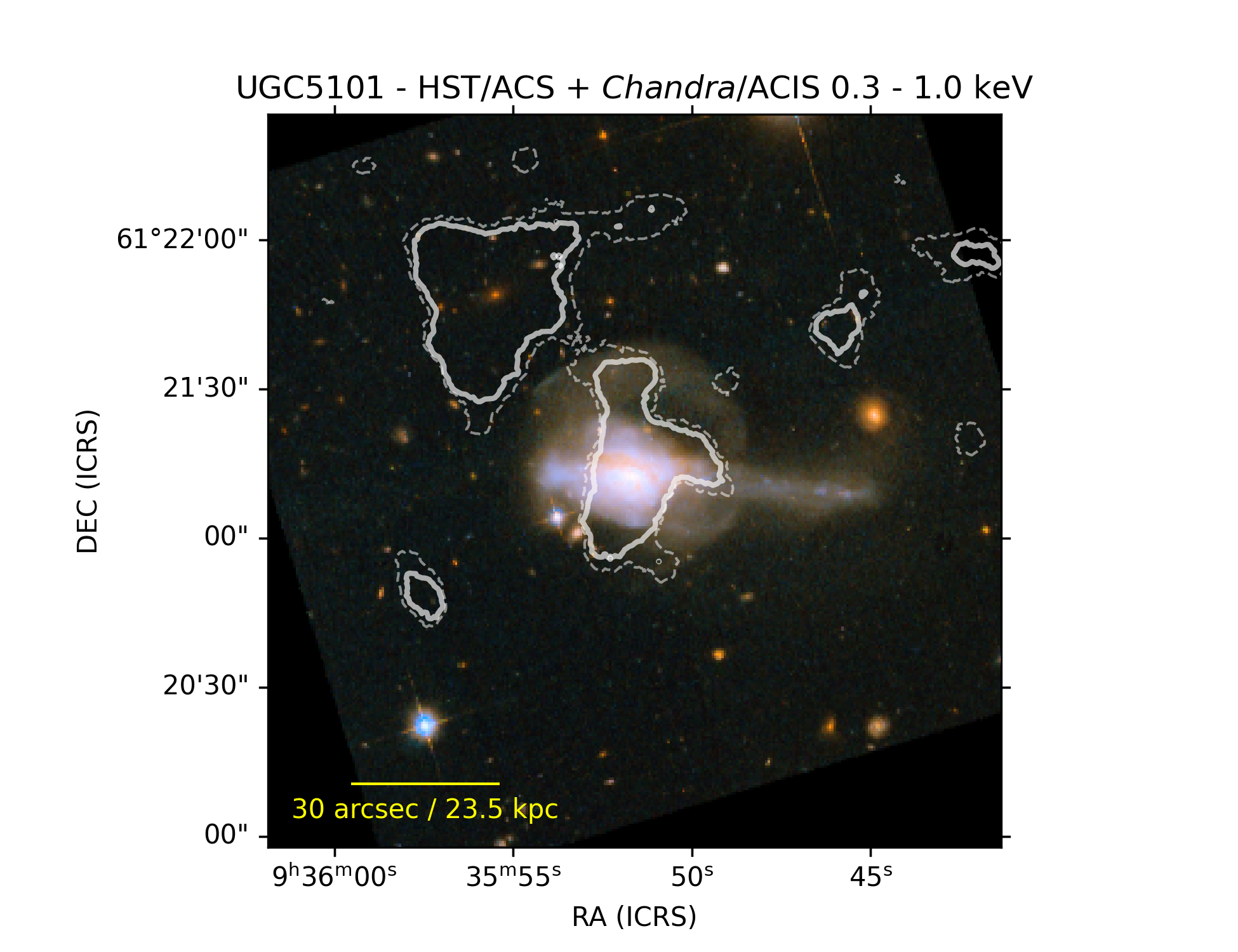}
\put(57,186.5){\color{white}{\textsf{HST/ACS}}}
\put(208,35){\color{cyan}{\textsf{F435W}}}
\put(167,25){\color{green}{\textsf{F435W + F814W}}}
\put(208,15){\color{red}{\textsf{F814W}}}
\end{overpic}
\begin{overpic}[trim={70 25 10 38}, clip, height=7cm]{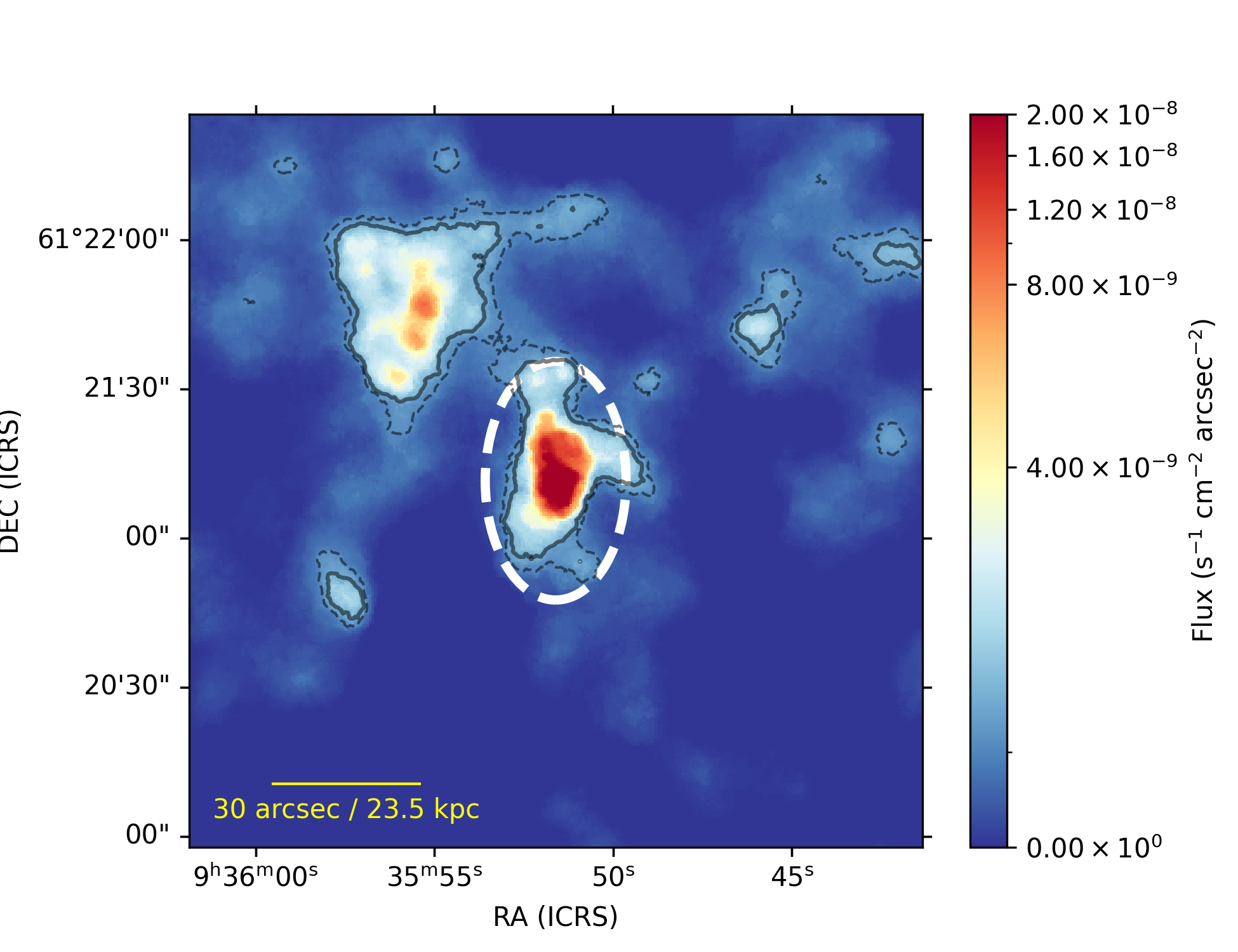}
\put(1.5,185.5){\color{black} \colorbox{white}{\textsf{Chandra/ACIS}}}
\put(120,90){\color{yellow}{\textsf{UGC\,5101}}}
\put(20,150){\color{yellow}{\textsf{X1}}}
\put(136,13){\color{black} \colorbox{white}{\textsf{0.3-1.0 keV}}}
\end{overpic}
}

\resizebox{\linewidth}{!}{
\begin{overpic}[trim={20 25 85 38}, clip, height=7cm]{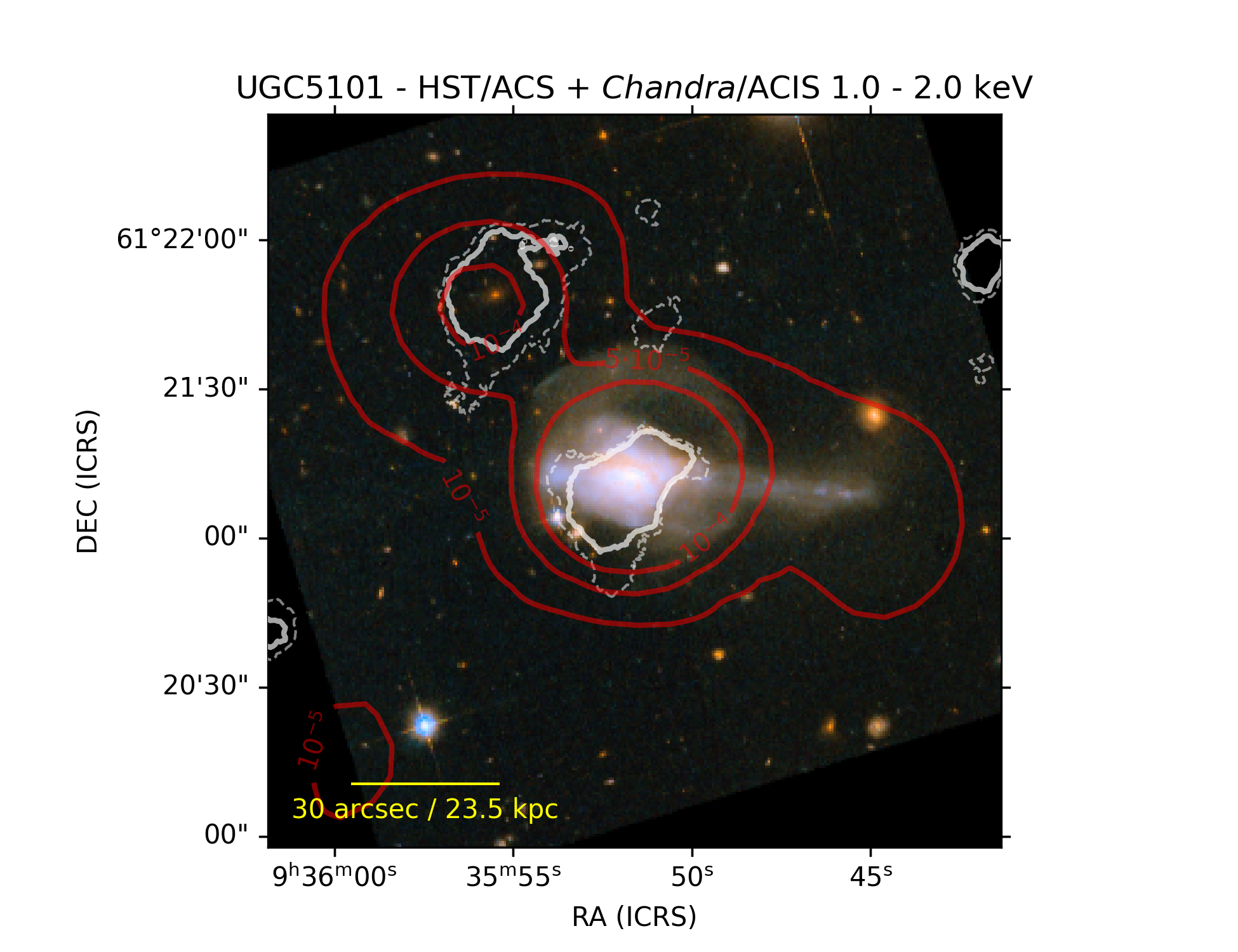}
\put(165,186.5){\color{red}{\textsf{GMRT 150 MHz}}}
\end{overpic}
\begin{overpic}[trim={70 25 10 38}, clip, height=7cm]{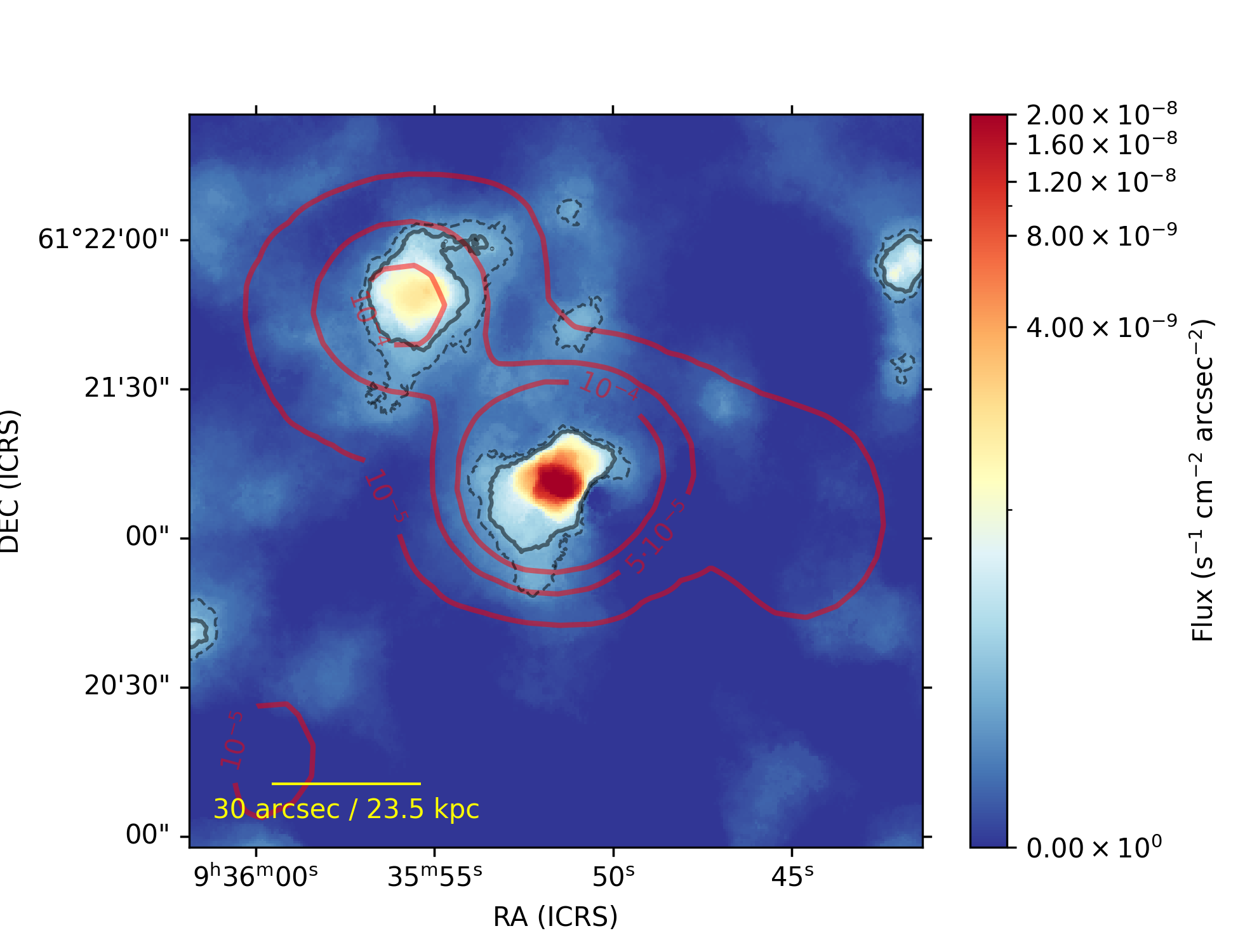}
\put(1.5,185.5){\color{black} \colorbox{white}{\textsf{Chandra/ACIS}}}
\put(136,13){\color{black} \colorbox{white}{\textsf{1.0-2.0 keV}}}
\end{overpic}
}

\resizebox{\linewidth}{!}{
\begin{overpic}[trim={20 0 85 38}, clip, height=7cm]{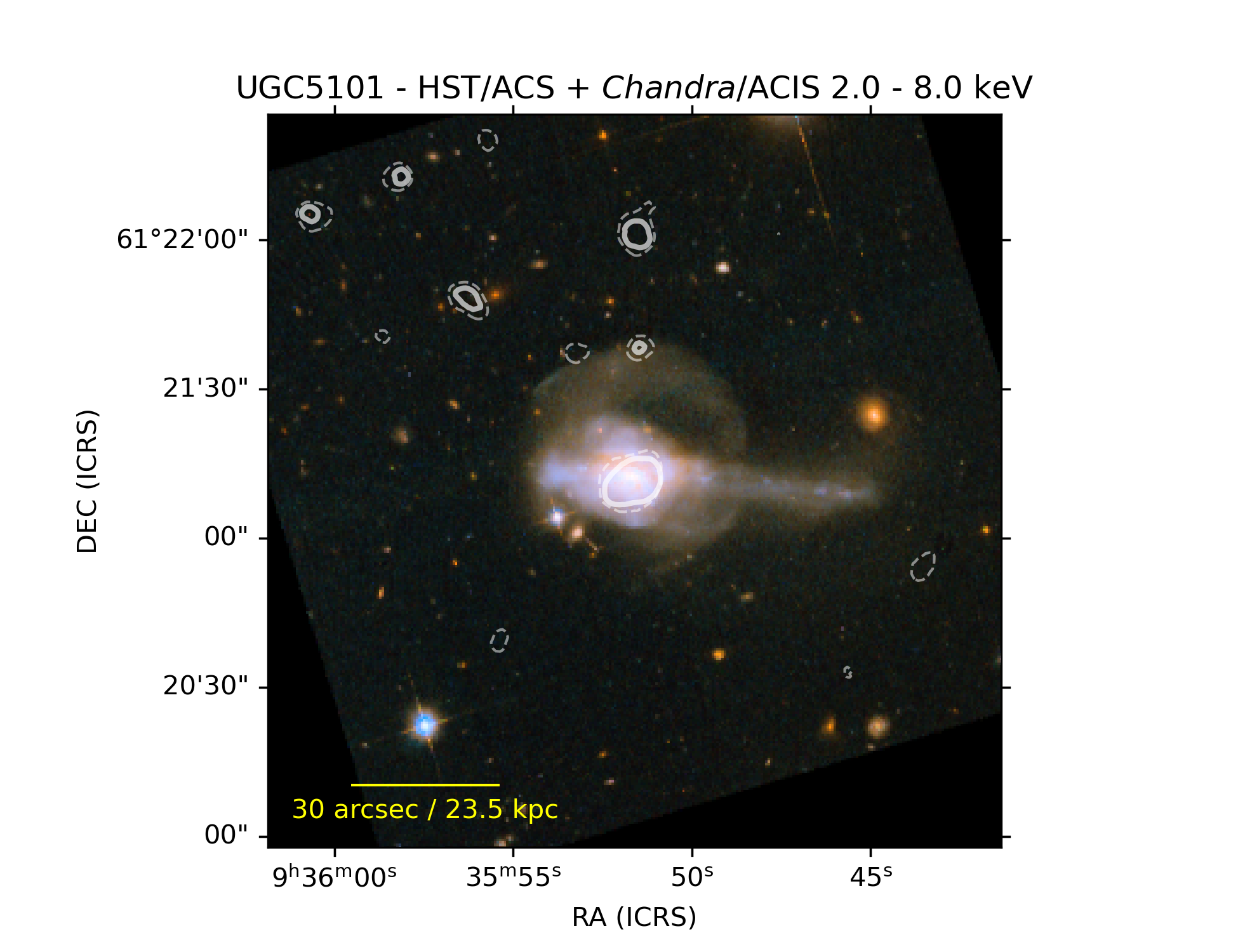}
\end{overpic}
\begin{overpic}[trim={70 0 10 38}, clip, height=7cm]{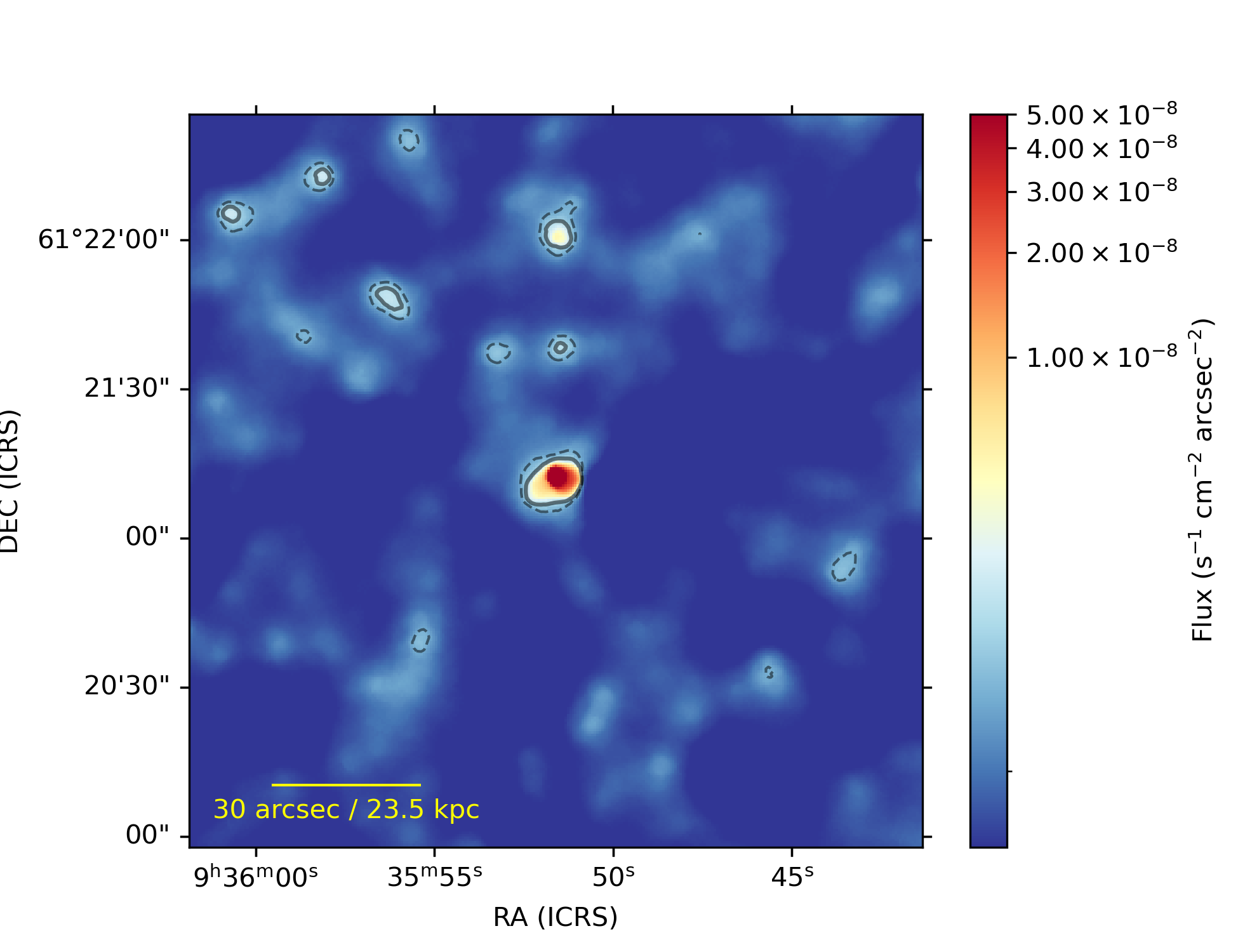}
\put(1.5,185.5){\color{black} \colorbox{white}{\textsf{Chandra/ACIS}}}
\put(121.5,27.8){\color{black} \colorbox{white}{\textsf{2.0-8.0 keV}}}
\end{overpic}
}
\vspace{-0.75cm}
\caption{Diffuse X-ray emission of UGC\,5101 as detected with \SAUNAS/\Chandra\ in the 0.3--1.0~keV band (top), 1.0--2.0~keV band (central), and 2.0--8.0 keV band (bottom). \emph{Left:} HST/ACS color image (red: \emph{F814W}, green: \emph{F435W + F814W}, blue: \emph{F435W}). \emph{Upper Right:} \SAUNAS\ map of the diffuse X-ray emission, corrected for PSF effect, point-sources, and background. Solid contours represent  $3\sigma$ detections and dotted contours the $2\sigma$ detection level of X-ray emission, represented in white (left panel) and black (right panel) for contrast. Solid red contours show GMRT~150~MHz data. White dashed ellipse represents the previous detection limits reported by \citet{smith+2019aj158_169} of UGC\,5101 in the same band.} 
\label{fig:UGC5101}
\end{center}
\end{figure*}
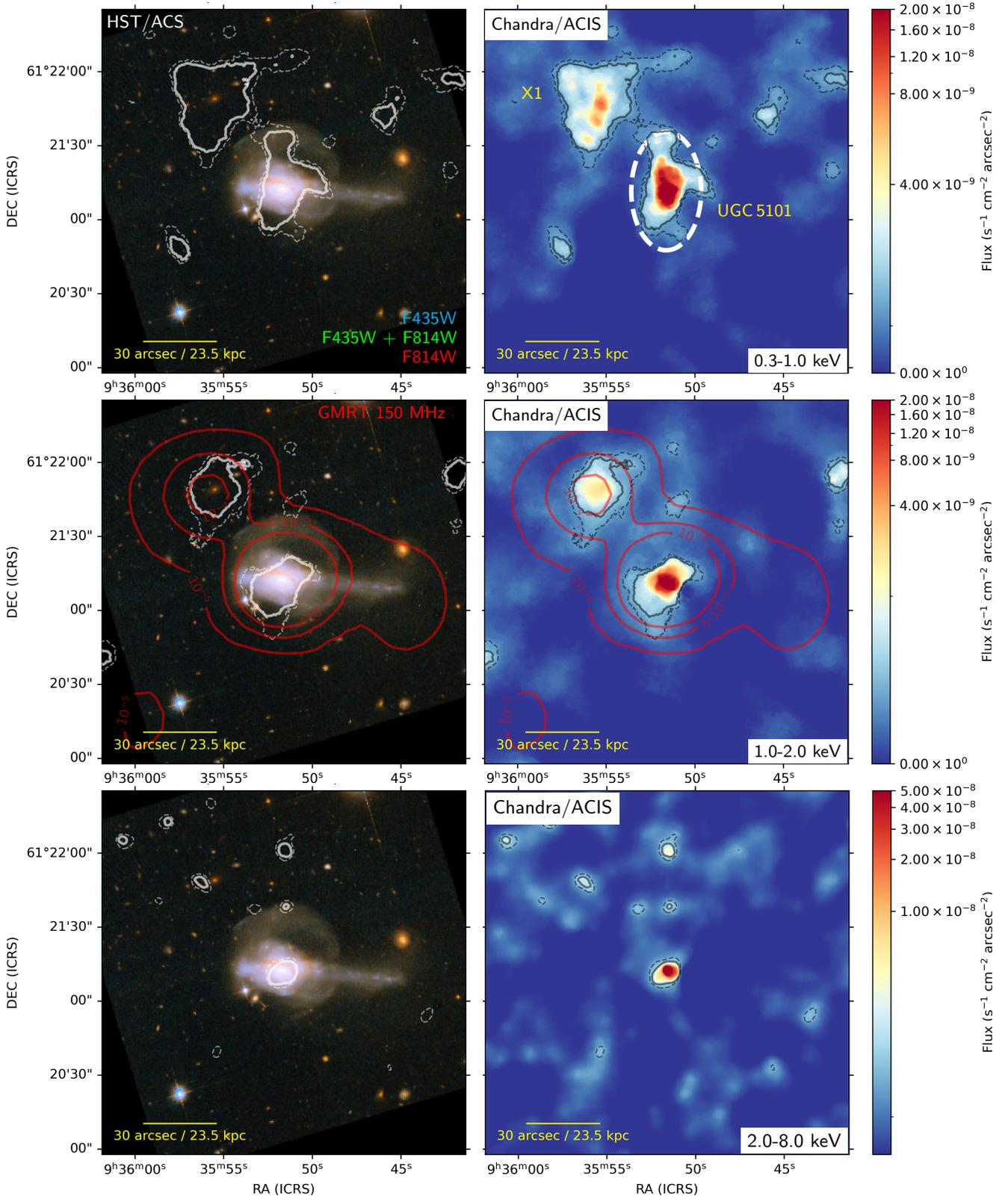

Given these known robust detections, we employ \SAUNAS\ in the characterization of the low surface brightness emission from UGC\,5101. Three bandpasses are used, to ensure a direct comparison to the analyses by  \citet{smith+2019aj158_169}: soft (0.3--1.0~keV), medium (1.0--2.0~keV), and hard (2.0--8.0~keV). The flux conservation ratio after PSF deconvolution in this exposure is $96.0\pm0.02$\% in the three bands. The processed X-ray emission maps are presented in Fig.\,\ref{fig:UGC5101}, in comparison with the optical/NIR observations from HST, as well as ancillary radio observations for reference. The PSFs and unprocessed events of the UGC\,5101 observations in the three bands analyzed are available in Figs.\,\ref{fig:UGC5101_psf} and \ref{fig:UGC5101_events} in Appendices \ref{Appendix:Observed_PSFs} and \ref{Appendix:Observed_events}, respectively.

The results are summarized in Fig.\,\ref{fig:UGC5101}. The analysis of the \Chandra/ACIS observations with \SAUNAS\ reveal that even after PSF deconvolution, the soft X-ray emission of UGC\,5101 still shows extended emission around its core. The 0.3--1.0 and 1.0--2.0~keV bands present X-ray emission with an elongated morphology, with a characteristic bright plume-like structure in the core, oriented in the north-south direction ($\mu_{\rm soft}$ = 1--2$\times$10$^{-8}$~s$^{-1}$~cm$^{-2}$~arcsec$^{-2}$), very similar to the results of \citet{smith+2018aj155_81}. In contrast, the hard band only shows a bright core in the center, compatible with an unresolved source. In the soft band, the diffuse X-ray emission is detectable down to levels of $\mu_{\rm soft} = $ 1.23$^{+1.02}_{-0.66}$$\times10^{-9}$~s$^{-1}$~cm$^{-2}$~arcsec$^{-2}$ ($2\sigma$), compared to the medium band level of $\mu_{\rm medium} = $ 1.25$^{+1.38}_{-0.73}$$\times10^{-9}$~s$^{-1}$~cm$^{-2}$~arcsec$^{-2}$ ($2\sigma$). Both soft and medium band emissions are centered over the main core of UGC\,5101, showing the same orientation as observed by \citet{smith+2019aj158_169}. The soft band emission extends up to 25~arcsec (20~kpc) to the north and 17~arcsec (13.5~kpc) to the south ($3\sigma$). 

\begin{figure*}[t!]
\begin{center}
\resizebox{\linewidth}{!}{
\begin{overpic}[trim={20 0 20 40}, clip, height=7cm]{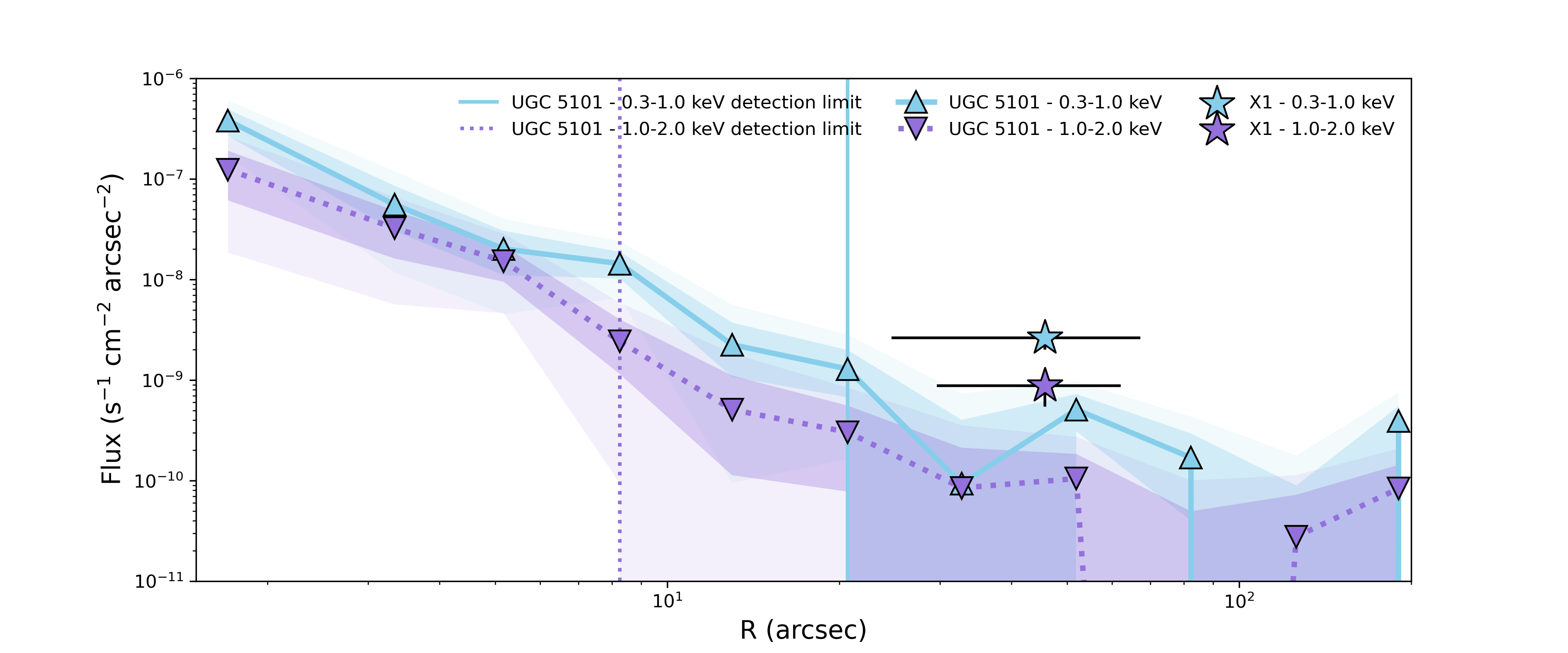}
\end{overpic}
}

\caption{Surface brightness profiles of the diffuse X-ray emission of UGC\,5101 and the extended diffuse north-east source (X1) detected with \SAUNAS/\Chandra\ in the 0.3--1.0 and 1.0--2.0~keV bands. Radially averaged surface brightness profile (blue upward triangles: 0.3--1.0~keV band, purple downward triangles: 1.0--2.0~keV band). Shaded areas represent the $1\sigma$ and $2\sigma$ error bars. Solid blue and dashed purple vertical lines represent the $2\sigma$ detection limits for the 0.3--1.0 keV and the 1.0--2.0 keV bands. Blue and purple stars show the average surface brightness of the north-east extended emission X1, represented at the measured galactocentric distance from UGC\,5101.} 
\label{fig:UGC5101_profiles}
\end{center}
\end{figure*}

The spatial distribution of X-ray emission around UGC\,5101 is generally comparable to that detected in previous works \citep{smith+2019aj158_169}. However, at approximately 40--60~arcsec radius to the north-east ($\alpha$, $\delta=143.980^{\circ}$, $+61.363^{\circ}$), the \SAUNAS\ map reveals a diffuse bridge connecting with UGC\,5101, at a $\sim2\sigma$ level ($\mu_{\rm soft}\sim6.2\times10^{-10}$~\escmarc\ in the soft band). For clarity, we will refer to this extended emission as X1. Fig.\,\ref{fig:UGC5101_profiles} displays surface brightness profile analysis results and associated comparisons with X1. The central surface brightness of X1 is $\mu_{\rm soft}$ = 4.2$^{+1.5}_{-1.3}\times10^{-9}$~\escmarc\ in the soft band and $\mu_{\rm medium}$ = 1.54$^{+0.76}_{-0.66}\times10^{-9}$~\escmarc\ in the medium band. The emission of X1 is detectable at a $3\sigma$ confidence level with a comparable angular area to UGC\,5101, but with a maximum surface brightness 20--30 times lower than the main object (see Fig.\,\ref{fig:UGC5101_profiles}).

Fig.\,27 in \citet{smith+2019aj158_169} shows a hint of what might be emission jutting to the North-East of UGC\,5101 where we see X1, but at a considerably lower detectability. The X1 feature has not been discussed previously in the literature as part of the UGC\,5101 system, but rather as a potential higher-$z$ galaxy cluster  \citep{clerc+2012mnras423_3561, koulouridis+2021aap652_12} in need of spectroscopic confirmation.

Observations of the Giant Metrewave Radio Telescope (GMRT)~150~MHz all-sky radio survey\footnote{TIFR~GMRT Sky Survey (TGSS) Archive: \url{https://vo.astron.nl/tgssadr/q_fits/imgs/form}} \citep[][see bottom left panel in Fig.\,\ref{fig:UGC5101}]{intema+2017aap598_78} confirm the detection of an adjacent source centered over the recovered X-ray emission, with a surface brightness of $\mu=10^{-4}$~Jy~arcsec$^{-2}$. The GMRT flux maps are shown as contours in Fig.\,\ref{fig:UGC5101}, revealing a peak of radio emission over the center of X1 in addition to UGC\,5101. GALEX~UV observations provide a near-ultraviolet (NUV) flux of $5.14\pm0.15\times10^{-6}$~Jy \citep{seibert+2012inproceedings_340} but only upper limits in the far-ultraviolet (FUV) band ($9.8\cdot10^{-6}$~Jy). Recent JWST observations (GO\,1717, PI: Vivian~U., MIRI) of UGC\,5101 were inspected for this work, but they suffer from extreme saturation of the bright core of the galaxy, and the outer X-ray emitting region lies outside the footprint, so they were discarded for this study. While investigating the nature of this extended X-ray emission is beyond the scope of this paper focused on the presentation of the \SAUNAS\ pipeline, we briefly discuss the main hypotheses (hot gas plume or high-$z$ galaxy cluster) in Sec.\,\ref{sec:discussion}.

\section{Discussion} \label{sec:discussion}

\subsection{Limitations} \label{subsec:general_discussion}

We have demonstrated the \SAUNAS\ methodology to be successful in recovering dim, extended surface brightness X-ray features under low signal-to-noise ratio conditions through performance tests using both synthetic (Section\,\ref{subsec:methods_test}) and real (Section\,\ref{sec:results}) X-ray datasets. 

There are, however, several limitations of \SAUNAS\ in its current form that will be addressed in future versions of the pipeline. Among them, \SAUNAS\ does not attempt to provide a quantitative separation between extended sources, such as a segmentation map. Deblending of extended X-ray sources is one of the main objectives of a complementary code, \texttt{EXSdetect} \citep{liu+2013aap549_143}, using a friend-of-friends algorithm. Other specialized pipelines for X-ray observations, such as \texttt{CADET}, based on machine-learning algorithms, allow for the identification of specific source morphologies, such as X-ray cavities \citep{plvsek+2024mnras527_3315}. The potential combination of \SAUNAS\ for generating low surface brightness detection maps with existing morphological identification and segmentation software will be explored in the future.

\begin{figure}[t!]
\begin{center}
\includegraphics[trim={0 0 0 0}, clip, width=\textwidth]{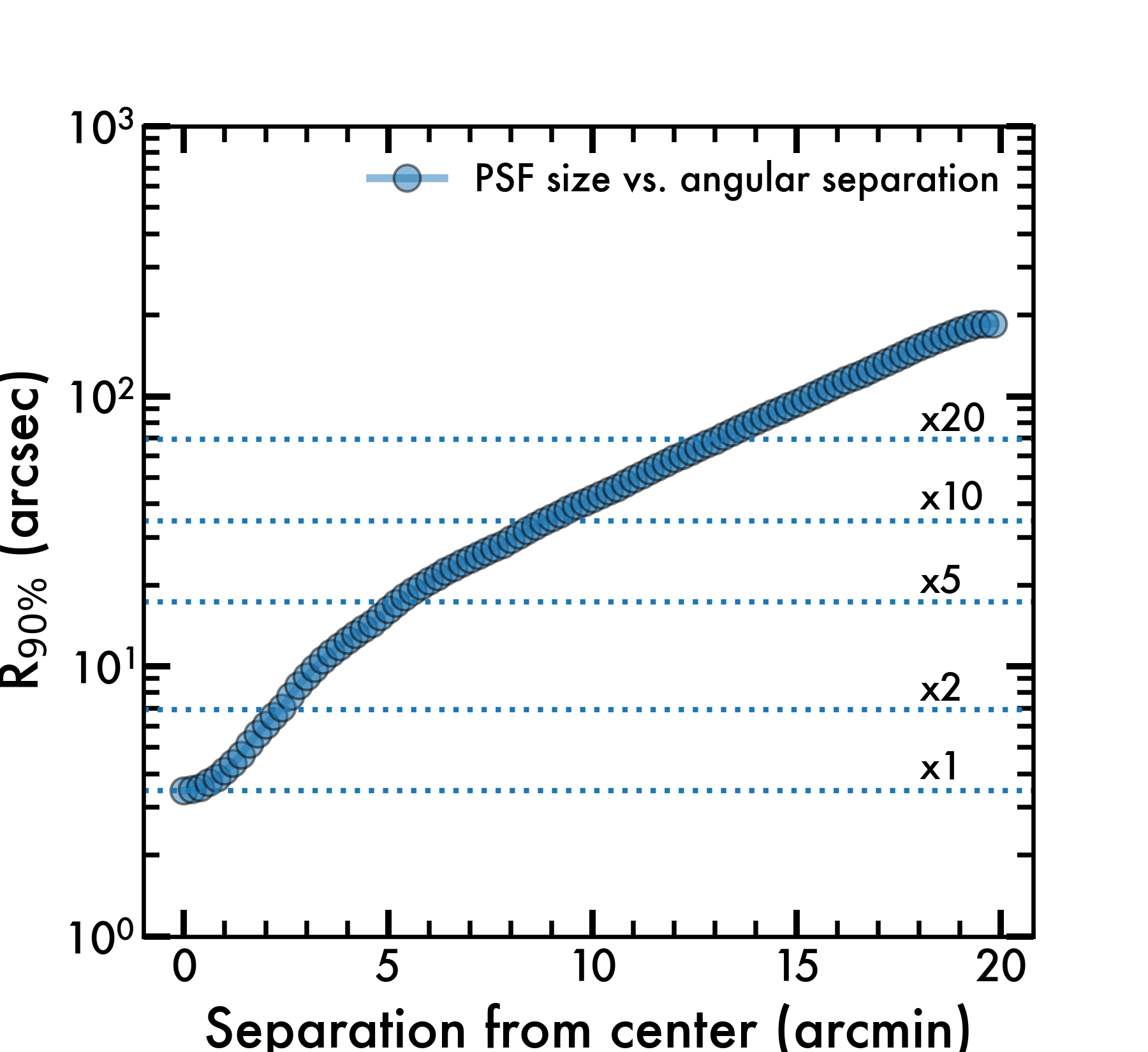}

\caption{Variation of the \Chandra/ACIS PSF size as a function of the angular separation to the center of the FOV. \emph{Vertical axis:} Radius enclosing 90\% of the flux from the PSF at 1.0 keV, based on the observations of UGC\,5101. \emph{Horizontal axis:} Angular separation to the center of the source, approximately the center optical axis. The horizontal dotted lines mark the PSF sizes that correspond to $\times2$, $\times5$, $\times10$, and $\times20$ the PSF size at its center ($\times1$).}
\label{fig:PSF_size_vs_distance}
\end{center}
\end{figure}

Another limitation of the \SAUNAS\ pipeline is the precision of the PSF. The generation of the \Chandra/ACIS PSFs depends on multiple factors, including, but not limited to, the position of the source on the detector, the SED of the source, or the specific parameters fed into the \texttt{MARX} simulation software (like the aspect blur).  For example, \LIRA\ deconvolution software only accepts one PSF for the whole image, and as a consequence, the shape of sources at high distances from the center of the image might be inaccurate. This phenomena can cause residuals if observations present bright sources at high angular distances from the center of the source, since the deconvolution will be based on the PSF at the center of the observation, but not at the location of the secondary contaminating source. As an attempt to quantify this effect, we estimate in Fig.\,\ref{fig:PSF_size_vs_distance} the variation of the PSF size ($R_{90 \%}$, radius that contains 90\% of the flux of a point source) vs.\, angular separation to the source using \ciao\/ \texttt{psfsize\_srcs}\footnote{\ciao/\texttt{psfsize\_srcs}: \url{https://cxc.cfa.harvard.edu/ciao/ahelp/psfsize_srcs.html}}, based on the \Chandra/ACIS observations on UGC\,5101. The results show that the PSF increases a factor of $\times$2 in $\sim$2 arcmin ($\times$10 in $\sim$10 arcmin). In our science cases, no bright object was observed in the environment of the main sources (NGC\,3079, UGC\,5101), so the main contributors to the scattered light are the sources for which the PSF was calculated. However, observers must be wary of strong residual PSF wings from nearby sources at $\sim$2 arcmin and longer distances. While a complete analysis of the uncertainties of the PSF in \Chandra\ is out of the scope of the current paper, we refer to the Appendix in \citet{ma+2023apj948_61} for a review in the field.

\subsection{NGC\,3079} 
\label{subsec:discussion_NGC3079}

The analysis of the \Chandra/ACIS observations in the field of NGC\,3079 revealed signs of X-ray wind out to galactocentric distances $R\sim30$~kpc, compatible with previous observations using XMM--Newton \citep{hodgeskluck+2020apj903_35}. While XMM--Newton is able to trace the extended X-ray emission out to larger distances ($\sim$40~kpc) in some directions, some considerations must be made in order to compare XMM--Newton results with the benchmark study provided here: 
\begin{enumerate}
    \item XMM--Newton observations of NGC\,3079 combine an $\sim$11 times longer exposure time (300.6~ks) than the usable time in \Chandra/ACIS (26.6~ks) observations. 

    \item XMM-Newton has a larger effective area (4650~cm$^2$ at 1~keV) than \Chandra\ ($555~$cm$^2$), at the expense of a lower spatial resolution\footnote{\url{https://xmm-tools.cosmos.esa.int/external/xmm_user_support/documentation/uhb/xmmcomp.html}} (XMM-Newton/FWHM $=$ 6~arcsec vs. \emph{Chandra}/FWHM $=$ 0.2~arcsec). While the aperture is smaller, proper masking of point sources improves detectability of dim structures by reducing the background noise.

    \item The analysis of the X-ray emission by \citet{hodgeskluck+2020apj903_35} is based on the inspection of the quadrant stacked images with a certain signal and radial threshold (see their Fig.\,4, central panel). The methodology they use  to calculate the limiting radius of the diffuse X-ray emission is not clearly stated in their analysis, making a direct and accurate comparison of results difficult. 
    
\end{enumerate}

Despite the differences of the detection methods, we conclude that \SAUNAS\ is able to recover extended, low surface brightness X-ray emission using \Chandra/ACIS X-ray observations of NGC\,3079, in   excellent agreement with the deeper exposure taken by XMM--\emph{Newton}.

\subsection{UGC\,5101} 
\label{subsec:discussion_UGC5101}

Section \ref{subsec:UGC5101} described evidence for extended low surface brightness emission (X1, $\mu_{\rm soft}$ = 4.2$^{+1.5}_{-1.3}\times10^{-9}$~s$^{-1}$~cm$^{-2}$~arcsec$^{-2}$, 0.3--1.0~keV) located in the north-east of the UGC\,5101 merging galaxy. X1 has been previously detected in X-ray by \citet{smith+2019aj158_169} but its emission was not discussed nor treated as part of UGC\,5101's outskirts. Other works \citep{clerc+2012mnras423_3561,koulouridis+2021aap652_12} tentatively classified X1 as a potential background galaxy cluster, but this feature remained unconfirmed as spectroscopic observations are unavailable. X1 is detected also in GMRT~150~Mhz observations as a secondary source adjacent to UGC\,5101, confirming the existence of a feature at this location. Two main hypotheses regarding the nature of X1 are:

\begin{enumerate}
    \item X1 is part of the extended X-ray emitting envelope of UGC\,5101.

    \item X1 is a background source, potentially the extended envelope of a higher-$z$ object, such as a massive early-type galaxy or a cluster.
\end{enumerate}

\begin{figure}[t!]
\begin{center}
\includegraphics[trim={20 0 80 40}, clip, width=\textwidth]{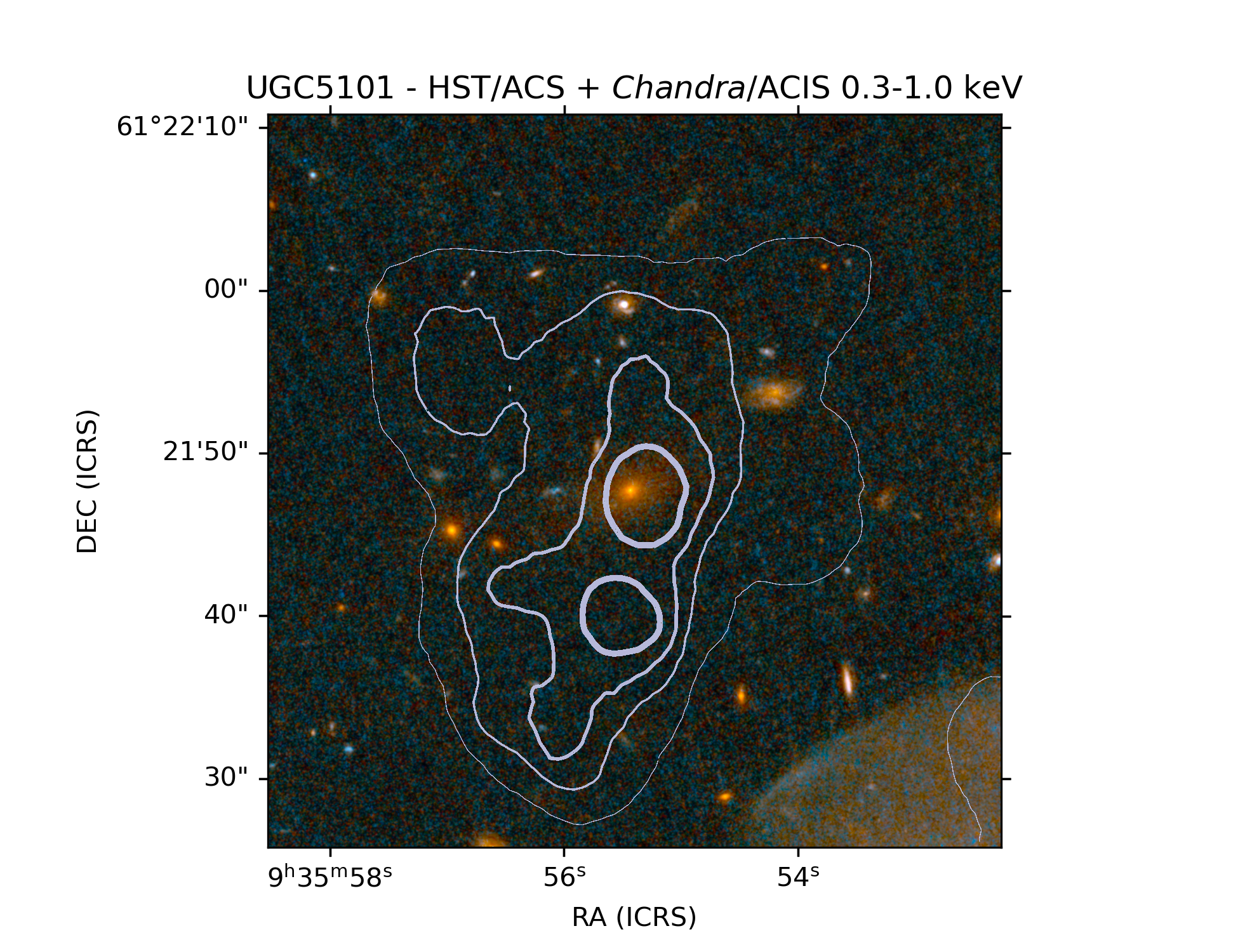}

\caption{\emph{Hubble} Space Telescope ACS imaging over the diffuse extended emission X1 found adjacent to UGC\,5101. Pseudo-RGB color combination: \emph{Blue:} F435W, \emph{Green:} F435W~$+$~F814W. \emph{Red:} F814W. Grey contours represent the (3, 5, 7, and 10) $\sigma$ detection levels obtained in the 0.3--1.0~keV band from \Chandra/ACIS observations, processed with \SAUNAS. Notice the merger shell structure of UGC\,5101 at the bottom right corner.} 
\label{fig:UGC5101_hst_RGB}
\end{center}
\end{figure}

Although the X-ray emission in the soft and medium bands of X1 is adjacent to that of UGC\,5101, and both objects have a dominant emission in the soft band compared to the medium and hard (see Figs.\, \ref{fig:UGC5101} and \ref{fig:UGC5101_profiles}, the emission could still be part of a hot gas halo at higher-$z$. In fact, the center of the \Chandra/ACIS X-ray emission overlaps remarkably well with that of a background galaxy. Fig.\,\ref{fig:UGC5101_hst_RGB} shows the HST/ACS imaging (bands) centered over X1, with the soft-band X-ray emission contours overlapped for reference. The peak of X-ray emission is coincident with the position of a background galaxy (WISE~J093555.43+612148.0). Unfortunately, WISE~J093555.43+612148.0 does not have spectroscopic or photometric redshifts available. 

While resolving the nature of X1 is beyond the scope of this paper, we conclude that the test performed with the \Chandra/ACIS observations of UGC\,5101 using \SAUNAS\ demonstrates the pipeline's capabilities in successfully producing adaptively smoothed, PSF-deconvolved X-ray images in different bands. The image reduction process presented here allows for a better calibration of the background to recover details at both high resolution and surface brightness (inner core structure of the merging galaxy) as well as extended ultra-low surface brightness regions, such as the previously unknown extended emission around UGC\,5101.

\section{Conclusions} 
\label{sec:conclusions}

In this paper we have presented \SAUNAS: a pipeline to detect extended, low surface brightness structures on \Chandra\ X-ray Observations. \SAUNAS\ automatically queries the \Chandra\ Archive, reduces the observations through the \ciao\ pipeline, generates PSF models and deconvolves the images, identifing and masking point sources, and generating adaptative smoothed surface brightness and detection SNR maps for the sources in the final mosaics. We have demonstrated through tests on simulated data and comparisons to published results that the \SAUNAS\ pipeline distinguishes itself from other existing X-ray pipelines by meeting the following main objectives:
\begin{enumerate}


    \item Generate X-ray detection maps for extended sources in a consistent, statistically reproducible way. 
    
    \item Provide a modular framework for reduction of \Chandra/ACIS observations focusing on the detection of faint extended sources, simplifying the access to X-ray archival observations for multi-wavelength studies.
\end{enumerate}

Our approach to meeting these objectives is to assess the statistical probability that signal in low-count areas is real. This strategy can both produce detections of previously-overlooked diffuse emission as well as minimize false positive detections of extended hot gas emission. In Sec.\,\ref{sec:results}, we compare \SAUNAS-processed archival \Chandra/ACIS data to published results. 
This section demonstrates that the proposed methodology succeeds in recovering the extended emission detected in a selection of local Universe targets. 
While the \ciao\ pipeline provides a canonical and highly efficient procedure to reduce the \Chandra\ observations, the secondary analysis of the resulting event files is usually performed in an independent way by the observers. Such a situation results in two suboptimal consequences: 1) Most X-ray studies are focused on single objects, or very small samples (three or four objects), and 2) most studies develop their own procedure to correct the PSF effects (if considered), to generate  smoothed maps, and to determine the significance of emission over the background. Planned future work includes an analysis of the extended emission of nearby galaxies using \Chandra/ACIS archival data, and releasing the tools to the astronomical community. In this first article, we made the processed maps available\footnote{The \SAUNAS\ X-ray surface brightness maps of NGC\,3079 and UGC\,5101 are publicly available in Zenodo: \url{https://zenodo.org/records/10892485}.} for the community through the Zenodo open repository. 

A benefit of the automated functionality provided by this tool is its provision of straightforward access to high-level archival \Chandra\ products and facilitation of their use in  multi-wavelength studies. In future works of this series (Borlaff et al. in prep.) we will explore the X-ray emission of a sample of targets using the \SAUNAS\ pipeline, focusing on the evolution of lenticular galaxies based on \Chandra/ACIS data in combination with \emph{Hubble} and \emph{Spitzer} observations. 
The serendipitous discovery presented in this work in one of the galaxies studied; UGC\,5101, an on-going merger galaxy, demonstrate that the combination of multi-wavelength legacy archives, such as those of \Chandra, GMRT, and \emph{Hubble}, may already hold the information to disentangle the impact of the different evolutionary drivers in galaxies.

\vfill\null

\begin{acknowledgments} 
The authors thank to the anonymous referee for the provided input that helped to improve this publication significantly. The \SAUNAS\ X-ray surface brightness and signal-to-noise maps of NGC\,3079 and UGC\,5101 are publicly available in \texttt{Zenodo} in \texttt{FITS} format: \dataset[DOI: 10.5281/zenodo.10892485]{https://zenodo.org/records/10892485}.  The list of Chandra datasets, obtained by the Chandra X-ray Observatory, are contained in~\dataset[DOI: 10.25574/cdc.225]{https://doi.org/10.25574/cdc.225}. A.B. acknowledges the tireless support from 
Nicholas Lee, Tara Gokas, Kenny Glotfelty, Catherine Cranmer, and the rest of the CXC Helpdesk team. Without your dedication, this project would have not been possible. This research has made use of data obtained from the Chandra Data Archive and the Chandra Source Catalog, and software provided by the Chandra X-ray Center (CXC) in the application packages \texttt{CIAO} \citep{fruscione+2006inproceedings_62701V} and \texttt{Sherpa} \citep{freeman+2001inproceedings_76}. A.S. acknowledge support from NASA contract to the Chandra X-ray Center NAS8-03060.
The work conducted at NASA Ames Research Center was funded through NASA's  NNH22ZDA001N Astrophysics Data and Analysis Program under Award  22-ADAP22-0118. This work was authored by an employee of Caltech/IPAC under Contract No. 80GSFC21R0032 with the National Aeronautics and Space Administration. This paper represents the views of the authors only and should not be interpreted as representing the views of ConstructConnect, Inc.
\end{acknowledgments}

%

\vspace{5mm}
\facilities{\Chandra, \Hubble}


\software{\texttt{Matplotlib} \citep{Hunter:2007}, \ciao, \texttt{astropy} \citep{collaboration+2018aj156_123, collaboration+2013aap558_33, collaboration+2022apj935_167}, \texttt{LIRA} \citep{donath+2022inproceedings_98}\footnote{\texttt{pyLIRA:} \url{https://github.com/astrostat/pylira}}, \texttt{VorBin} \citep{cappellari+2003mnras342_345}}



\appendix

\section{PSF deconvolution efficiency test}
\label{Appendix:SAUNAS_psf_deconvolution_test}

In this section, a set of synthetic observations generated with \ciao/MARX\footnote{Using MARX to Simulate an Existing Observation: \url{https://cxc.cfa.harvard.edu/ciao/threads/marx_sim/}} are used to evaluate the reliability of the \SAUNAS\ algorithm when applied to a simple point source. \SAUNAS's ability to accurately recover diffuse emission is significantly governed by limitations imposed by \LIRA, the associated deconvolution tool. \SAUNAS\ could have instead utilized the widely-used and proven \texttt{arestore} tool, which can restore emission structures down to scales comparable to the \Chandra/ACIS resolution (0.492\arcsec). 
Here we benchmark these two PSF deconvolution methodologies using simulated observations of an unresolved object constructed by convolving a point source with a highly off-axis PSF from \Chandra/ACIS, associated with the observations of 3C\,264 (NGC\,3862, $\alpha=176.2709^{\circ}$, $\delta=+19.6063^{\circ}$, Obs. ID: 514). The simulated observations processed with \SAUNAS\ (Sec.\,\ref{subsec:methods_saunas}) are  compared to the results produced by standard application of \texttt{arestore}. Both methods use the same number of iterations ($N_{\rm{iter}}=1000$). For each method, surface brightness profiles are constructed from  Voronoi binning of the deconvolved data and compared to that of the model point source. 

\begin{figure*}[t!]
\begin{center}

\begin{overpic}[trim={0 0 0 0}, clip, width=0.49\textwidth]{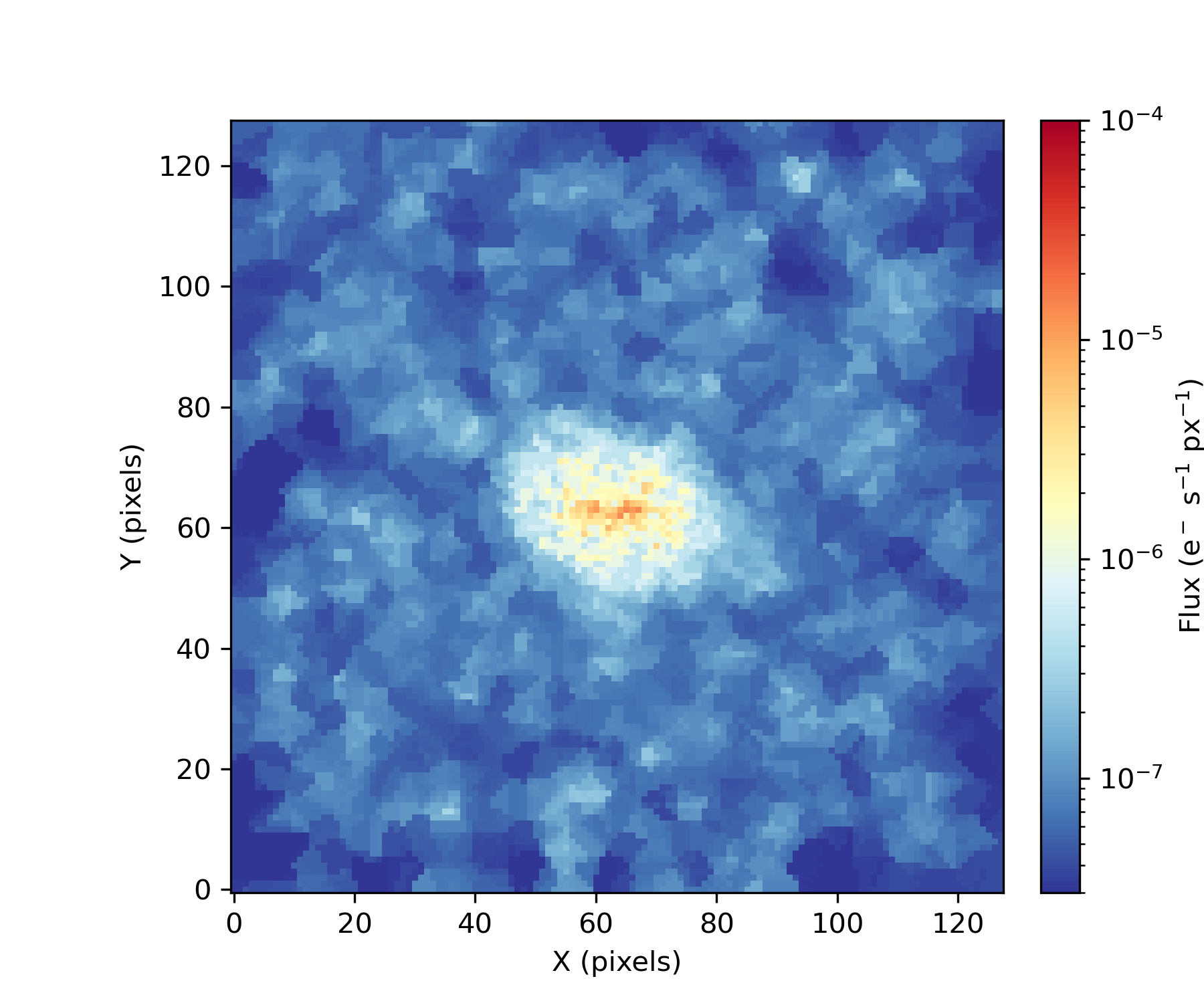}
\put(49,175){\color{black} \colorbox{white}{\textsf{PSF convolved point source}}}
\end{overpic}
\begin{overpic}[trim={0 0 0 0}, clip, width=0.49\textwidth]{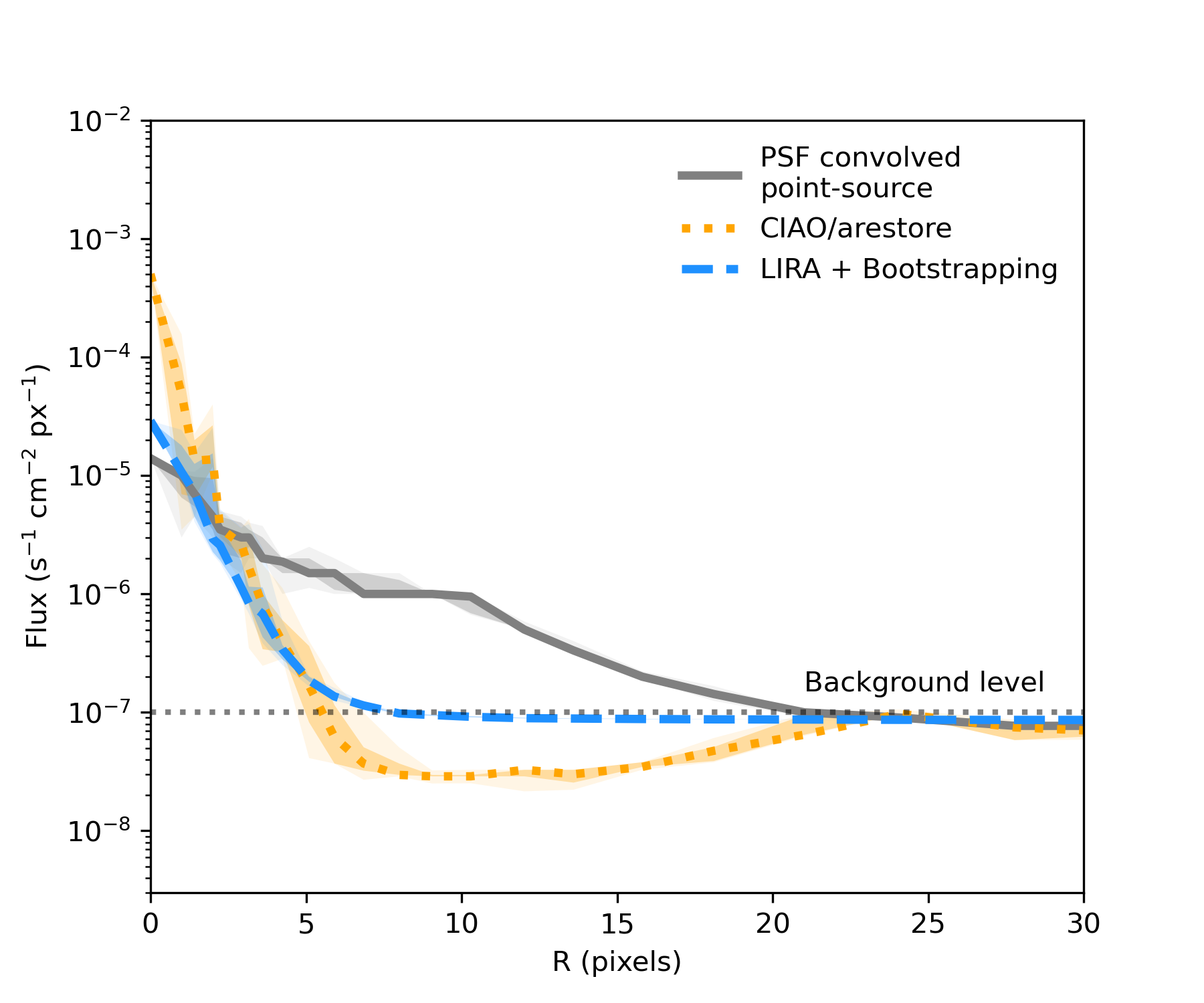}
\end{overpic}

\begin{overpic}[trim={0 0 0 0}, clip, width=0.49\textwidth]{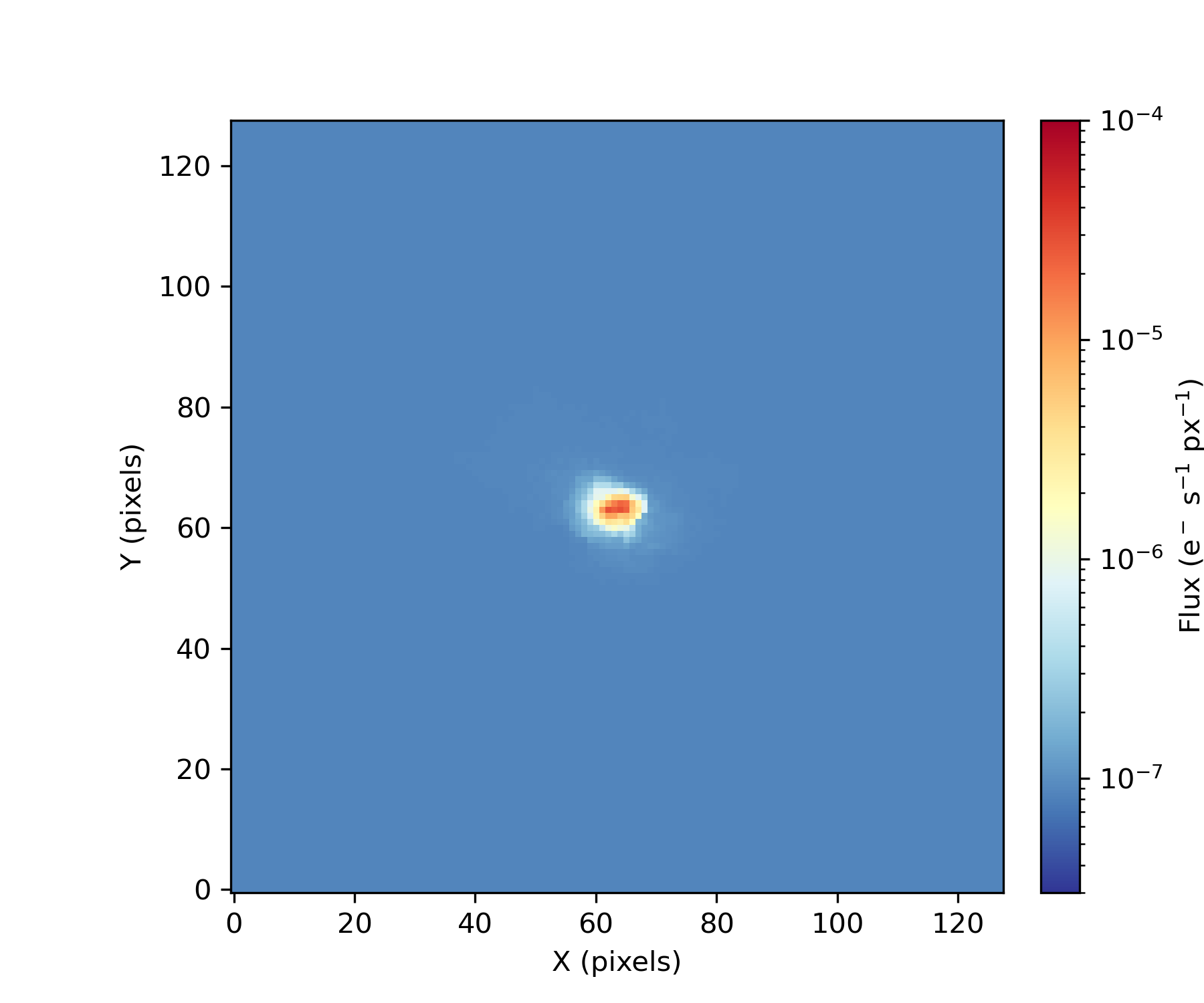}
\put(49,175){\color{black} \colorbox{white}{\textsf{SAUNAS}}}
\end{overpic}
\begin{overpic}[trim={0 0 0 0}, clip, width=0.49\textwidth]{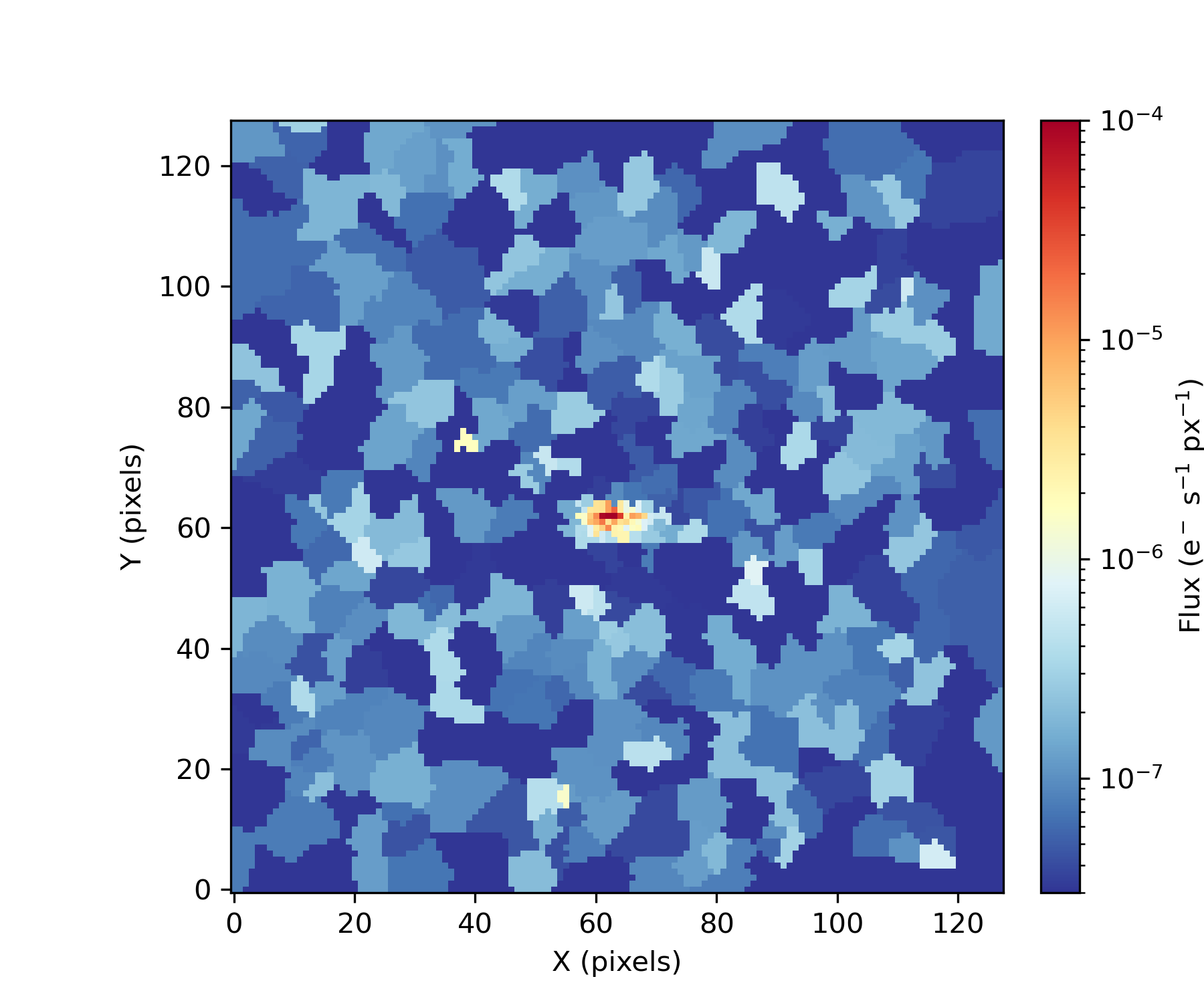}
\put(49,174){\color{black} \colorbox{white}{\textsf{CIAO/arestore}}}
\end{overpic}

\caption{Point-source PSF deconvolution test. \emph{Top left panel:} Emission from a point-source object and a background level of $\mu=$10$^{-7}$~s$^{-1}$~px$^{-1}$ convolved with a reference \Chandra/ACIS PSF (NGC\,3862, $\alpha=176.2709^{\circ}$, $\delta=+19.6063^{\circ}$, Obs.\,ID:\,514). \emph{Top right panel:} Surface brightness profiles of the convolved source (grey solid line), \SAUNAS\ deconvolved image (blue dashed), and \texttt{CIAO/arestore} deconvolved (orange dotted) images. The horizontal dotted line represents the sky background of the model. \emph{Bottom left panel:} \SAUNAS\ deconvolved image. \emph{Bottom right panel:}  \texttt{CIAO/arestore} deconvolved image. Note that the convolved image (events map) and the \texttt{CIAO/arestore} deconvolved image have been processed using Voronoi binning for visualization purposes of the surface brightness. See the legend and colorbar in the figure.} 
\label{fig:example_deco_point_source}
\end{center}
\end{figure*}

The results are shown in Fig.\,\ref{fig:example_deco_point_source}. The PSF convolved point source shows the characteristic elliptical shape of the off-axis PSF from \Chandra/ACIS. The surface brightness profiles obtained from the images show that \texttt{CIAO/arestore} provides output images with more flux at their core than \SAUNAS. However, \texttt{CIAO/arestore}'s deconvolved image has a higher noise in the surroundings of the center ($R=[0-10]$ px) than \SAUNAS, including some clear signs of oversubtraction (see the Voronoi bins at the bottom right image) around the center of the object. In addition, \texttt{CIAO/arestore} leave a characteristic residual at larger distances ($R=[10-20]$ px) that could easily be confused with a shell of extended X-ray emission. In contrast, \SAUNAS\ provides a deconvolved image with less central flux but a smoother transition to the background level and without the presence of residual halos of emission or oversubtraction. We conclude that \texttt{CIAO/arestore} concentrates more signal into a single point source at the expense of higher noise in the resulting images when compared to the methodology utilized by \SAUNAS\, described in Sec.\,\ref{subsec:methods_saunas}. 

\section{\SAUNAS\ extended test models}\label{Appendix:SAUNAS_extended_psf_deconvolution_test}

Appendix \ref{Appendix:SAUNAS_psf_deconvolution_test} demonstrated that the combination of \texttt{LIRA} + Bootstrapping methods adopted in the \SAUNAS\ pipeline provides a more accurate representation of the real distribution of light compared to \texttt{CIAO/arestore}, including avoiding \texttt{arestore}'s PSF over-subtraction. Given that the main aim of \SAUNAS\ is the detection of extended sources,  we extend the analysis from Appendix \ref{Appendix:SAUNAS_psf_deconvolution_test} to \SAUNAS\ processing of an extended source model. 

\begin{figure*}[t!]
\begin{center}

\begin{overpic}[trim={0 0 0 0}, clip, width=0.49\textwidth]{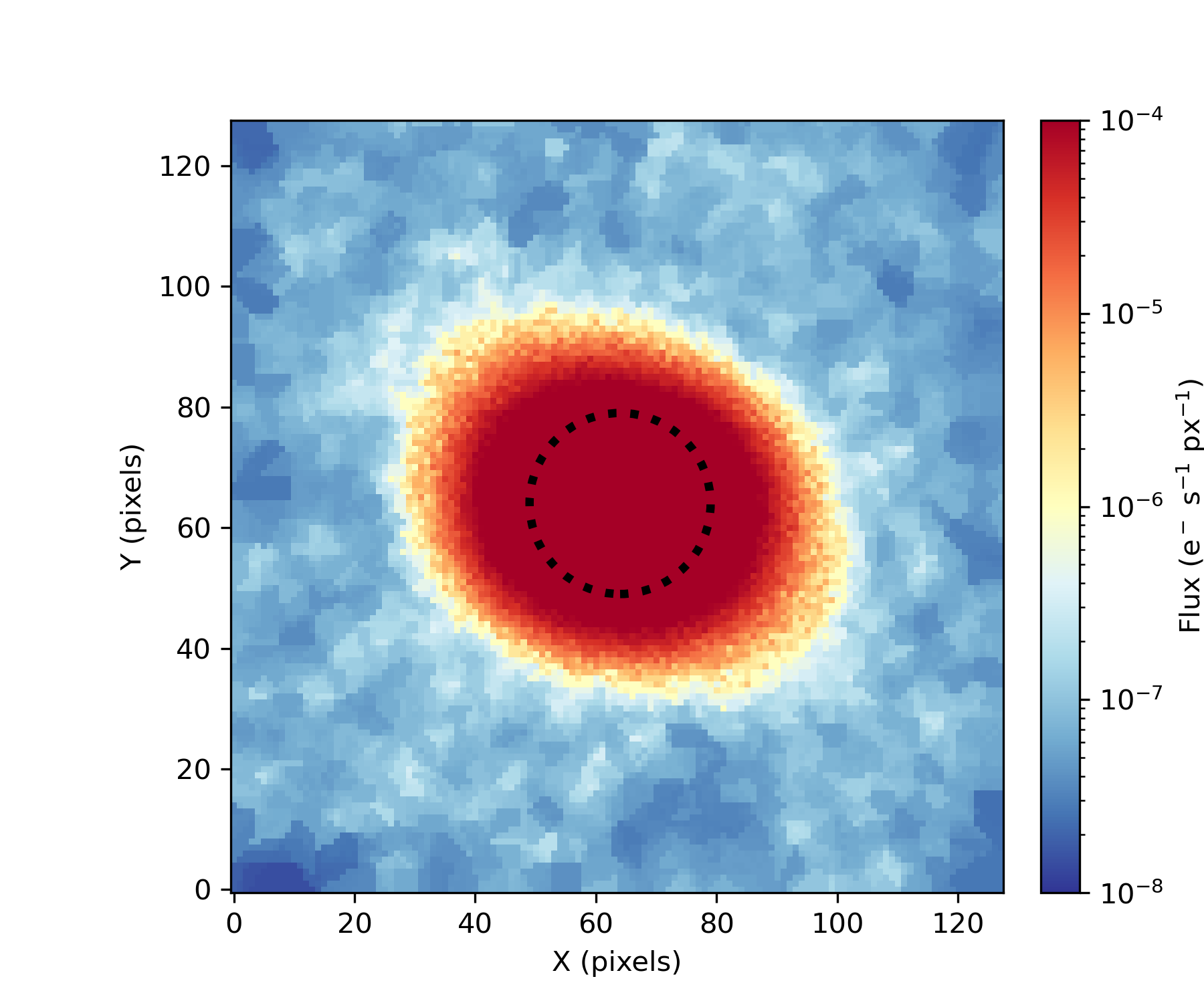}
\put(49,175){\color{black} \colorbox{white}{\textsf{PSF convolved circular source}}}
\end{overpic}
\begin{overpic}[trim={0 0 0 0}, clip, width=0.49\textwidth]{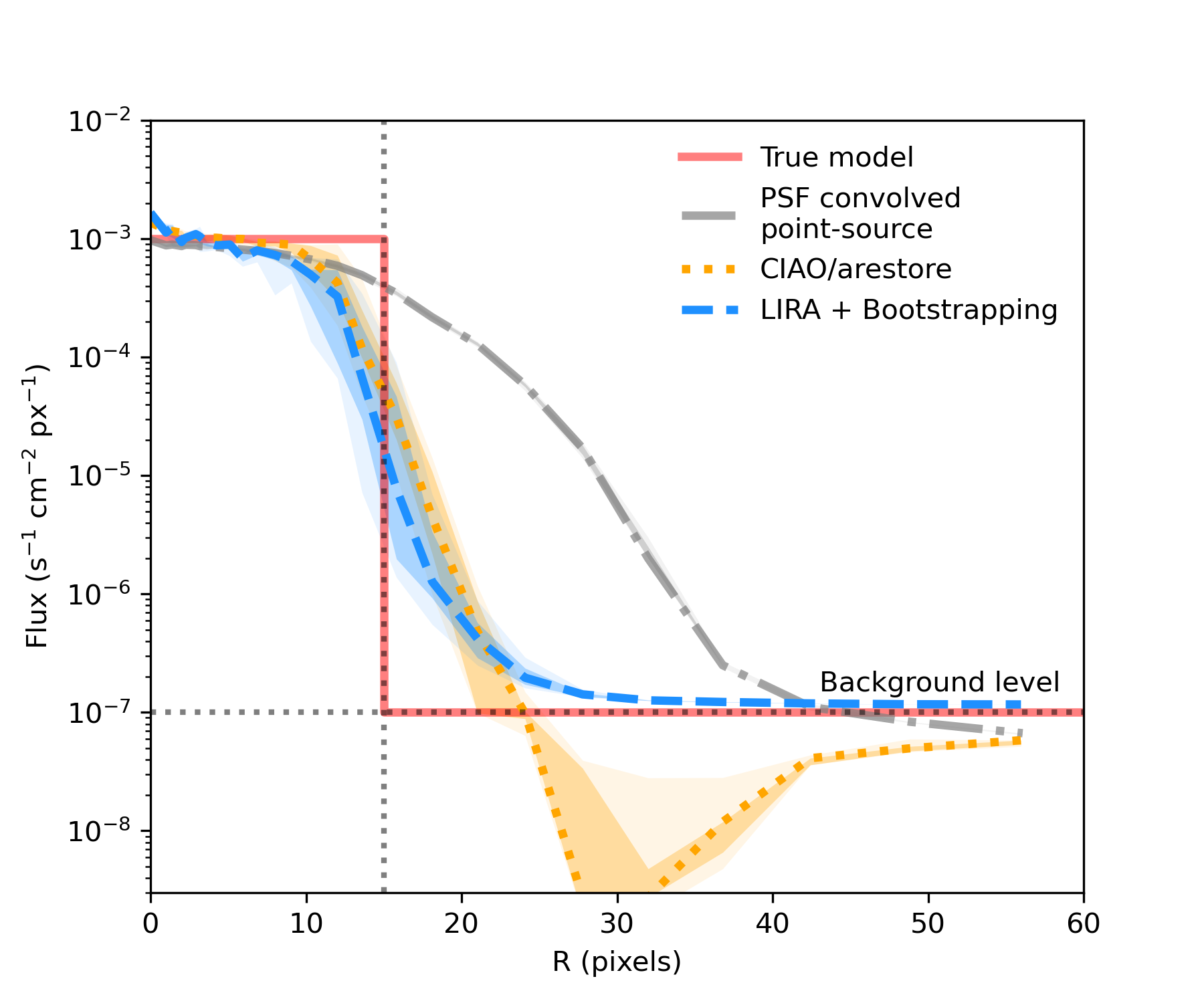}
\end{overpic}

\begin{overpic}[trim={0 0 0 0}, clip, width=0.49\textwidth]{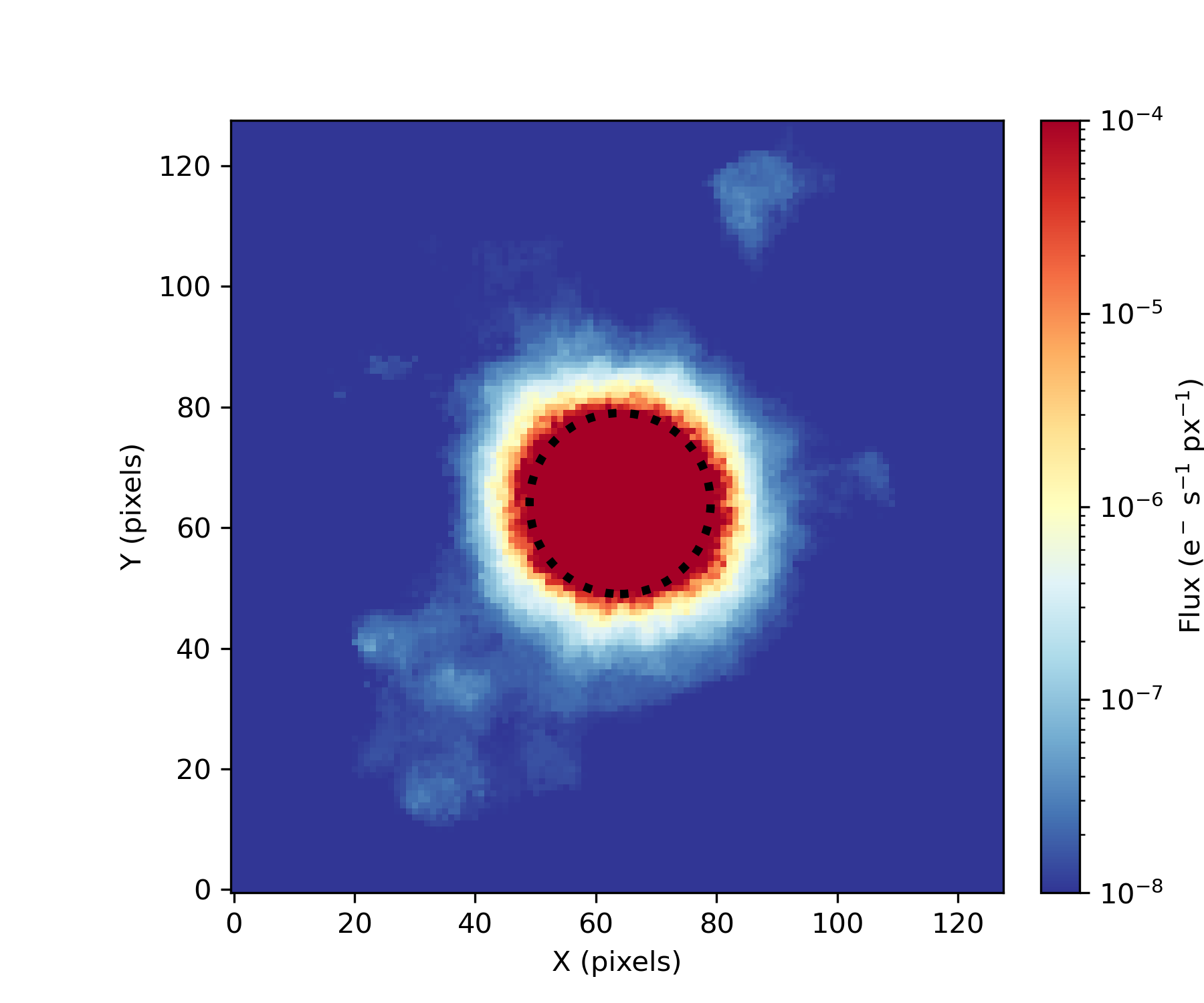}
\put(49,175){\color{black} \colorbox{white}{\textsf{SAUNAS}}}
\end{overpic}
\begin{overpic}[trim={0 0 0 0}, clip, width=0.49\textwidth]{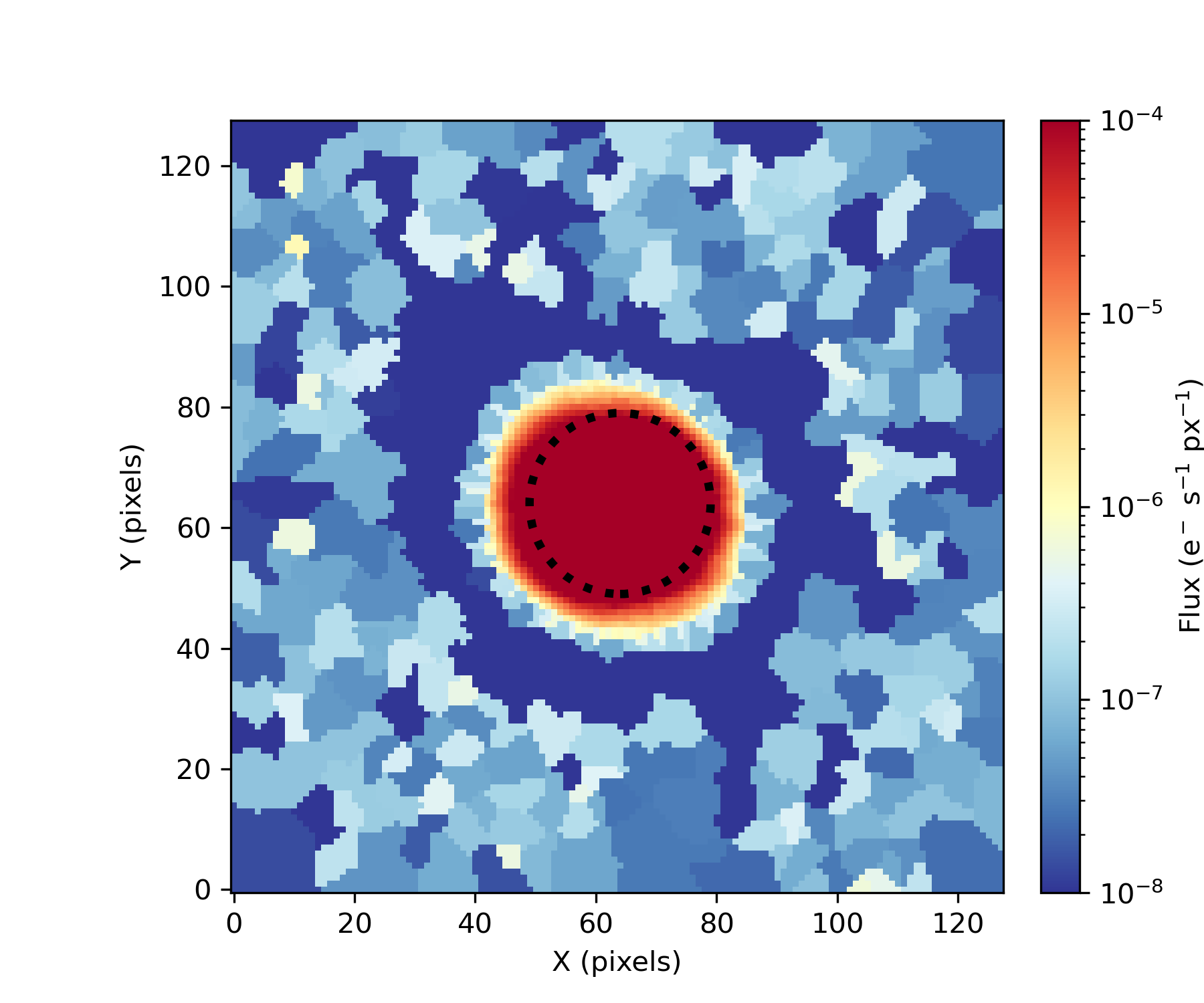}
\put(49,174){\color{black} \colorbox{white}{\textsf{CIAO/arestore}}}
\end{overpic}

\caption{Extended source PSF deconvolution test. \emph{Top left panel:} Emission from a circular source object with a central surface brightness of $\mu=$10$^{-3}$~s$^{-1}$~px$^{-1}$ and a background level of $\mu=$10$^{-7}$~s$^{-1}$~px$^{-1}$ convolved with a reference \Chandra/ACIS PSF (NGC\,3862, $\alpha=176.2709^{\circ}$, $\delta=+19.6063^{\circ}$, Obs.\,ID:\,514). \emph{Top right panel:} Surface brightness profiles of the ground-truth (non-PSF convolved) test source (red solid line), the PSF-convolved source (grey dashed-dotted line), \SAUNAS\ deconvolved image (blue dashed), and \texttt{CIAO/arestore} deconvolved (orange dotted) images. The horizontal dotted line represents the sky background of the model, and the vertical dotted line represents the radial limit of the circular test source ($R=15$ pixels). \emph{Bottom left panel:} \SAUNAS\ deconvolved image. \emph{Bottom right panel:}  \texttt{CIAO/arestore} deconvolved image. Note that the convolved image (events map) and the \texttt{CIAO/arestore} deconvolved image have been processed using Voronoi binning for visualization purposes of the surface brightness. See the legend and colorbar in the figure.} 
\label{fig:example_deco_circle}
\end{center}
\end{figure*}

Figure \ref{fig:example_deco_circle} shows the result from this analysis. A simulated source with a central surface brightness of $\mu=$10$^{-3}$~s$^{-1}$~px$^{-1}$ and a background level of $\mu=$10$^{-7}$~s$^{-1}$~px$^{-1}$ is convolved with the same PSF used by the point source tests described in Appendix \ref{Appendix:SAUNAS_psf_deconvolution_test}. The resulting event file of convolved data is then processed by \SAUNAS\ and deconvolved by a standard application of \texttt{CIAO/arestore}. A comparison of the associated surface brightness profiles  provides both quantitative and qualitative assessments of the different light reconstruction methods.

The top right panel of Fig.\,\ref{fig:example_deco_circle} shows that the methodology adopted in \SAUNAS\ produces a result that is more closely aligned with our science-driven requirements.  Proper treatment of the fainter regions surrounding objects is a  critical factor for the detection of faint extended emission, such as hot gas X-ray halos around galaxies. While \SAUNAS\ produces a well-behaved profile that smoothly transitions to the background level at large radii, \texttt{CIAO/arestore} manufactures an over-subtracted background region surrounding the object, similar to its treatment of point sources (Appendix \ref{Appendix:SAUNAS_psf_deconvolution_test}). 

Figures \ref{fig:FNFP_test_CenA_1} and \ref{fig:FNFP_test_CenA_2} show the results of the false positive / false negative quality test described in Sec.\,\ref{subsubsec:quality_FPFN} for the double jet model. In Figs.\,\ref{fig:FNFP_test_Cave_1} and \ref{fig:FNFP_test_Cave_2} the equivalent results are shown for the cavity model. Each row represents different equivalent exposure times, from $\tau_{\rm exp} = 5\times10^7$\scm to $\tau_{\rm exp} = 5\times10^4$\scm. We refer to the caption in the figures for details. 

\begin{figure*}[t!]
\begin{center}

\begin{overpic}[trim={25 32 110 30}, clip, height=0.25\textwidth]{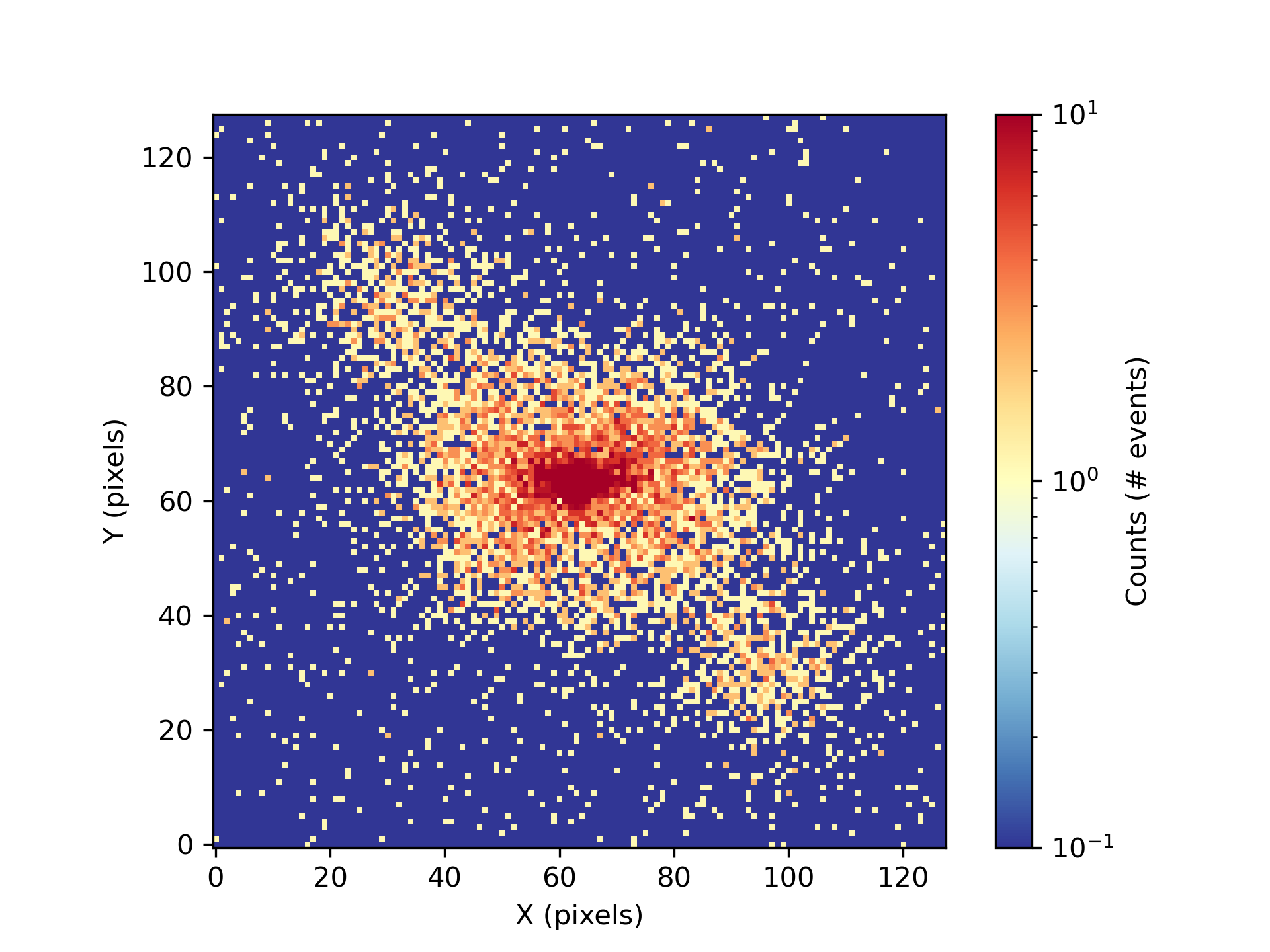}
\put(50,131){\color{black} \colorbox{white}{\textsf{$\tau_{\rm exp} = 10^7$\scm}}}
\put(19,130){\color{black} \colorbox{white}{\textsf{Events}}}
\end{overpic}
\begin{overpic}[trim={70 32 110 30}, clip, height=0.25\textwidth]{SAUNAS_CenA_5E6_events.png}
\put(50,131){\color{black} \colorbox{white}{\textsf{$\tau_{\rm exp} = 5\times10^6$\scm}}}
\end{overpic}
\begin{overpic}[trim={70 32 25 30}, clip, height=0.25\textwidth]{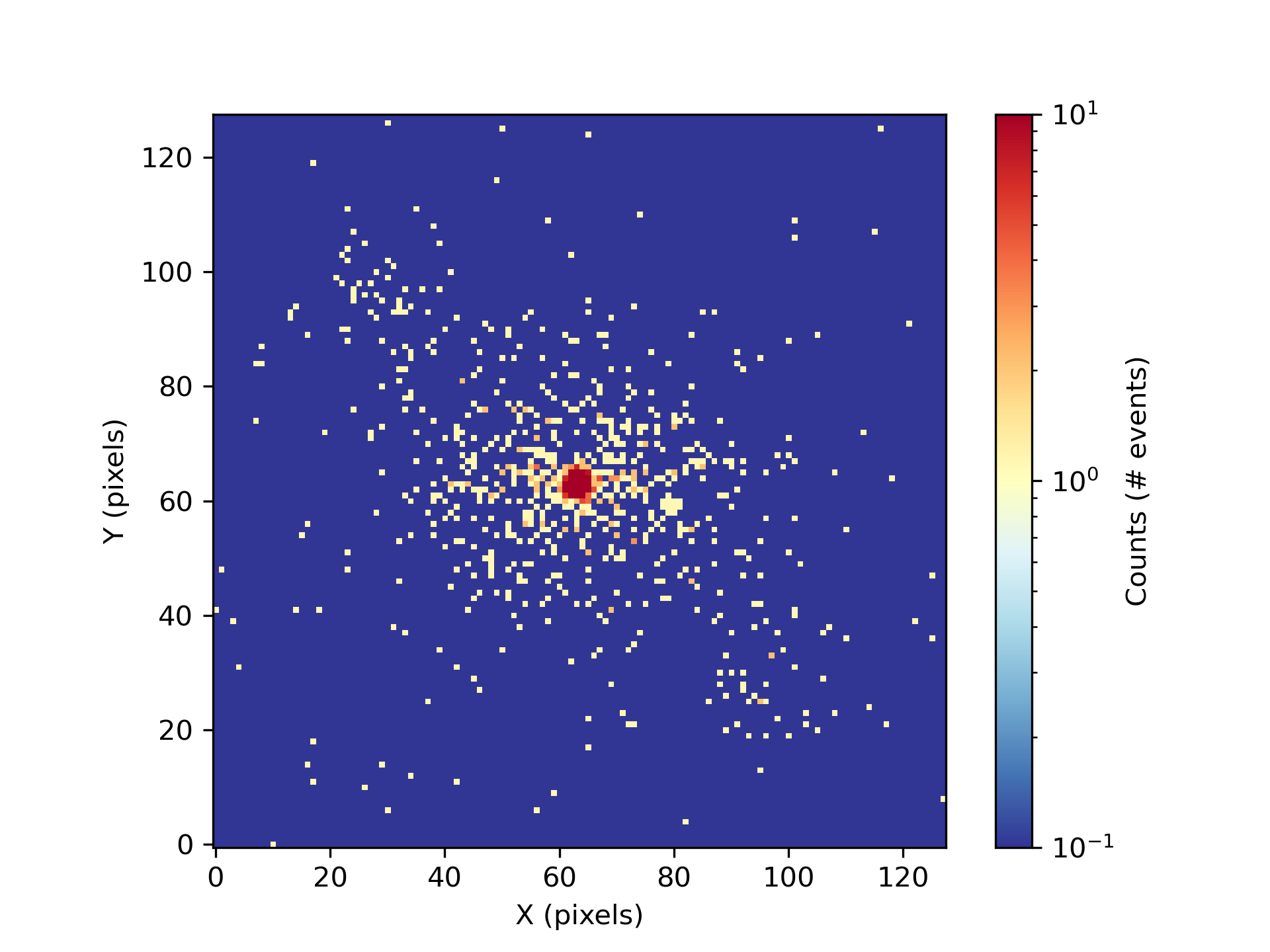}
\put(50,131){\color{black} \colorbox{white}{\textsf{$\tau_{\rm exp} = 10^6$\scm}}}
\end{overpic}

\begin{overpic}[trim={25 32 110 30}, clip, height=0.25\textwidth]{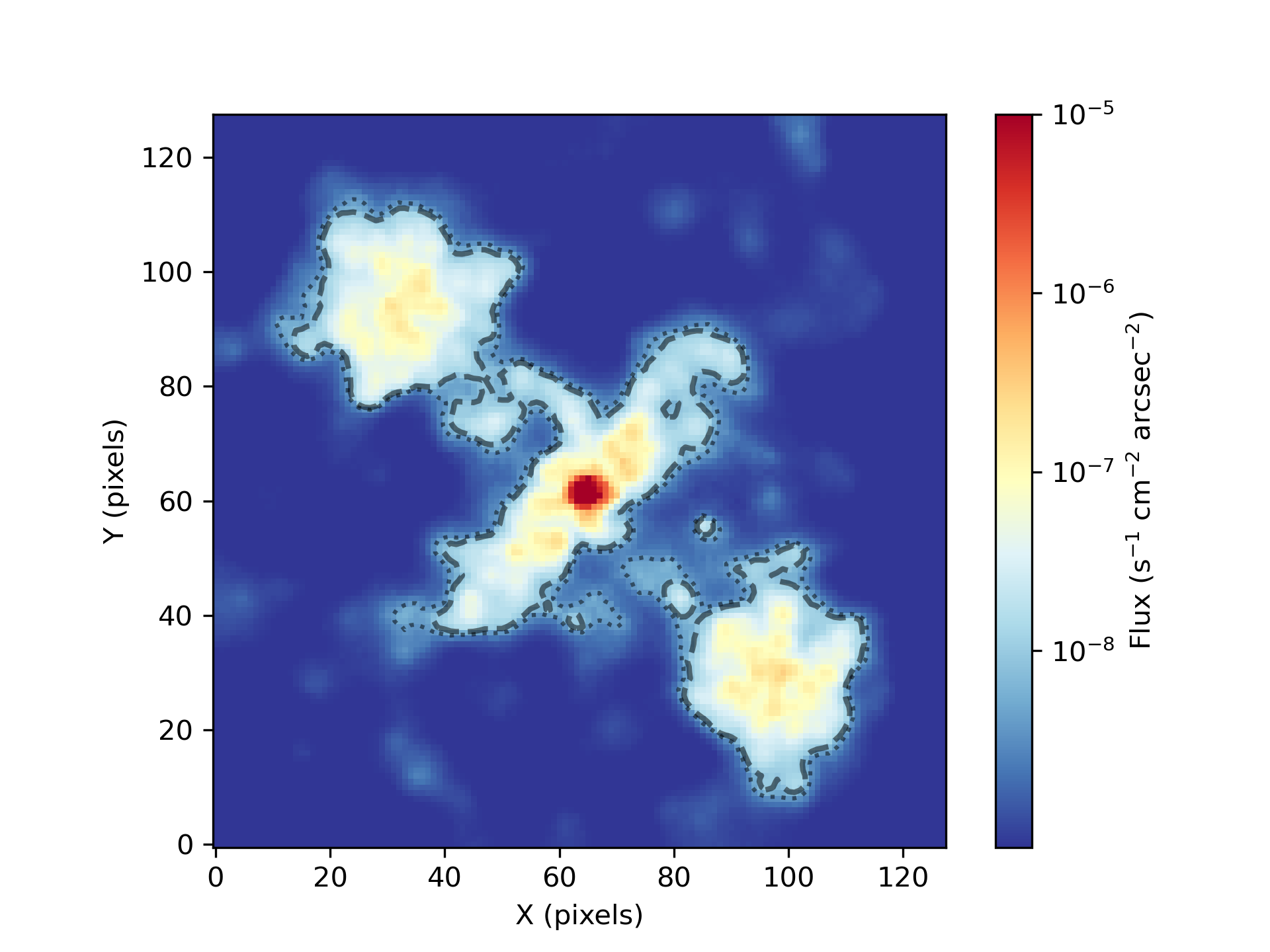}
\put(24,113.5){\color{black} \colorbox{white}{\textsf{Final}}}
\end{overpic}
\begin{overpic}[trim={70 32 110 30}, clip, height=0.25\textwidth]{SAUNAS_CenA_5E6_voronoi.png}
\end{overpic}
\begin{overpic}[trim={70 32 25 30}, clip, height=0.25\textwidth]{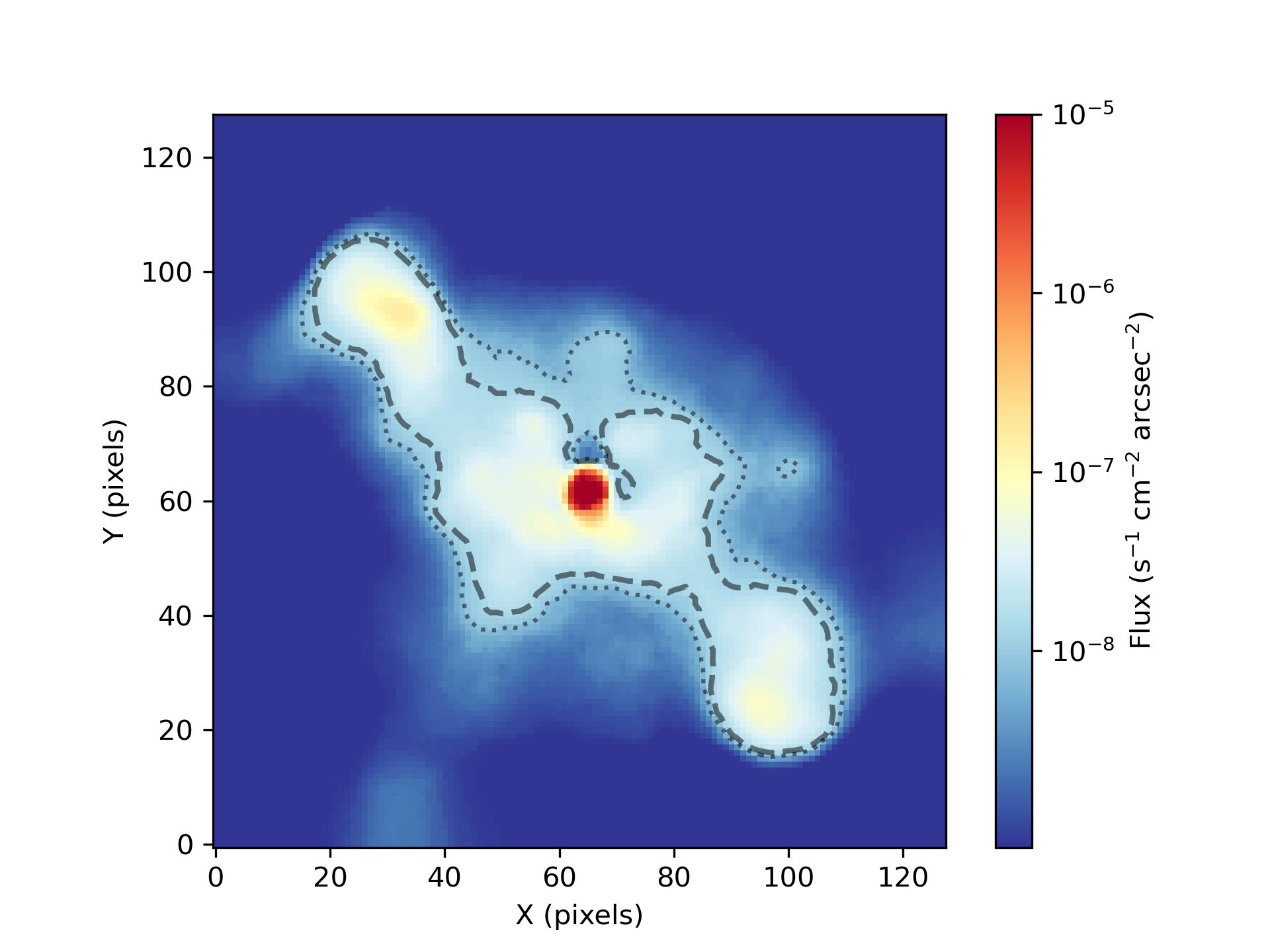}
\end{overpic}

\begin{overpic}[trim={62 0 80 30}, clip, height=0.277\textwidth]{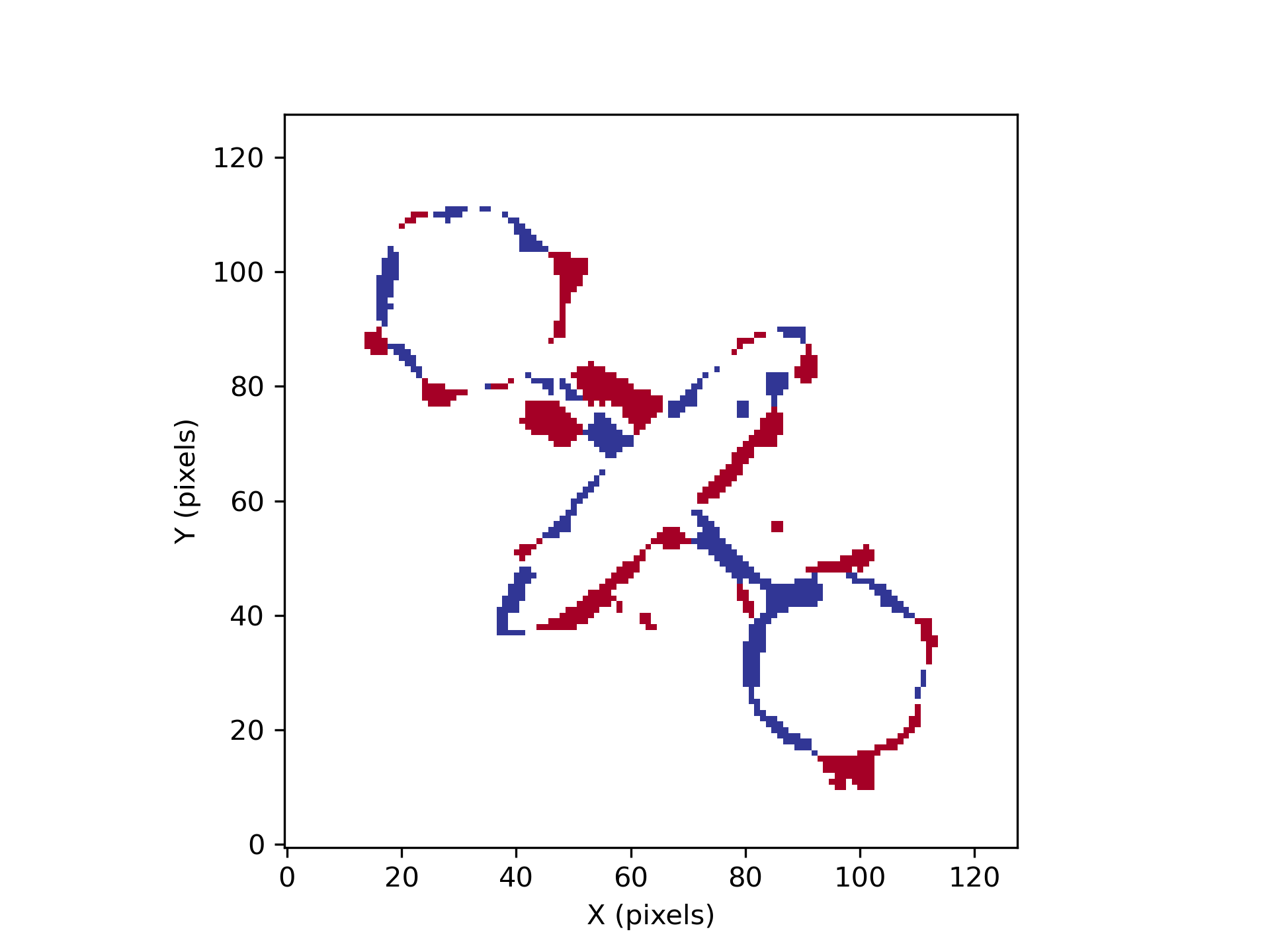}
\put(19,127){\color{black} \colorbox{white}{\textsf{False positive/negative}}}
\end{overpic}
\begin{overpic}[trim={102 0 75 30}, clip, height=0.277\textwidth]{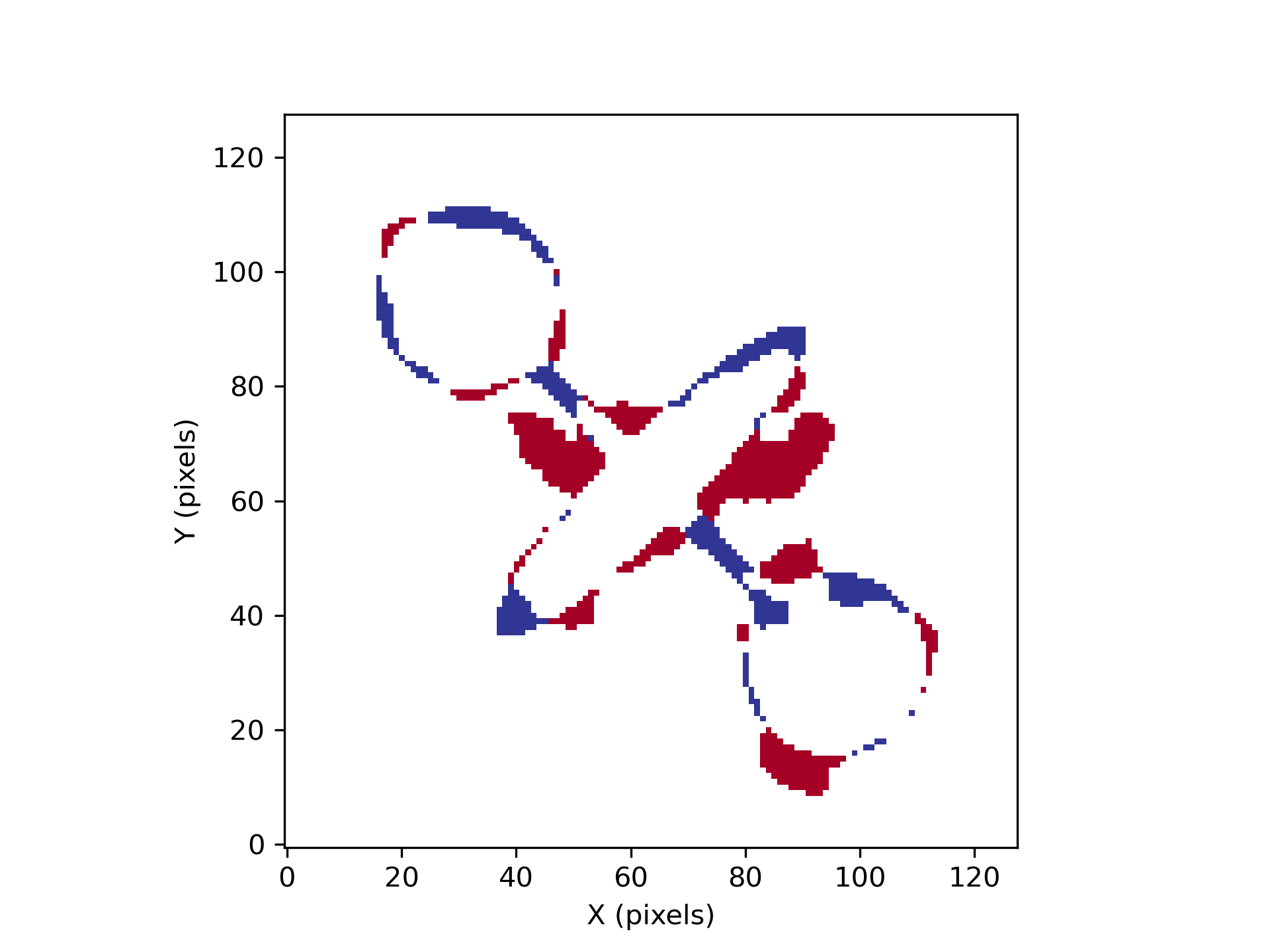}
\end{overpic}
\begin{overpic}[trim={100 0 25 30}, clip, height=0.277\textwidth]{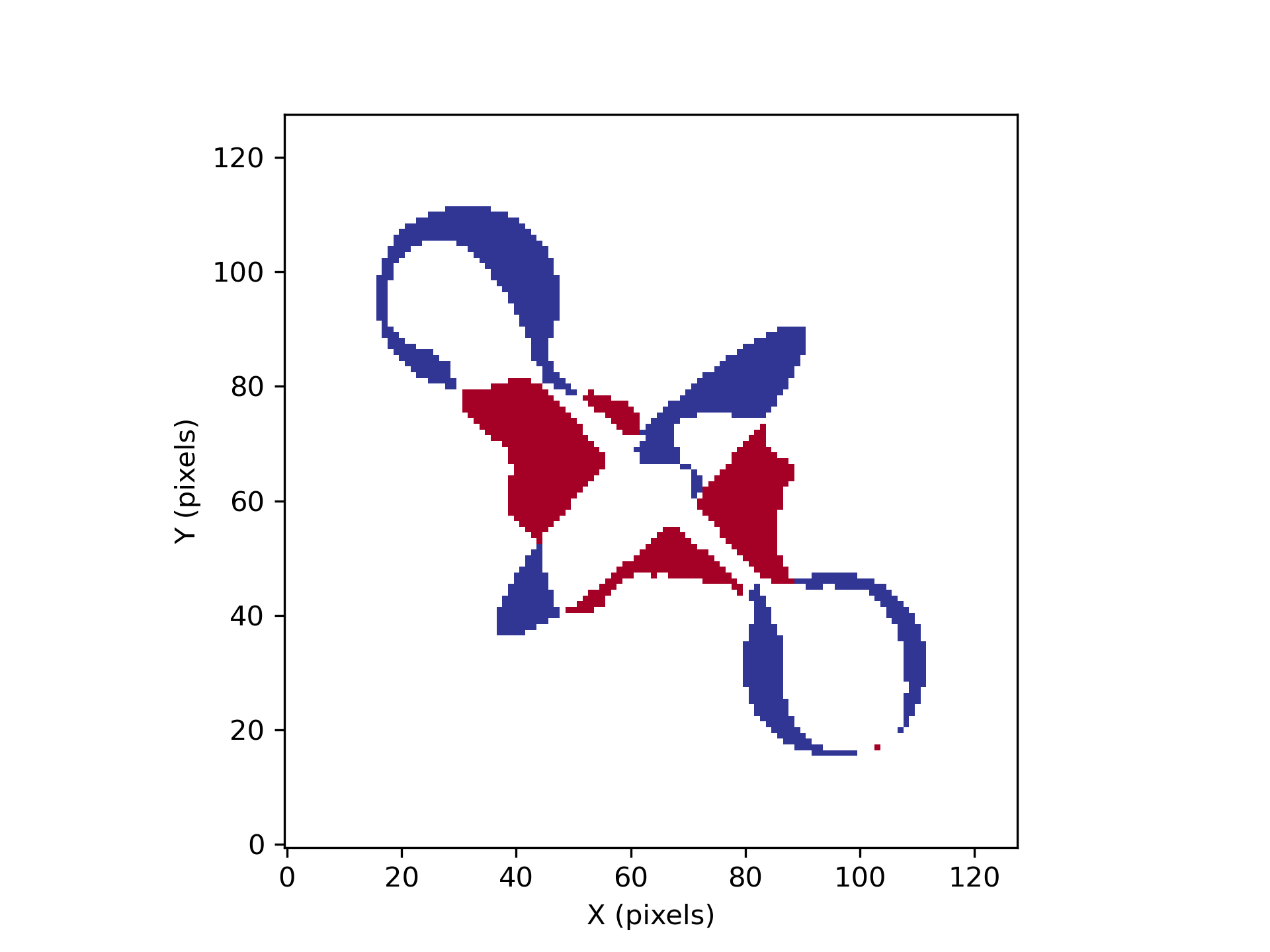}
\end{overpic}

\caption{\SAUNAS\ processing test using the double jet model as a function of the equivalent exposure time (see Sec.\,\ref{subsubsec:quality_FPFN} and Table \ref{tab:test_model_properties}). Top row shows the simulated event images. Middle row shows the final recovered surface brightness maps after processing with \SAUNAS. Dashed contours represent the $3\sigma$ and dotted contours the $2\sigma$ detection level of X-ray emission. Bottom row represents the false positive (red) and false negative (blue) detection maps for each simulation (see Sec.\,\ref{subsubsec:quality_FPFN}). From left to right columns, the equivalent exposure times are $\tau_{\rm exp}=10^7$, $5\times10^6$, and $10^6$\scm. See the labels in the panels. Colorbars represent the number of events per pixel (event images) and the surface brightness flux (final mosaics).} 
\label{fig:FNFP_test_CenA_1}
\end{center}
\end{figure*}

\begin{figure*}[t!]
\begin{center}
\begin{overpic}[trim={25 32 110 30}, clip, height=0.25\textwidth]{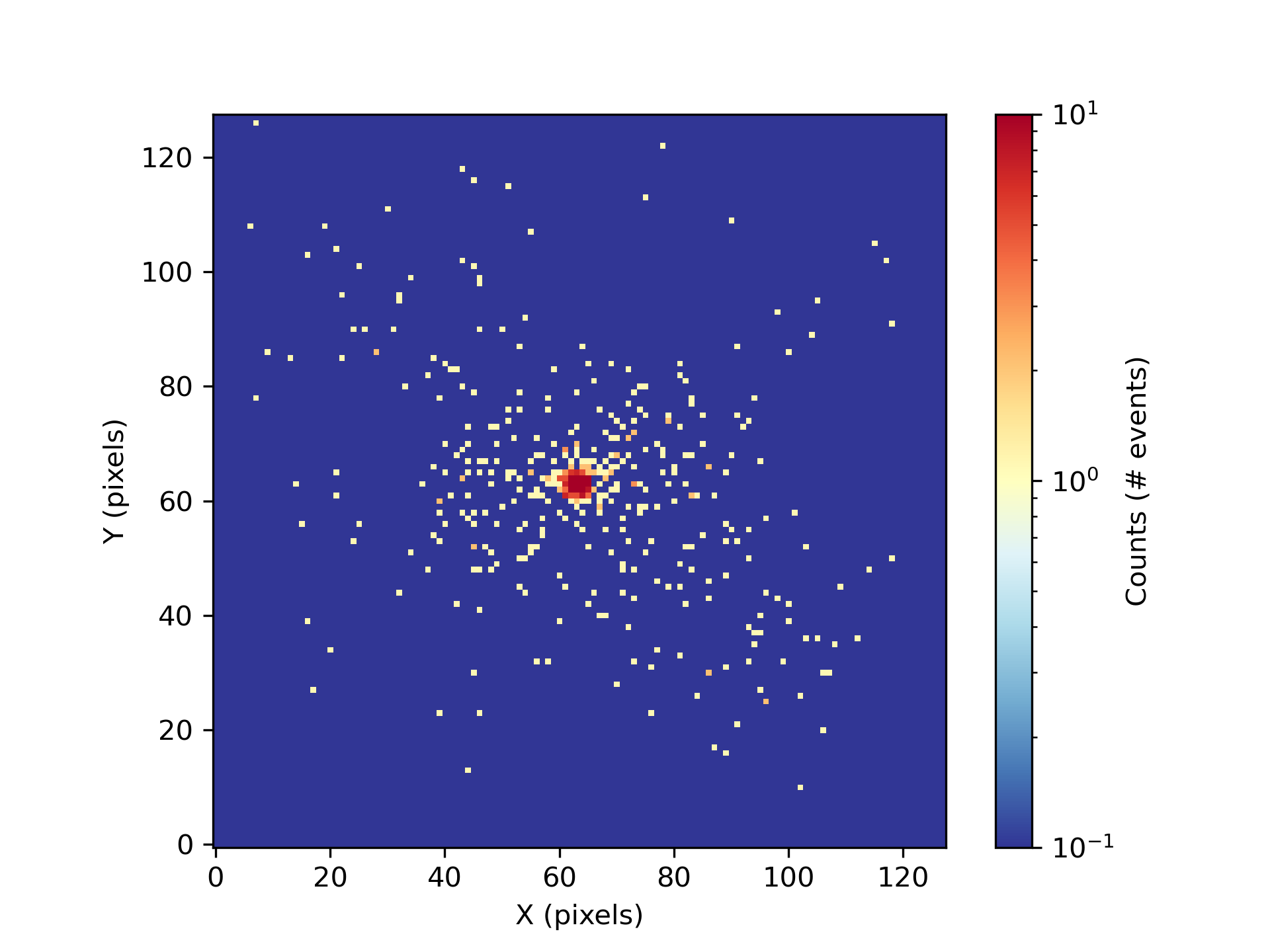}
\put(50,131){\color{black} \colorbox{white}{\textsf{$\tau_{\rm exp} = 5\times10^5$\scm}}}
\put(19,130){\color{black} \colorbox{white}{\textsf{Events}}}
\end{overpic}
\begin{overpic}[trim={70 32 110 30}, clip, height=0.25\textwidth]{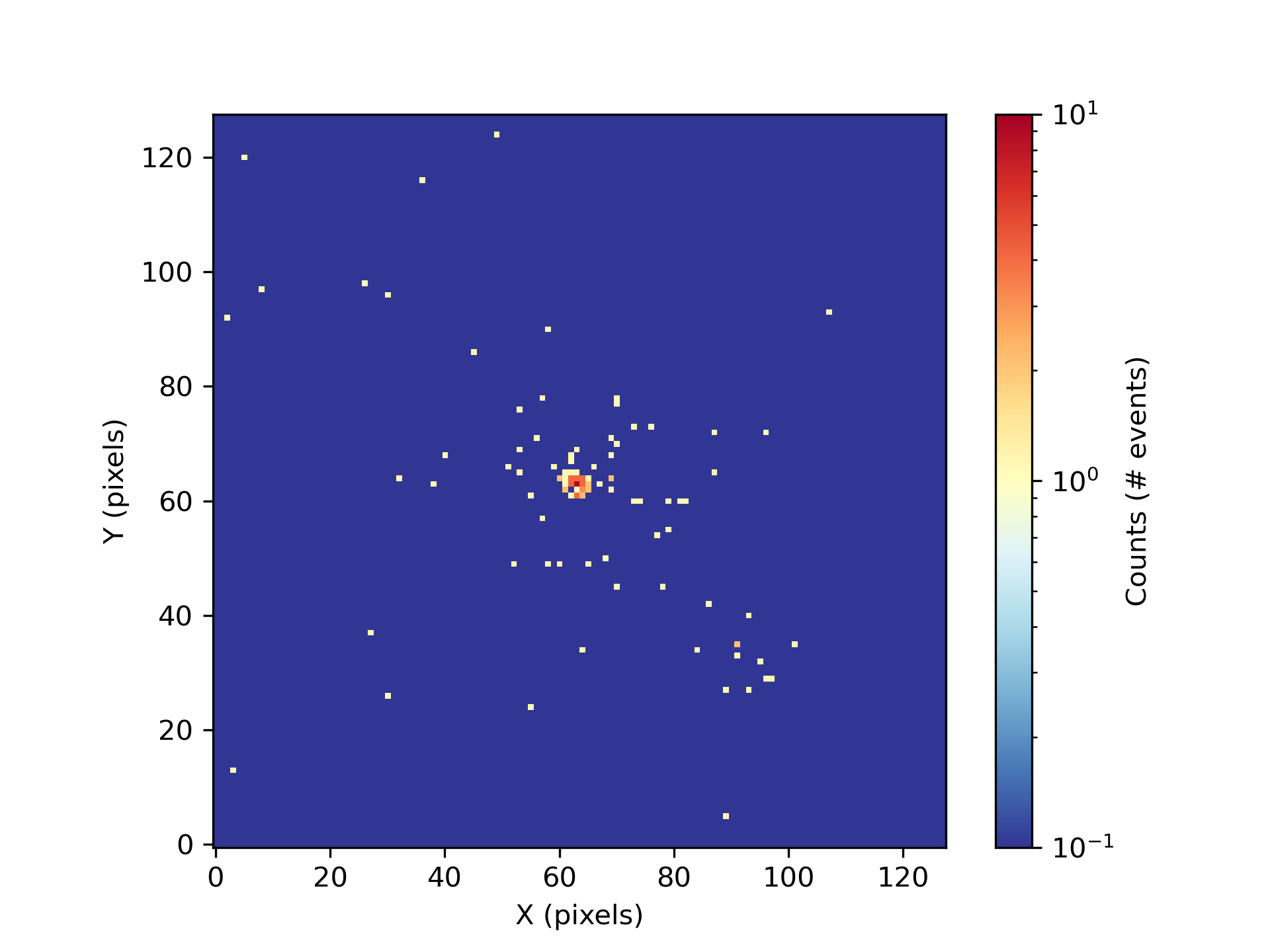}
\put(50,131){\color{black} \colorbox{white}{\textsf{$\tau_{\rm exp} = 10^5$~s cm$^{2}$}}}
\end{overpic}
\begin{overpic}[trim={70 32 25 30}, clip, height=0.25\textwidth]{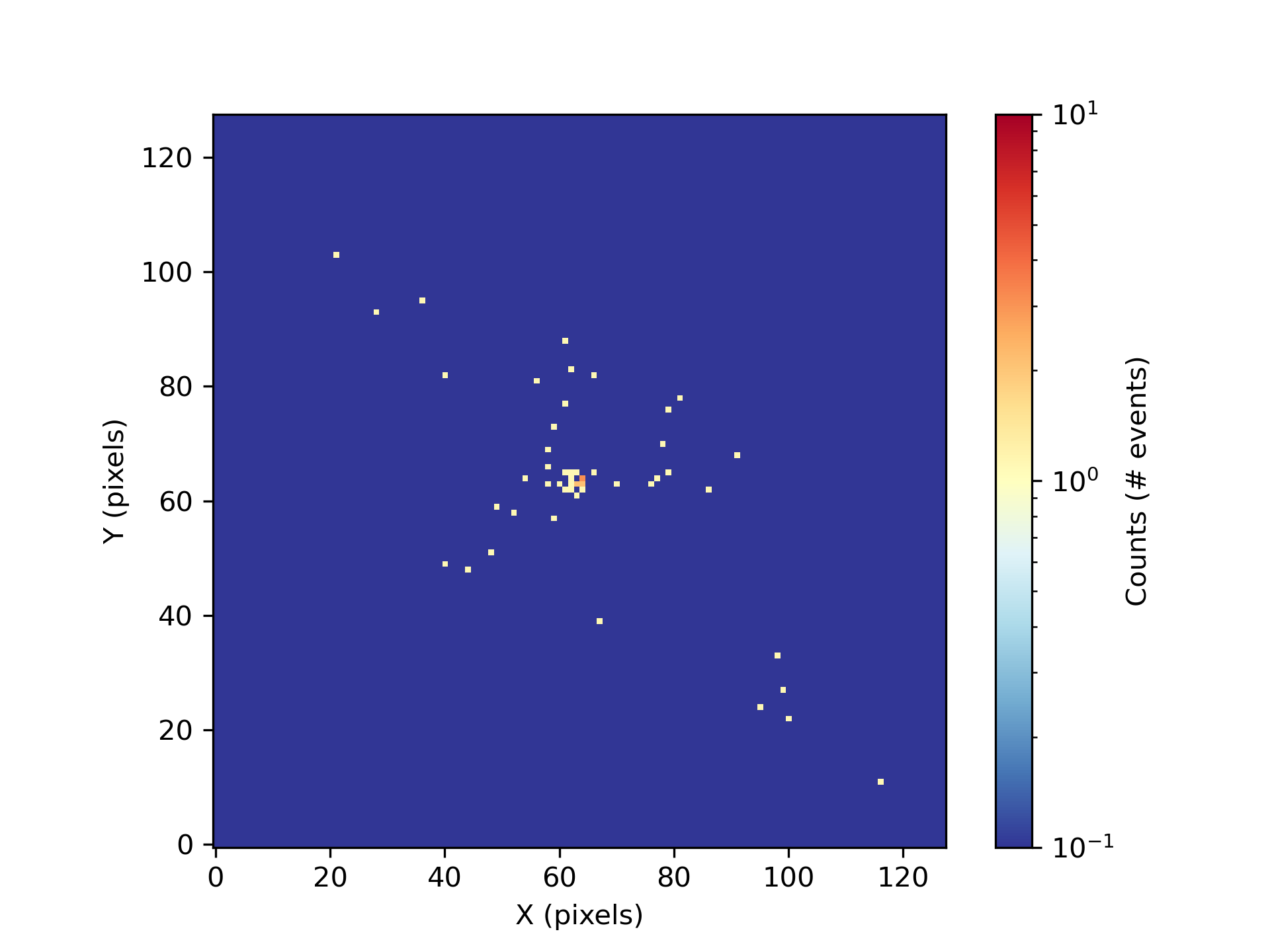}
\put(50,131){\color{black} \colorbox{white}{\textsf{$\tau_{\rm exp} = 5\times10^4$\scm}}}
\end{overpic}

\begin{overpic}[trim={25 32 110 30}, clip, height=0.25\textwidth]{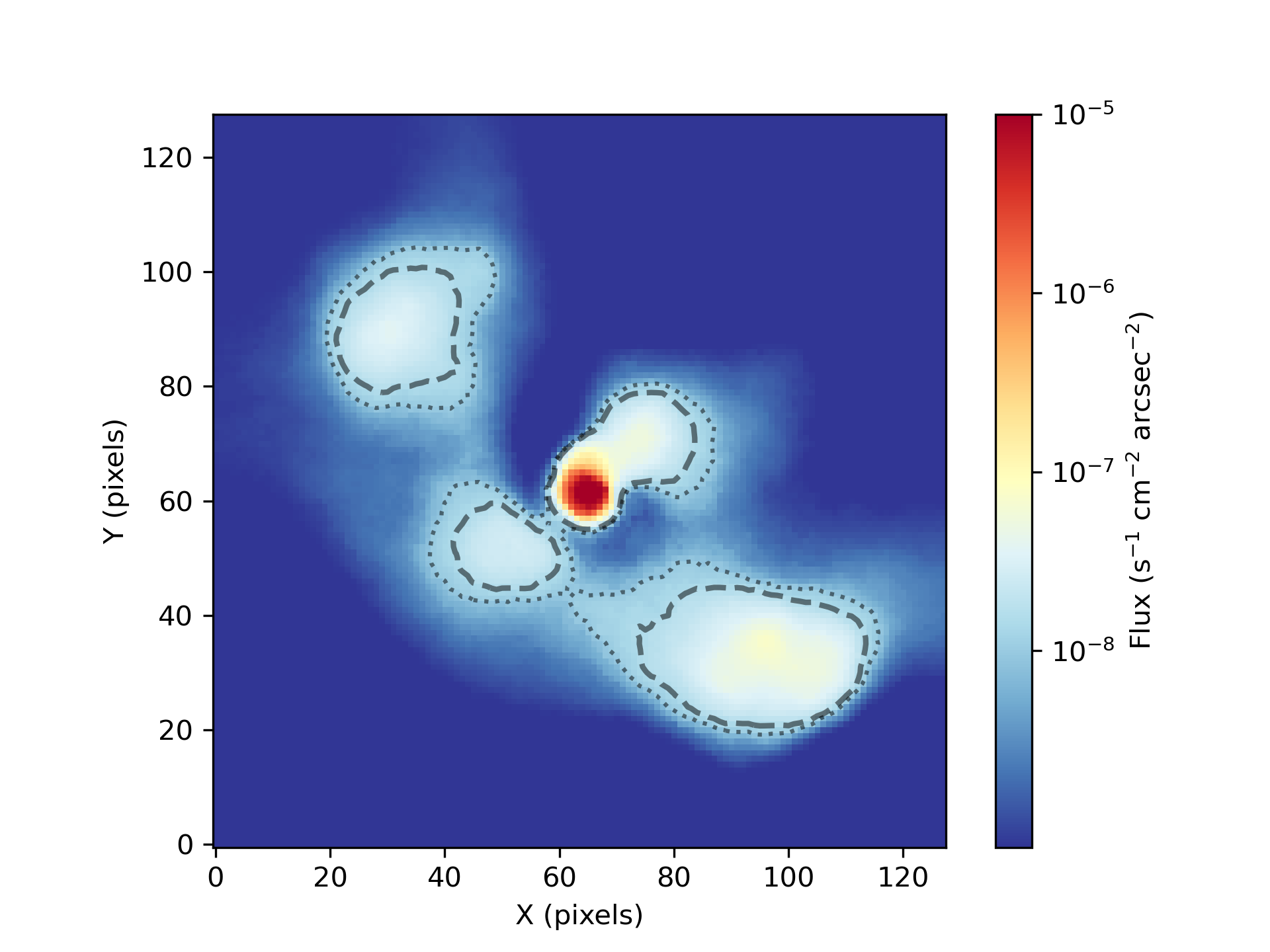}
\put(24,113.5){\color{black} \colorbox{white}{\textsf{Final}}}
\end{overpic}
\begin{overpic}[trim={70 32 110 30}, clip, height=0.25\textwidth]{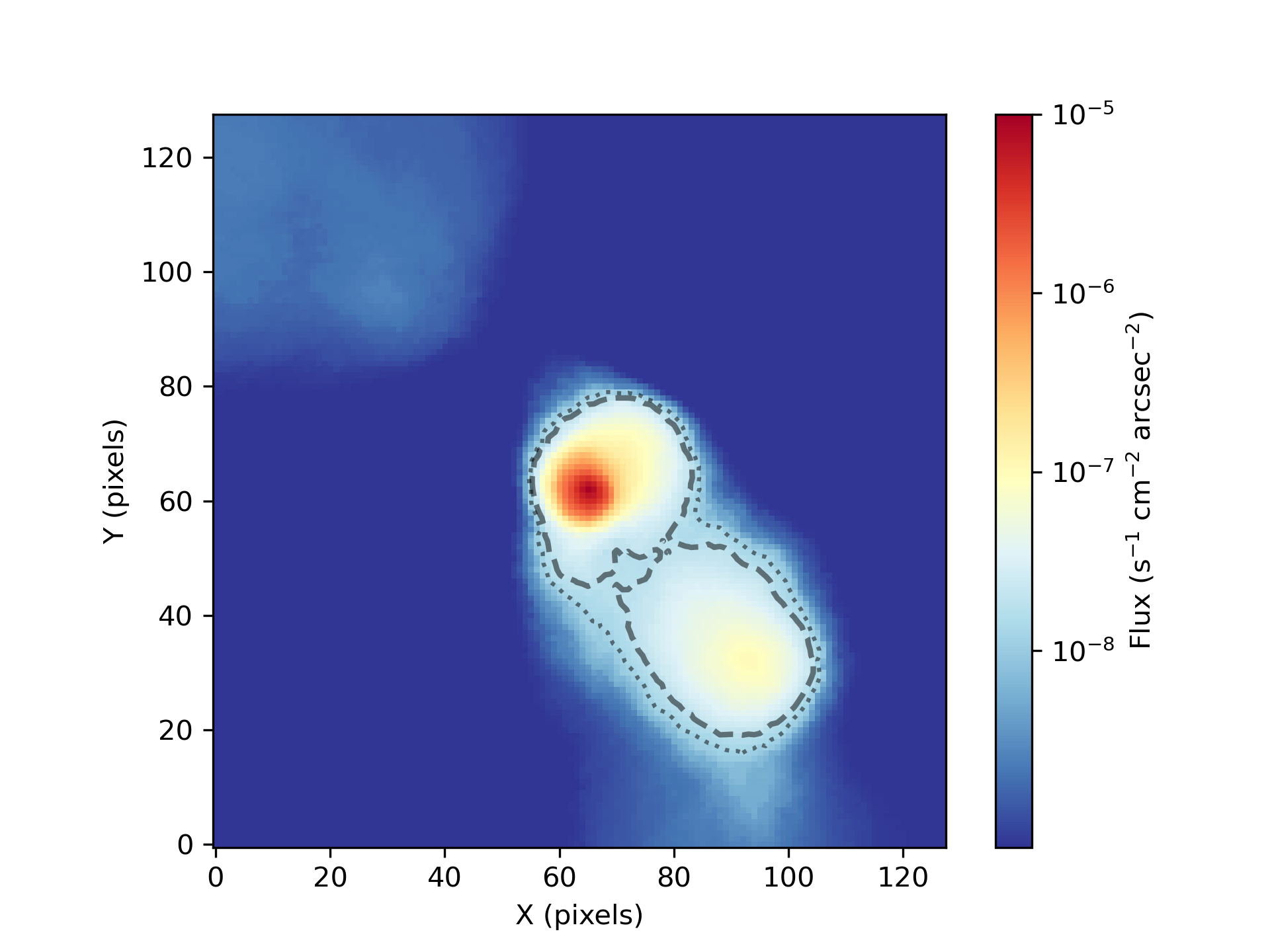}
\end{overpic}
\begin{overpic}[trim={70 32 25 30}, clip, height=0.25\textwidth]{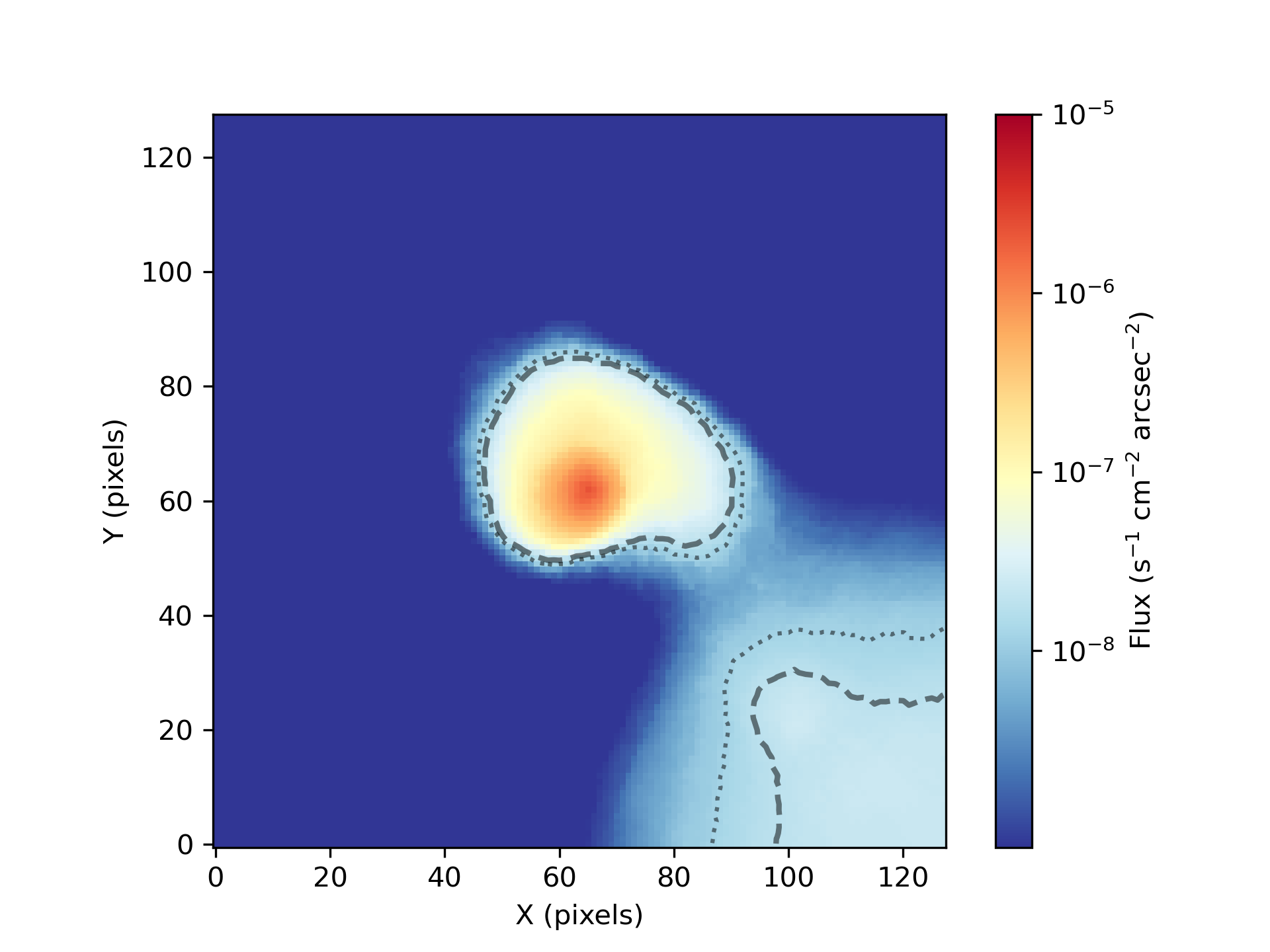}
\end{overpic}

\begin{overpic}[trim={62 0 80 30}, clip, height=0.277\textwidth]{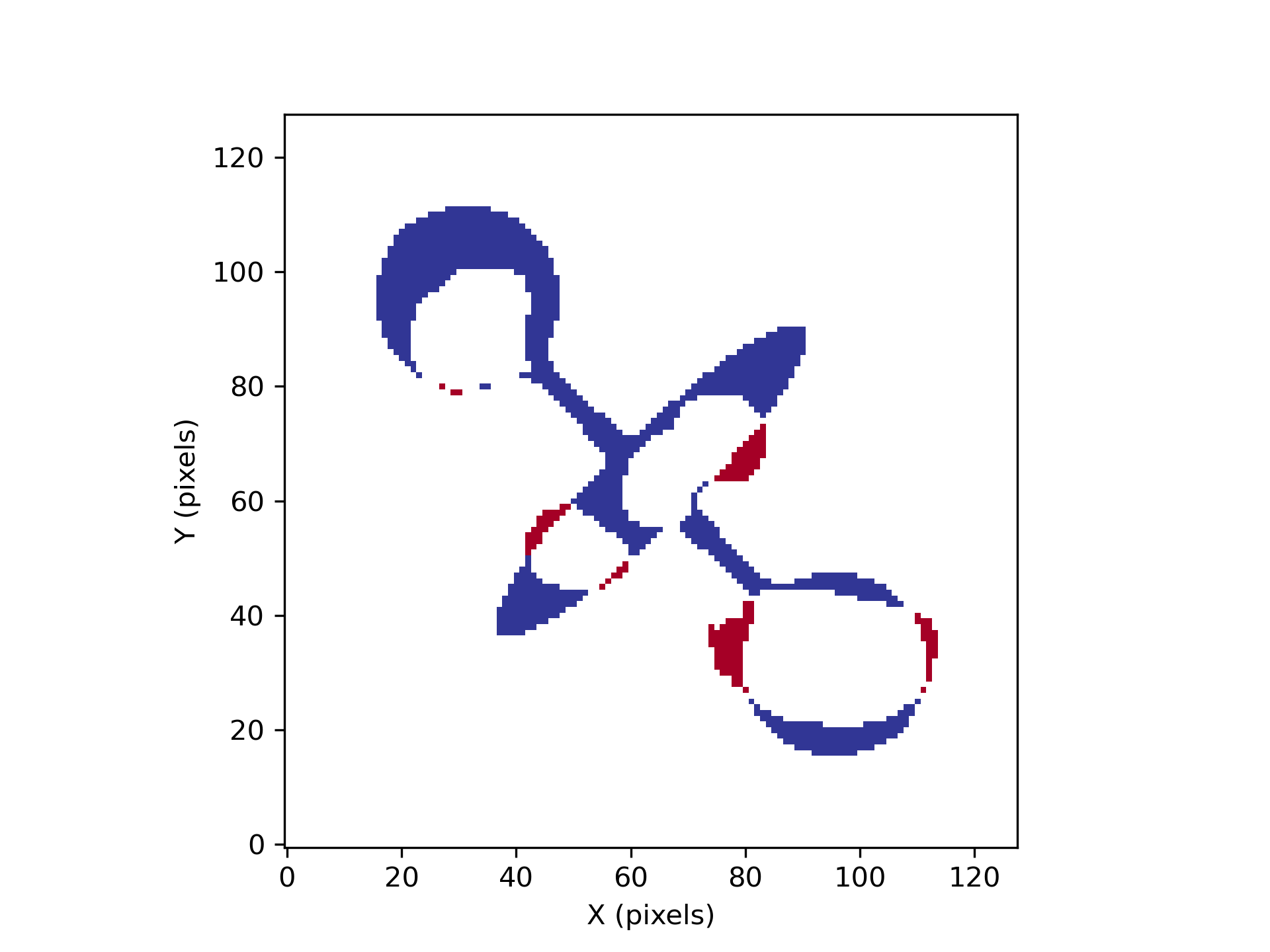}
\put(19,127){\color{black} \colorbox{white}{\textsf{False positive/negative}}}
\end{overpic}
\begin{overpic}[trim={102 0 75 30}, clip, height=0.277\textwidth]{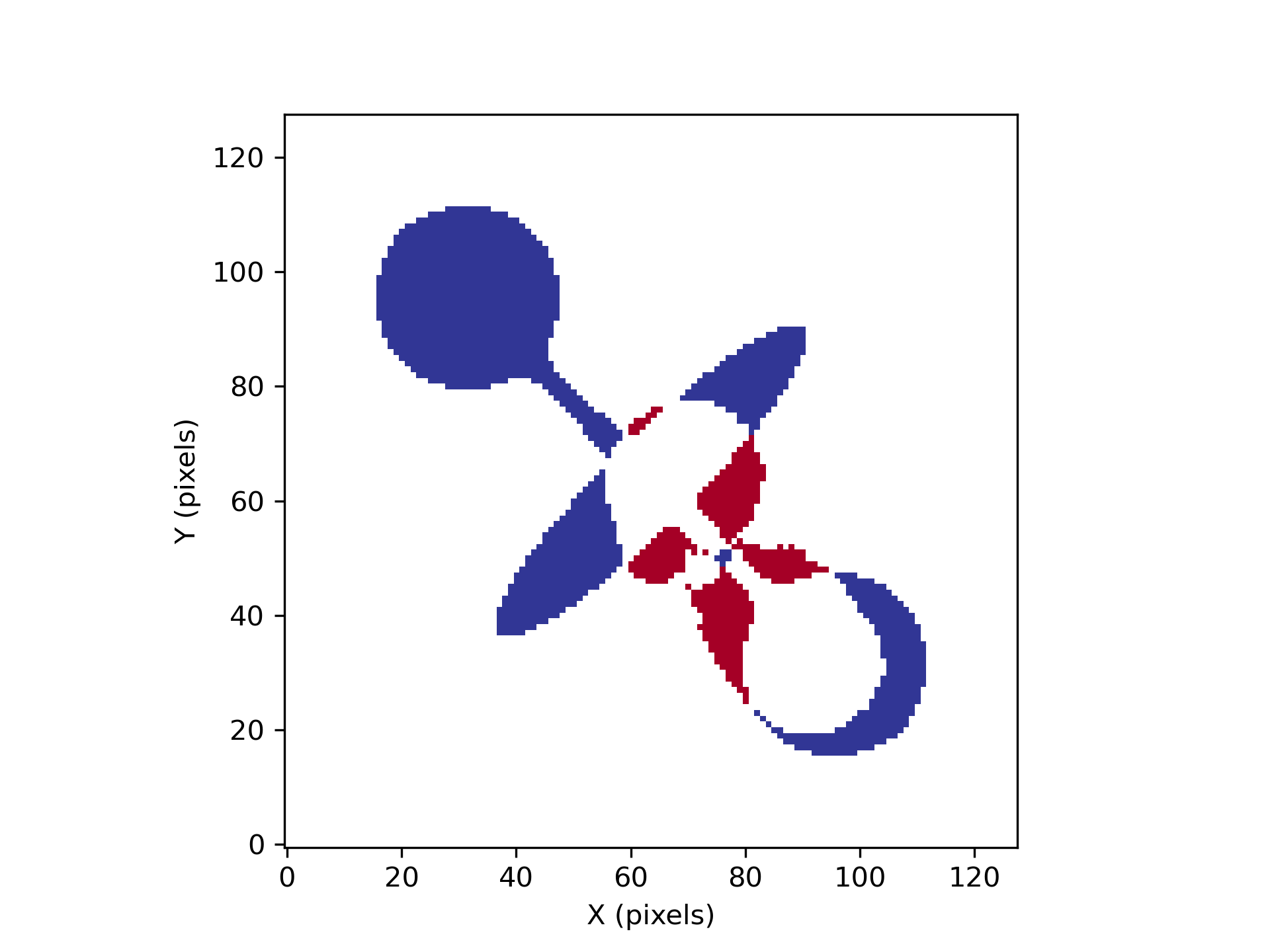}
\end{overpic}
\begin{overpic}[trim={100 0 25 30}, clip, height=0.277\textwidth]{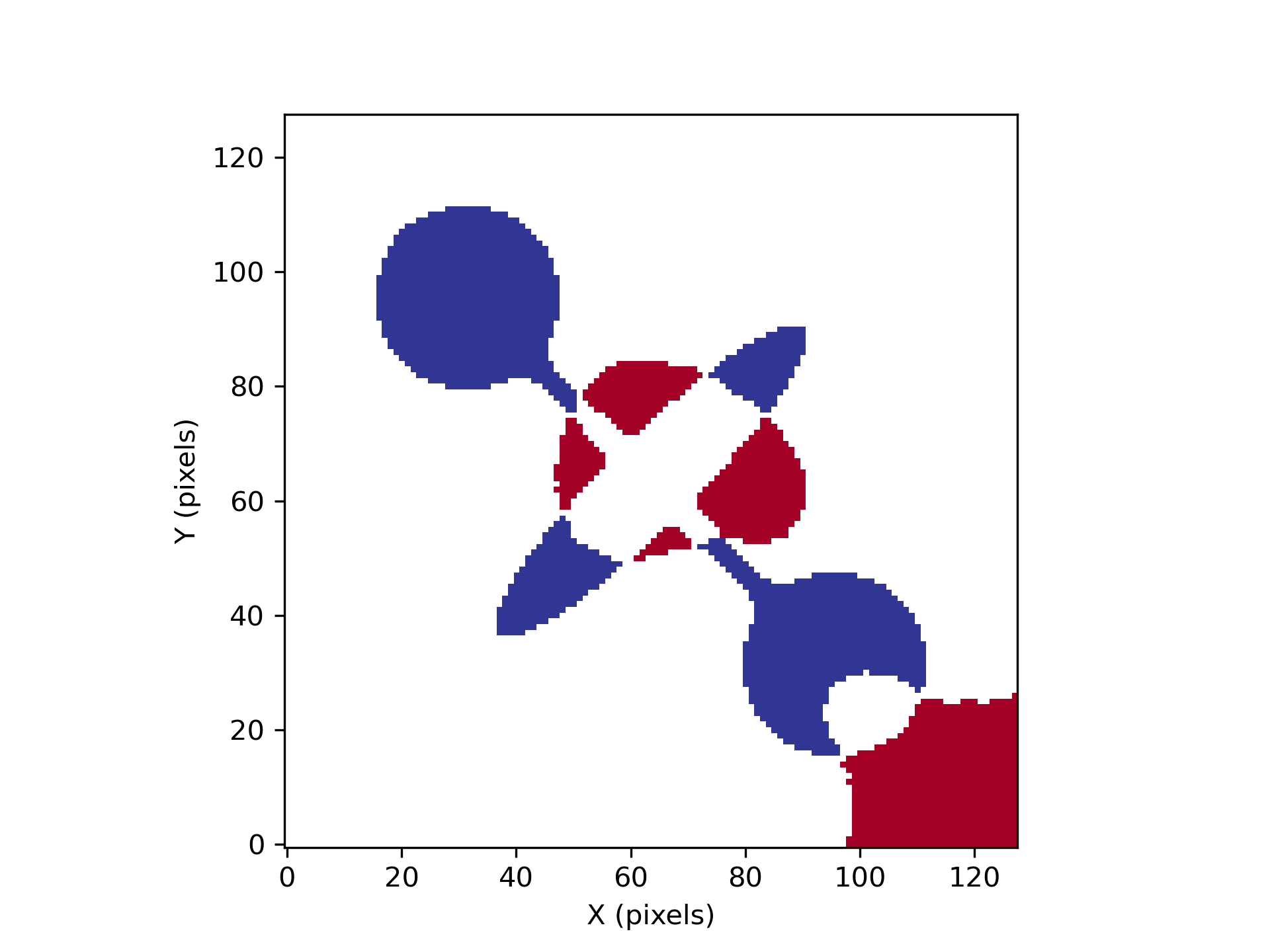}
\end{overpic}

\caption{(Continuation of Fig.\,\ref{fig:FNFP_test_CenA_1}) \SAUNAS\ processing test using the double jet model as a function of the equivalent exposure time (see Sec.\,\ref{subsubsec:quality_FPFN} and Table \ref{tab:test_model_properties}). Top row shows the simulated event images. Middle row shows the final recovered surface brightness maps after processing with \SAUNAS. Dashed contours represent the $3\sigma$ and dotted contours the $2\sigma$ detection level of X-ray emission. Bottom row represents the false positive (red) and false negative (blue) detection maps for each simulation (see Sec.\,\ref{subsubsec:quality_FPFN}). From left to right columns, the equivalent exposure times are $\tau_{\rm exp}=5\cdot10^5$, $10^5$, and $5\times10^4$\scm. See the labels in the panels. Colorbars represent the number of events per pixel (event images) and the surface brightness flux (final mosaics).} 
\label{fig:FNFP_test_CenA_2}
\end{center}
\end{figure*}

\begin{figure*}[t!]
\begin{center}

\begin{overpic}[trim={25 32 110 30}, clip, height=0.25\textwidth]{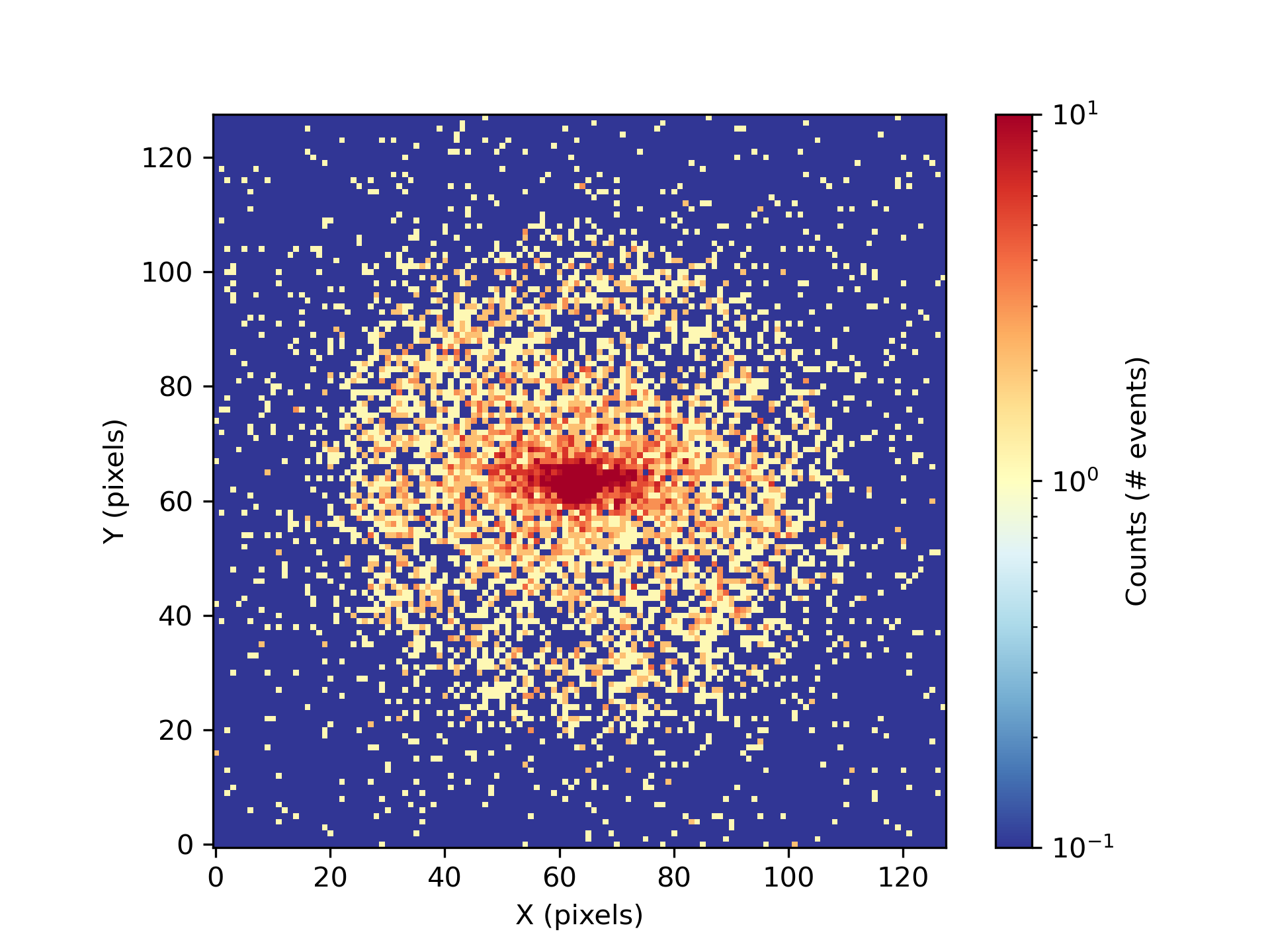}
\put(50,131){\color{black} \colorbox{white}{\textsf{$\tau_{\rm exp} = 10^7$\scm}}}
\put(19,130){\color{black} \colorbox{white}{\textsf{Events}}}
\end{overpic}
\begin{overpic}[trim={70 32 110 30}, clip, height=0.25\textwidth]{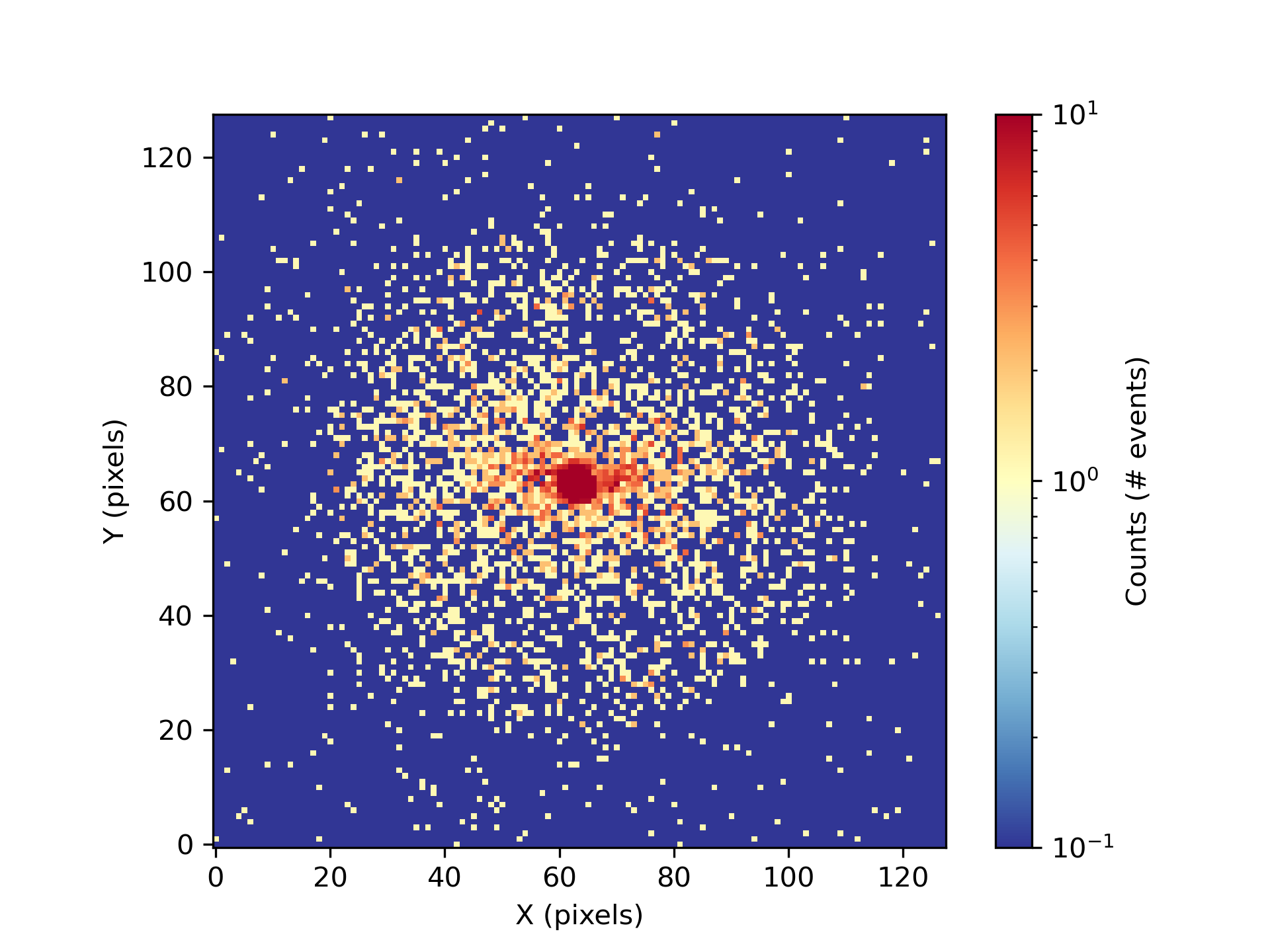}
\put(50,131){\color{black} \colorbox{white}{\textsf{$\tau_{\rm exp} = 5\times10^6$\scm}}}
\end{overpic}
\begin{overpic}[trim={70 32 25 30}, clip, height=0.25\textwidth]{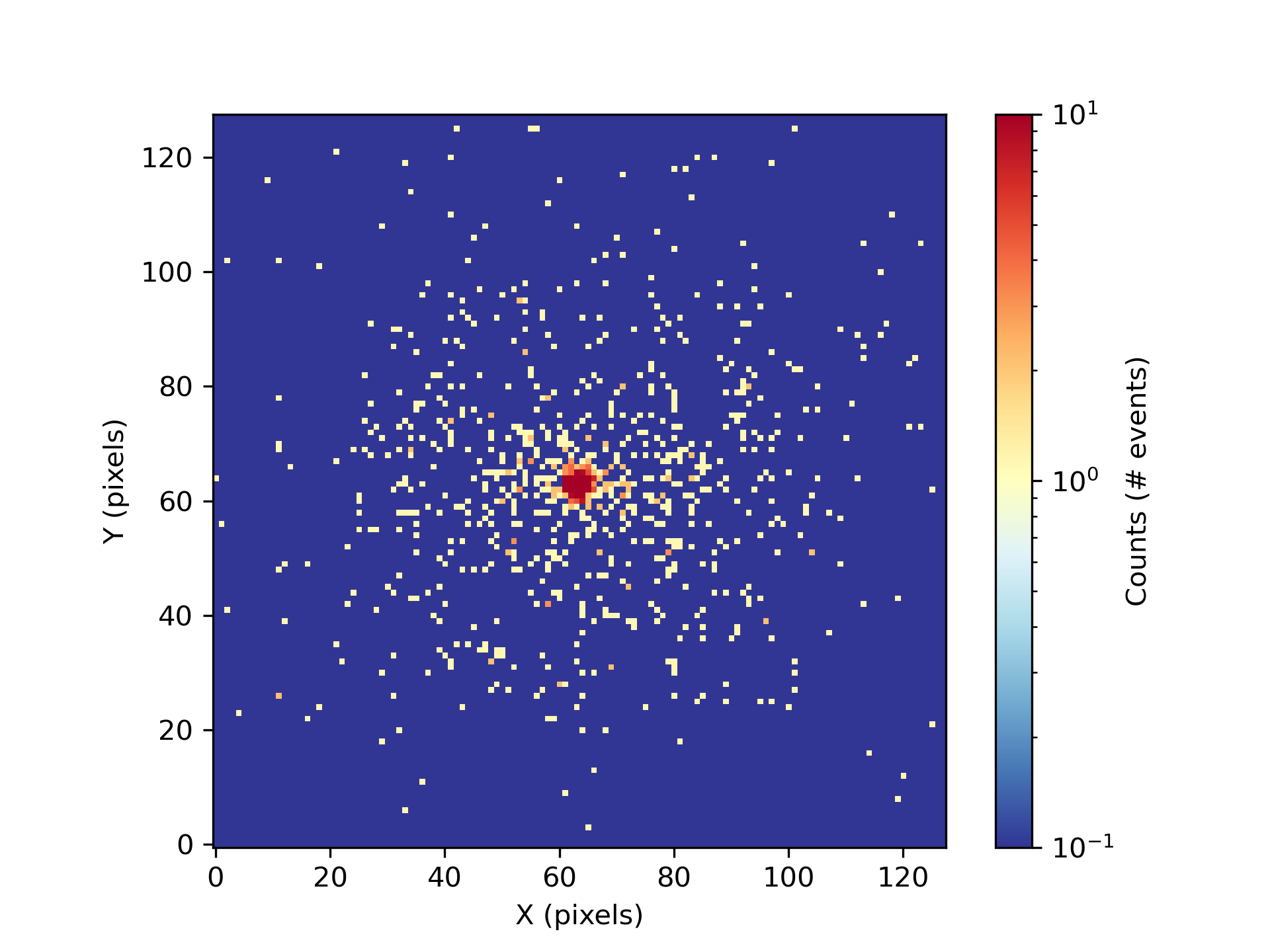}
\put(50,131){\color{black} \colorbox{white}{\textsf{$\tau_{\rm exp} = 10^6$\scm}}}
\end{overpic}

\begin{overpic}[trim={25 32 110 30}, clip, height=0.25\textwidth]{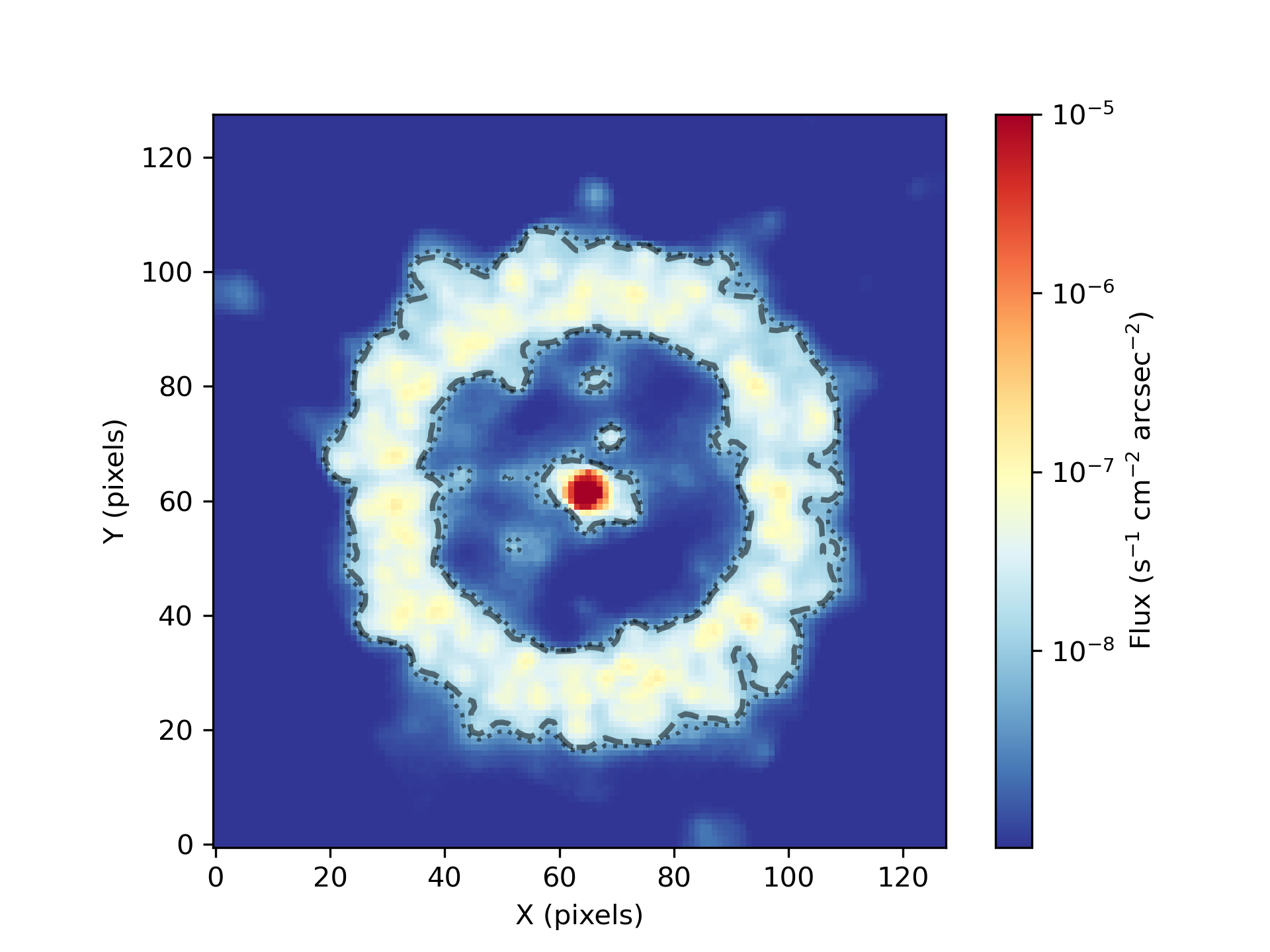}
\put(24,113.5){\color{black} \colorbox{white}{\textsf{Final}}}
\end{overpic}
\begin{overpic}[trim={70 32 110 30}, clip, height=0.25\textwidth]{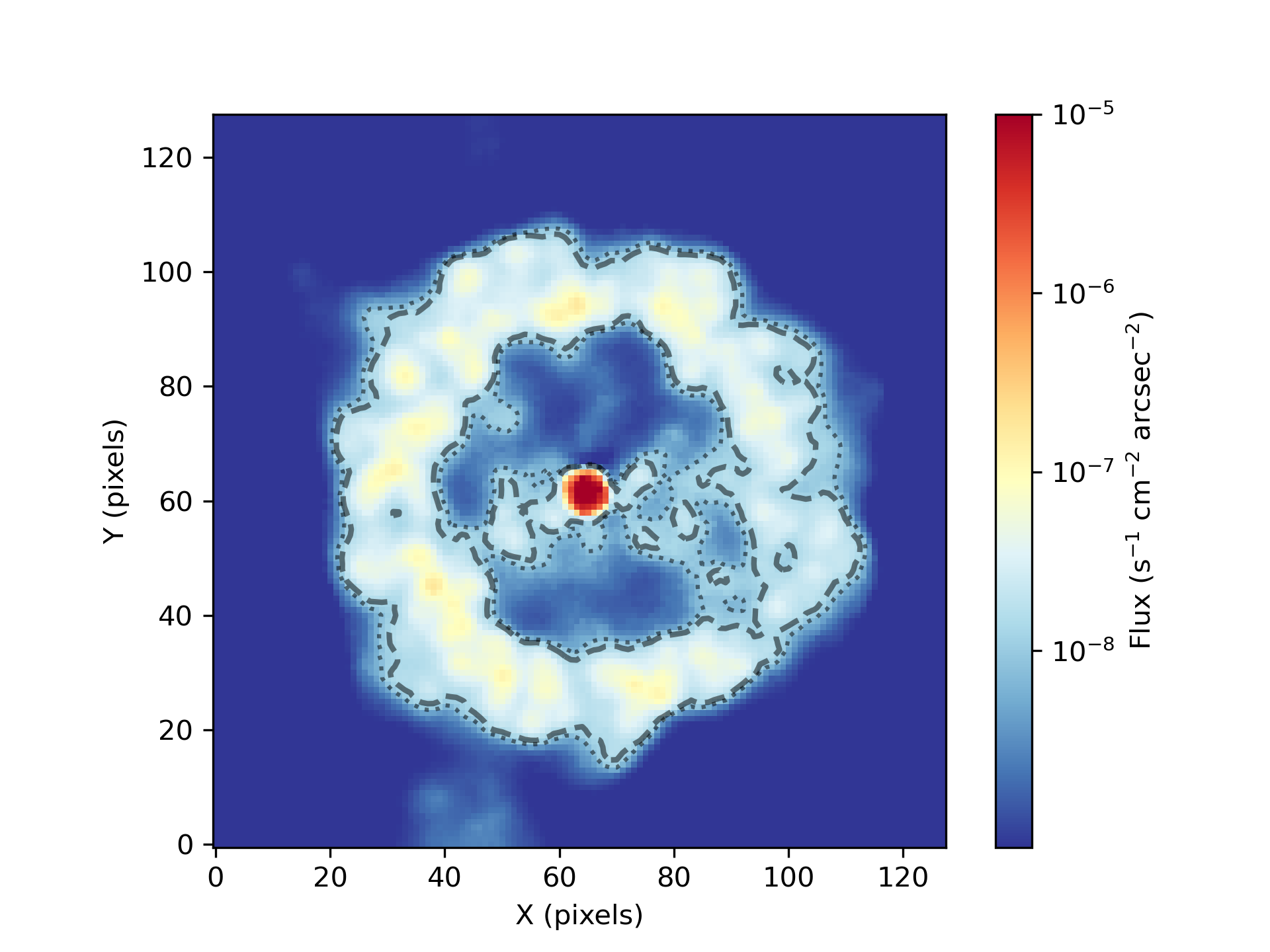}
\end{overpic}
\begin{overpic}[trim={70 32 25 30}, clip, height=0.25\textwidth]{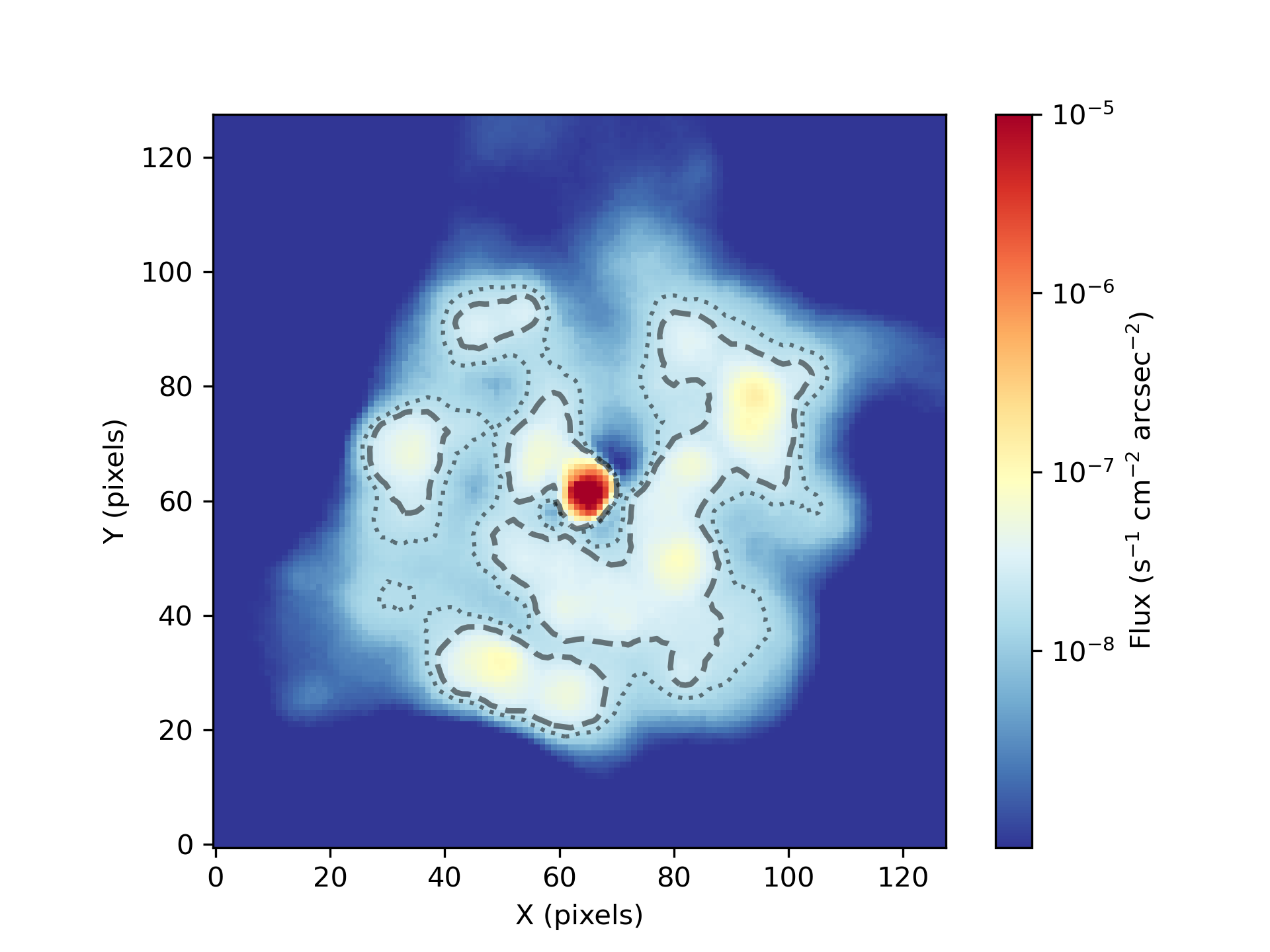}
\end{overpic}

\begin{overpic}[trim={62 0 80 30}, clip, height=0.277\textwidth]{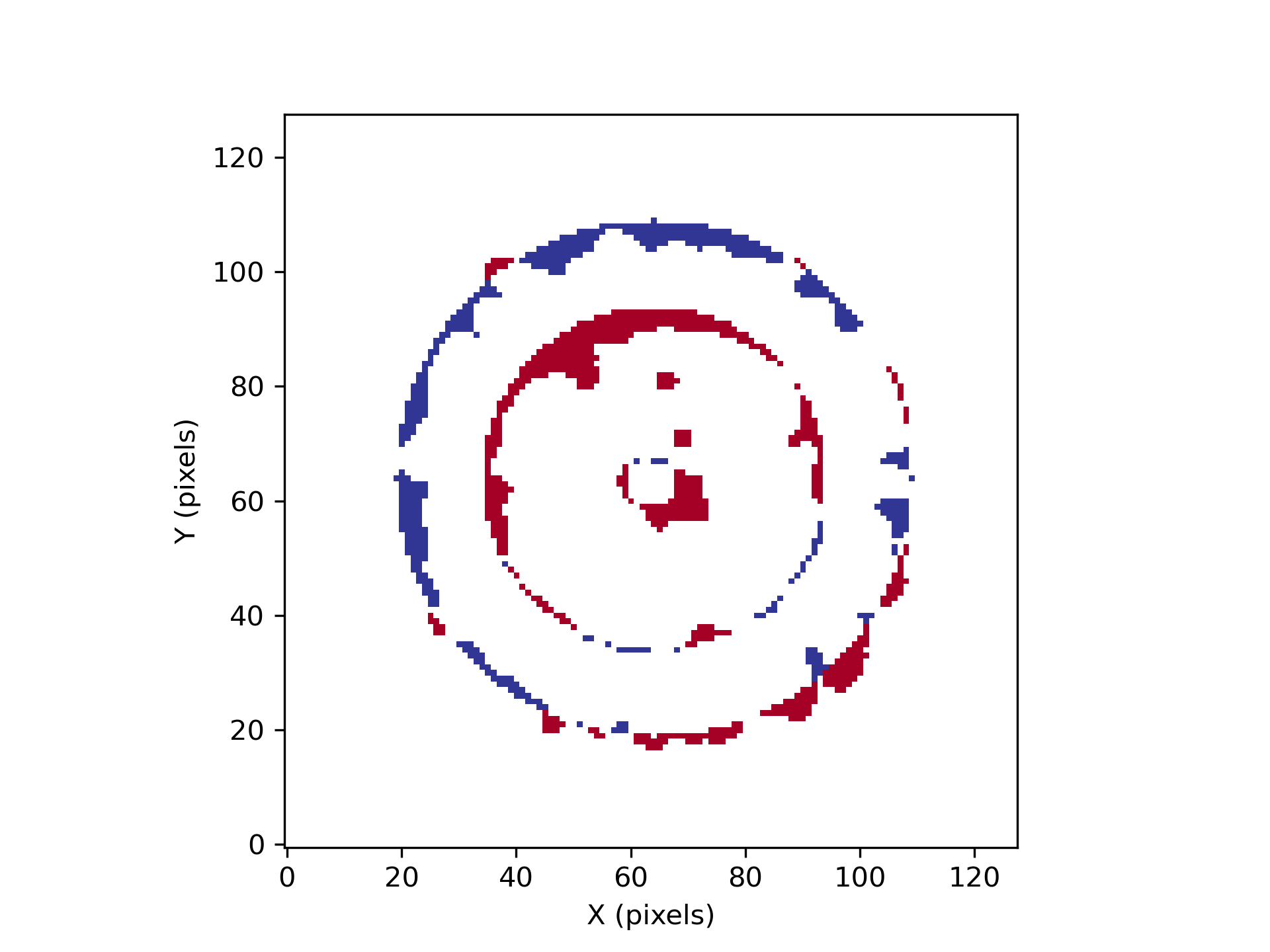}
\put(19,127){\color{black} \colorbox{white}{\textsf{False positive/negative}}}
\end{overpic}
\begin{overpic}[trim={102 0 75 30}, clip, height=0.277\textwidth]{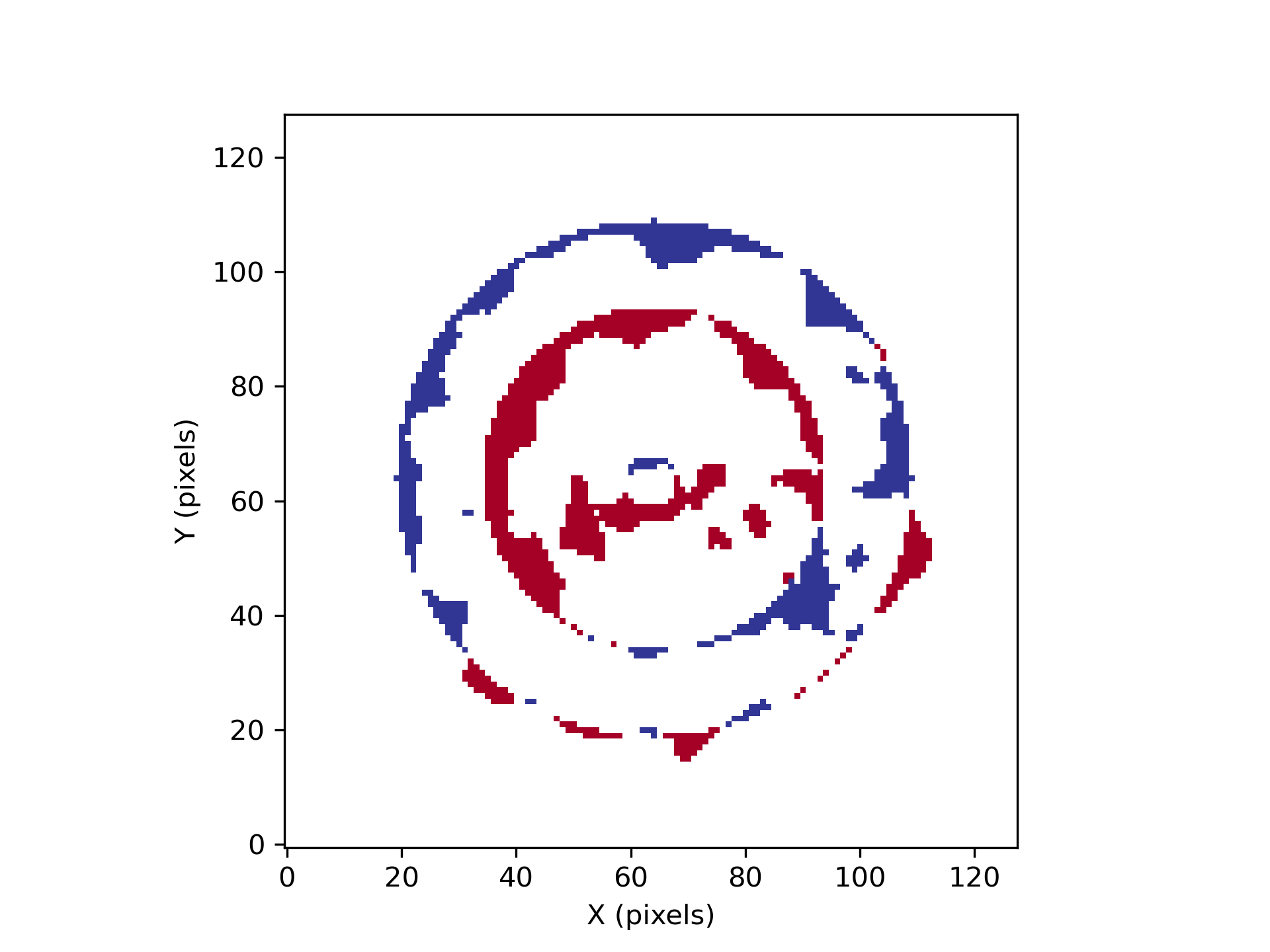}
\end{overpic}
\begin{overpic}[trim={100 0 25 30}, clip, height=0.277\textwidth]{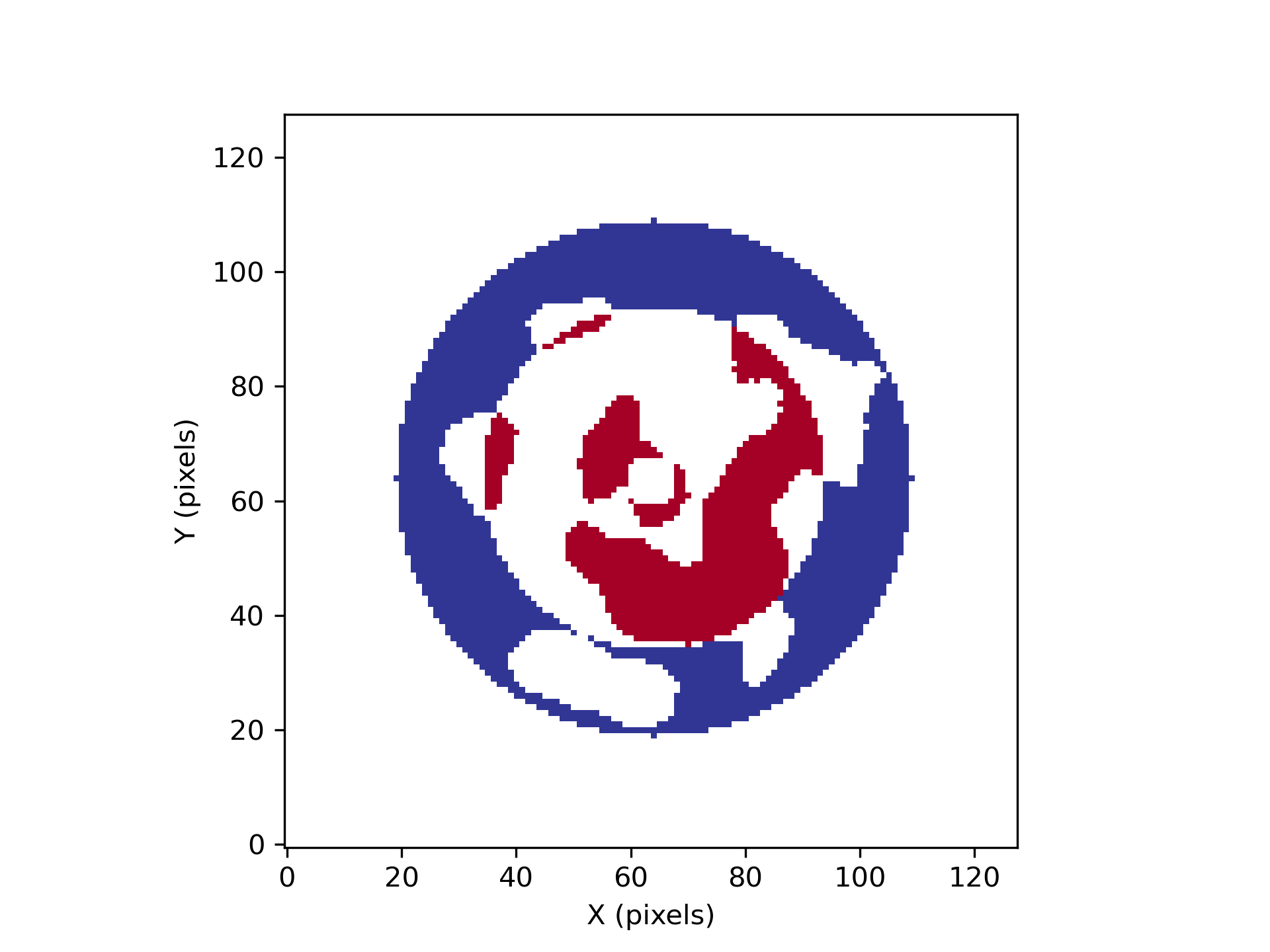}
\end{overpic}

\caption{\SAUNAS\ processing test using the cavity model as a function of the equivalent exposure time (see Sec.\,\ref{subsubsec:quality_FPFN} and Table \ref{tab:test_model_properties}). Top row shows the simulated event images. Middle row shows the final recovered surface brightness maps after processing with \SAUNAS. Dashed contours represent the $3\sigma$ and dotted contours the $2\sigma$ detection level of X-ray emission. Bottom row represents the false positive (red) and false negative (blue) detection maps for each simulation (see Sec.\,\ref{subsubsec:quality_FPFN}). From left to right columns, the equivalent exposure times are $\tau_{\rm exp}=10^7$, $5\times10^6$, and $10^6$\scm. See the labels in the panels. Colorbars represent the number of events per pixel (event images) and the surface brightness flux (final mosaics).} 
\label{fig:FNFP_test_Cave_1}
\end{center}
\end{figure*}

\begin{figure*}[t!]
\begin{center}
\begin{overpic}[trim={25 32 110 30}, clip, height=0.25\textwidth]{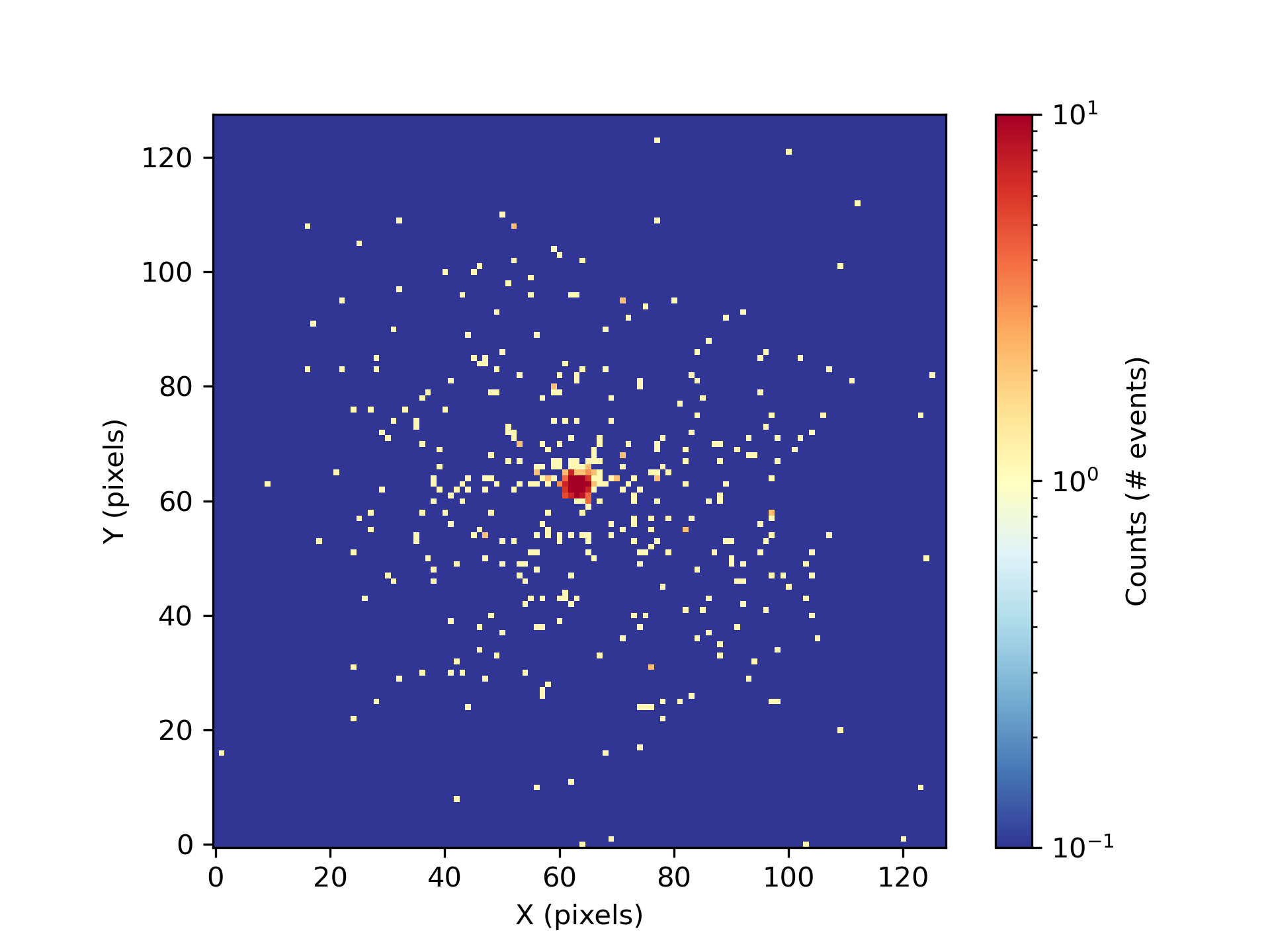}
\put(50,131){\color{black} \colorbox{white}{\textsf{$\tau_{\rm exp} = 5\cdot10^5$\scm}}}
\put(19,130){\color{black} \colorbox{white}{\textsf{Events}}}
\end{overpic}
\begin{overpic}[trim={70 32 110 30}, clip, height=0.25\textwidth]{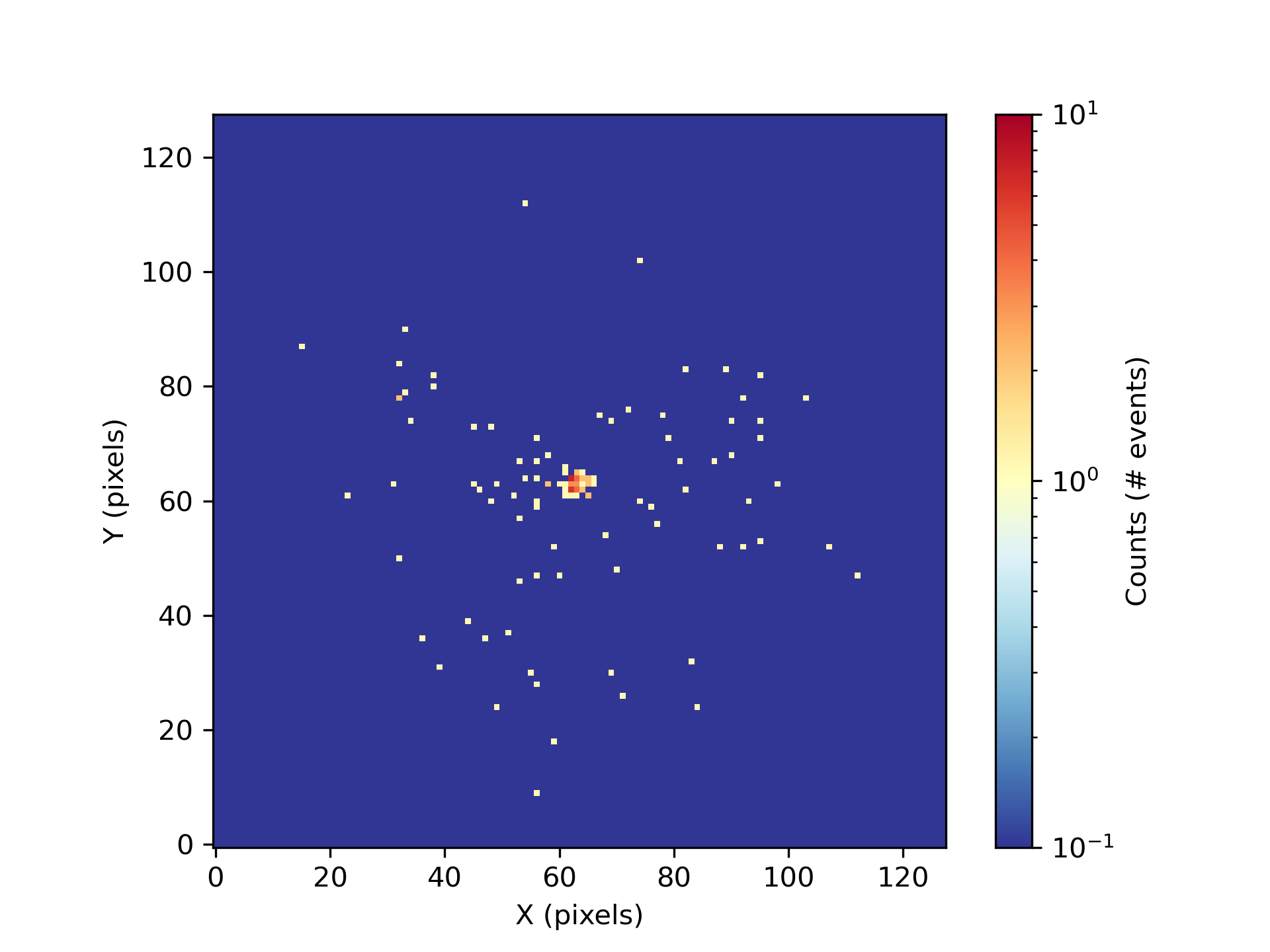}
\put(50,131){\color{black} \colorbox{white}{\textsf{$\tau_{\rm exp} = 10^5$\scm}}}
\end{overpic}
\begin{overpic}[trim={70 32 25 30}, clip, height=0.25\textwidth]{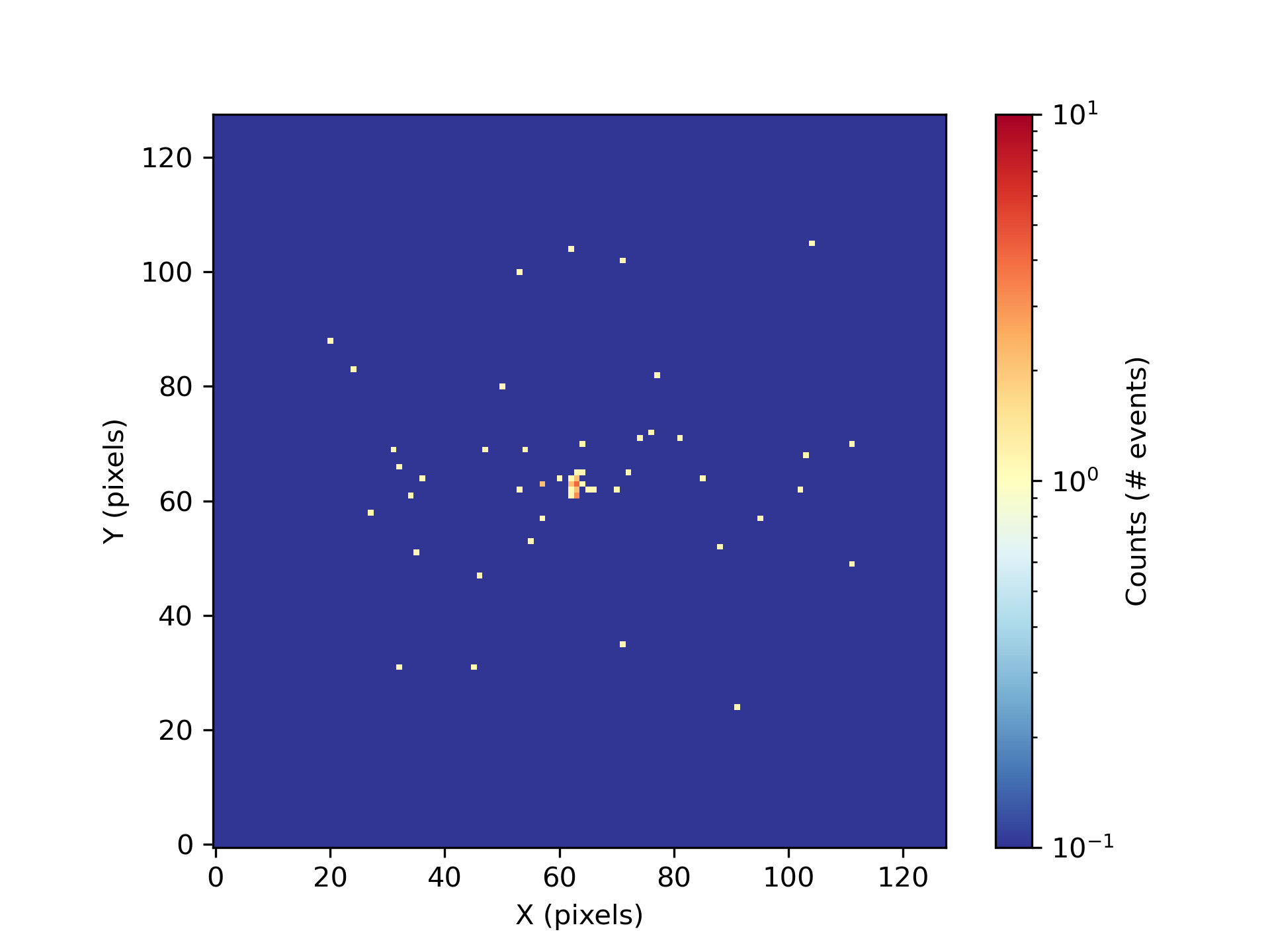}
\put(50,131){\color{black} \colorbox{white}{\textsf{$\tau_{\rm exp} = 5\cdot10^4$\scm}}}
\end{overpic}

\begin{overpic}[trim={25 32 110 30}, clip, height=0.25\textwidth]{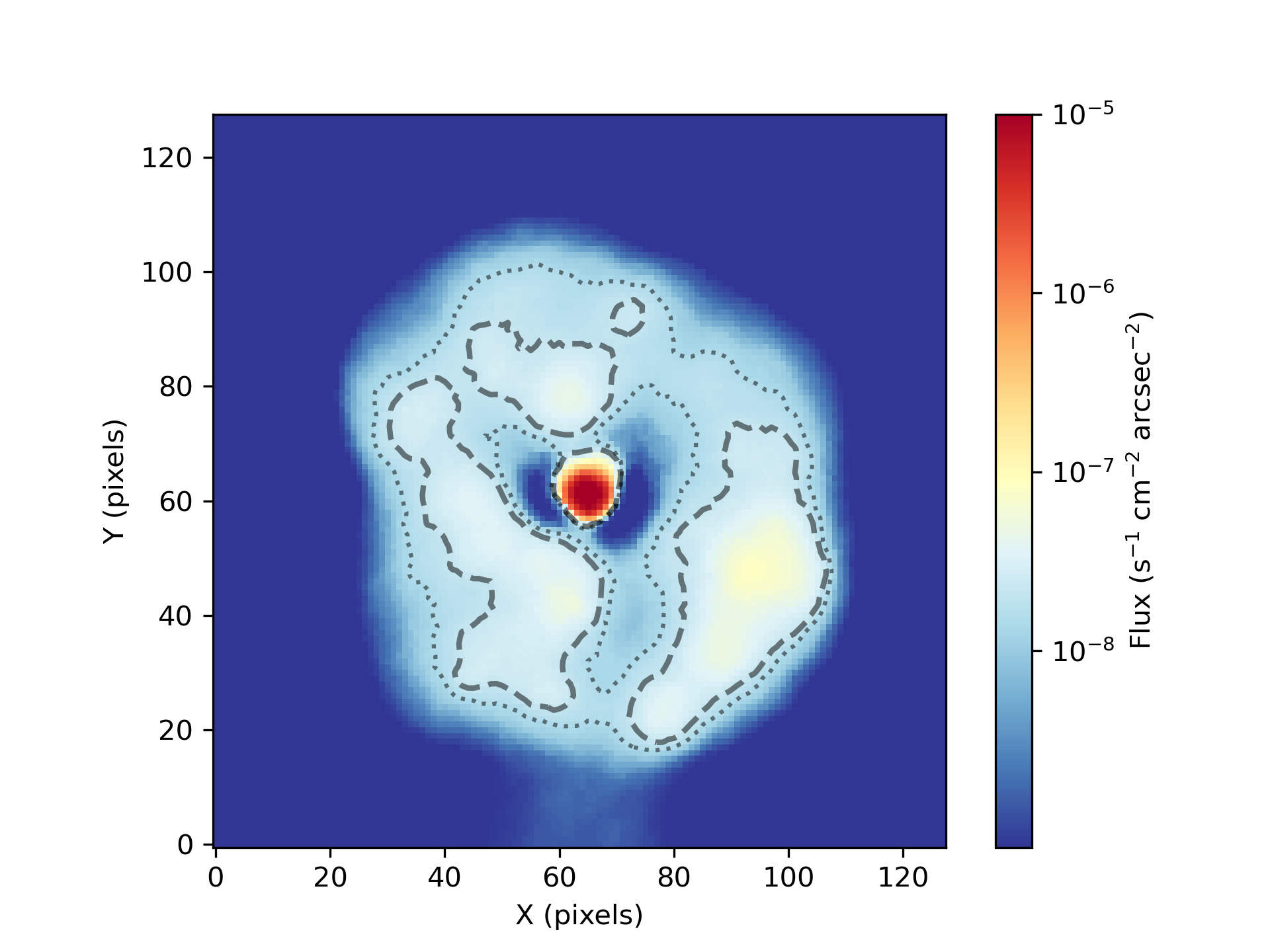}
\put(24,113.5){\color{black} \colorbox{white}{\textsf{Final}}}
\end{overpic}
\begin{overpic}[trim={70 32 110 30}, clip, height=0.25\textwidth]{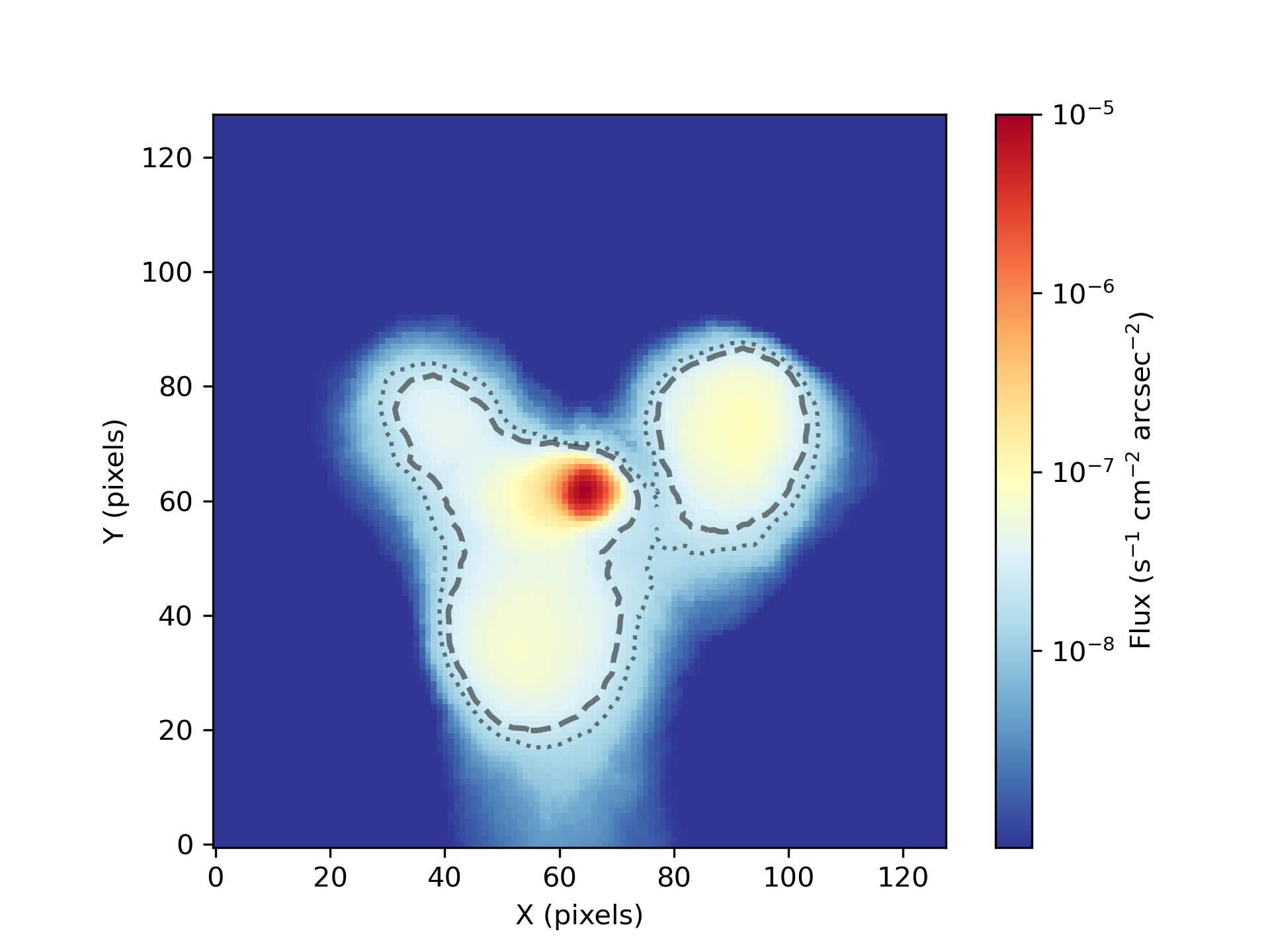}
\end{overpic}
\begin{overpic}[trim={70 32 25 30}, clip, height=0.25\textwidth]{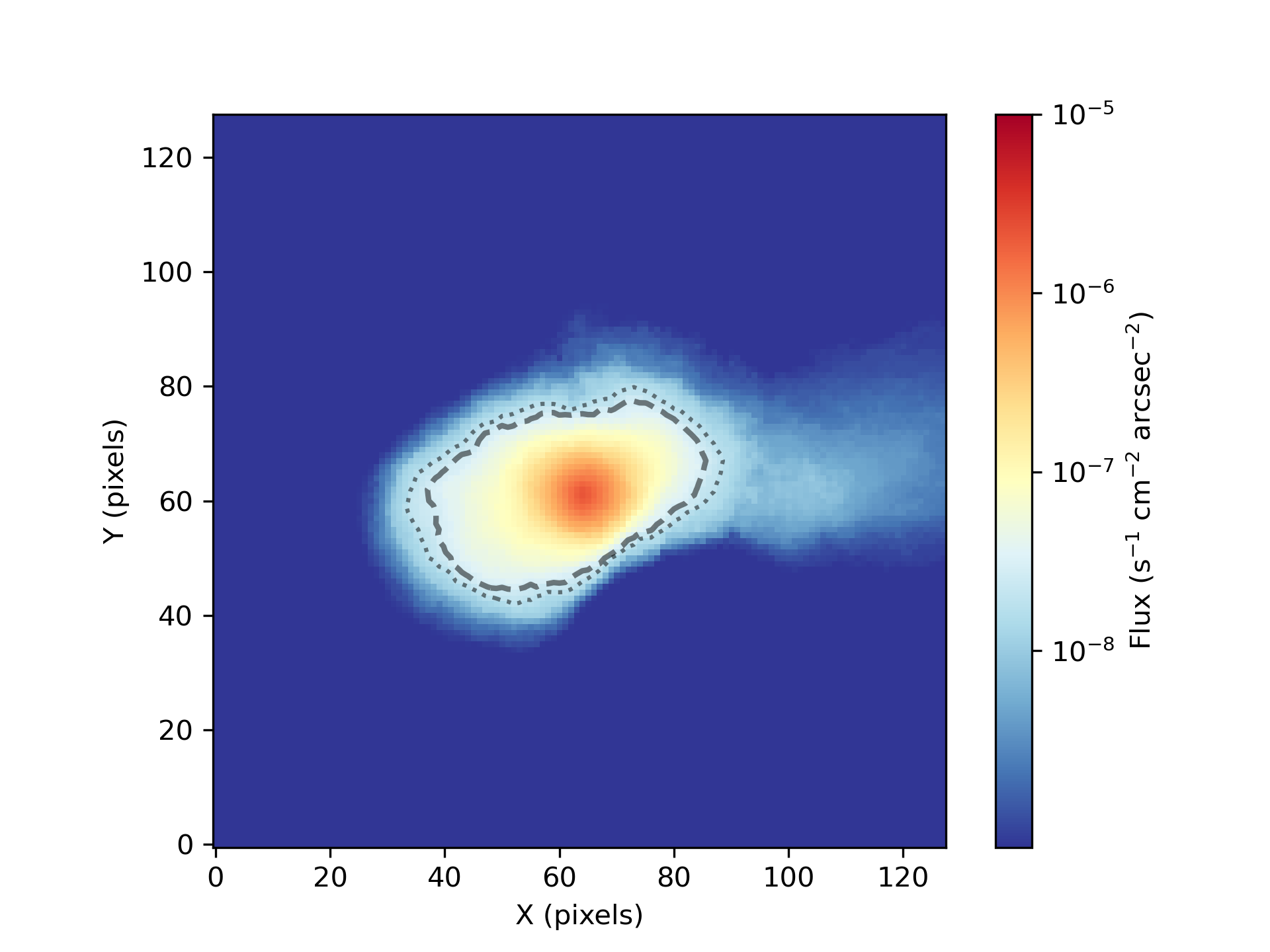}
\end{overpic}

\begin{overpic}[trim={62 0 80 30}, clip, height=0.277\textwidth]{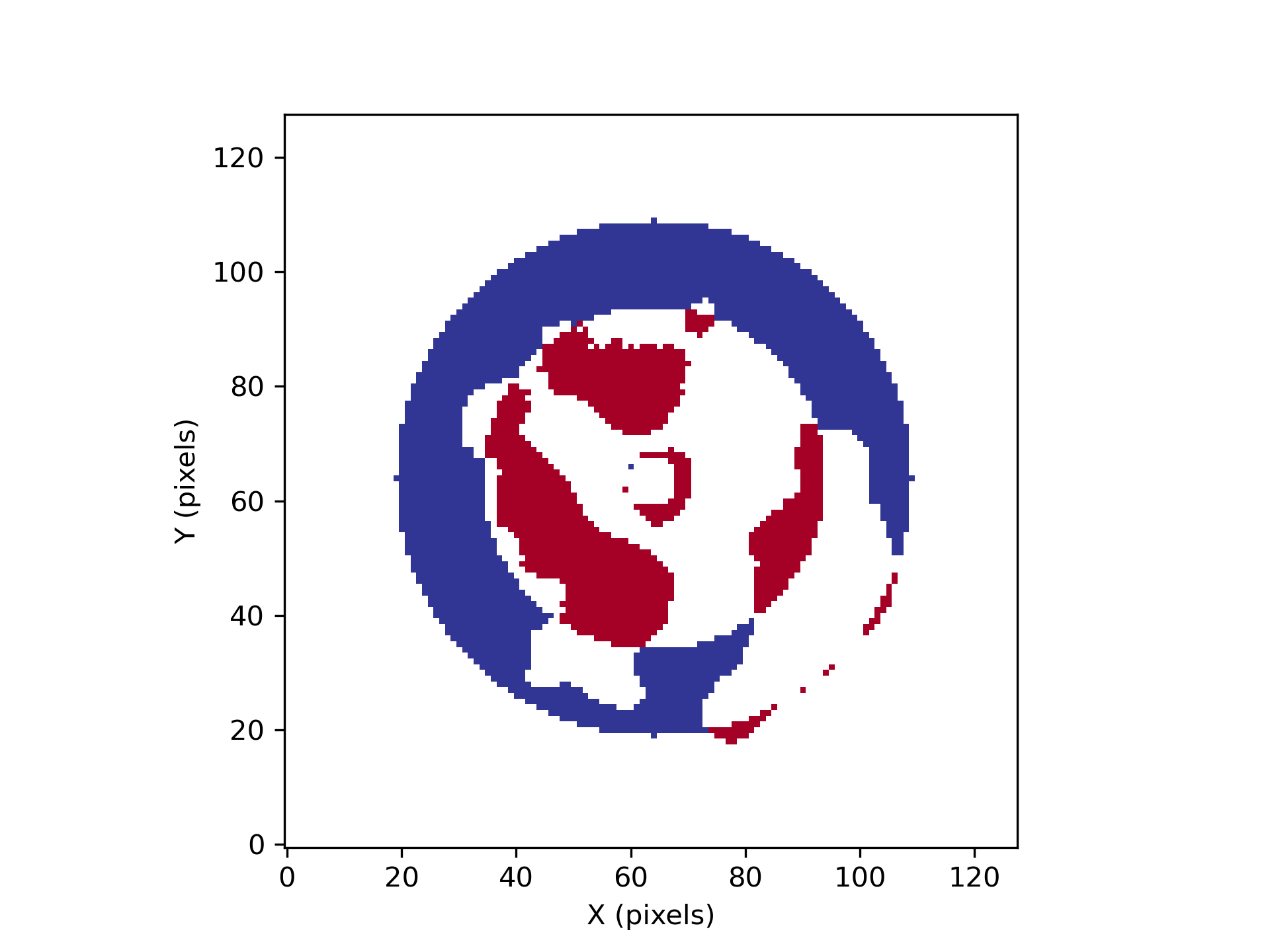}
\put(19,127){\color{black} \colorbox{white}{\textsf{False positive/negative}}}
\end{overpic}
\begin{overpic}[trim={102 0 75 30}, clip, height=0.277\textwidth]{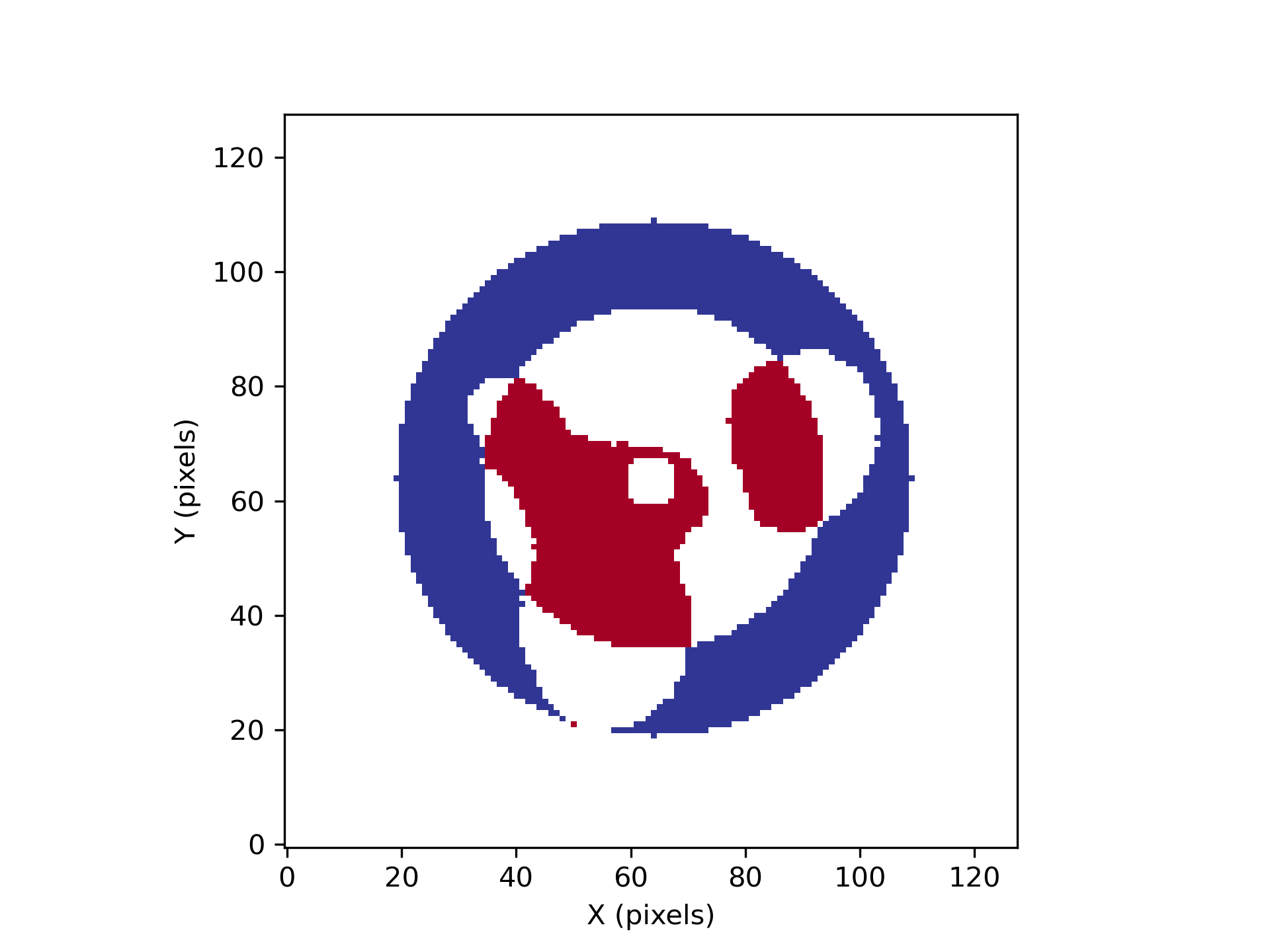}
\end{overpic}
\begin{overpic}[trim={100 0 25 30}, clip, height=0.277\textwidth]{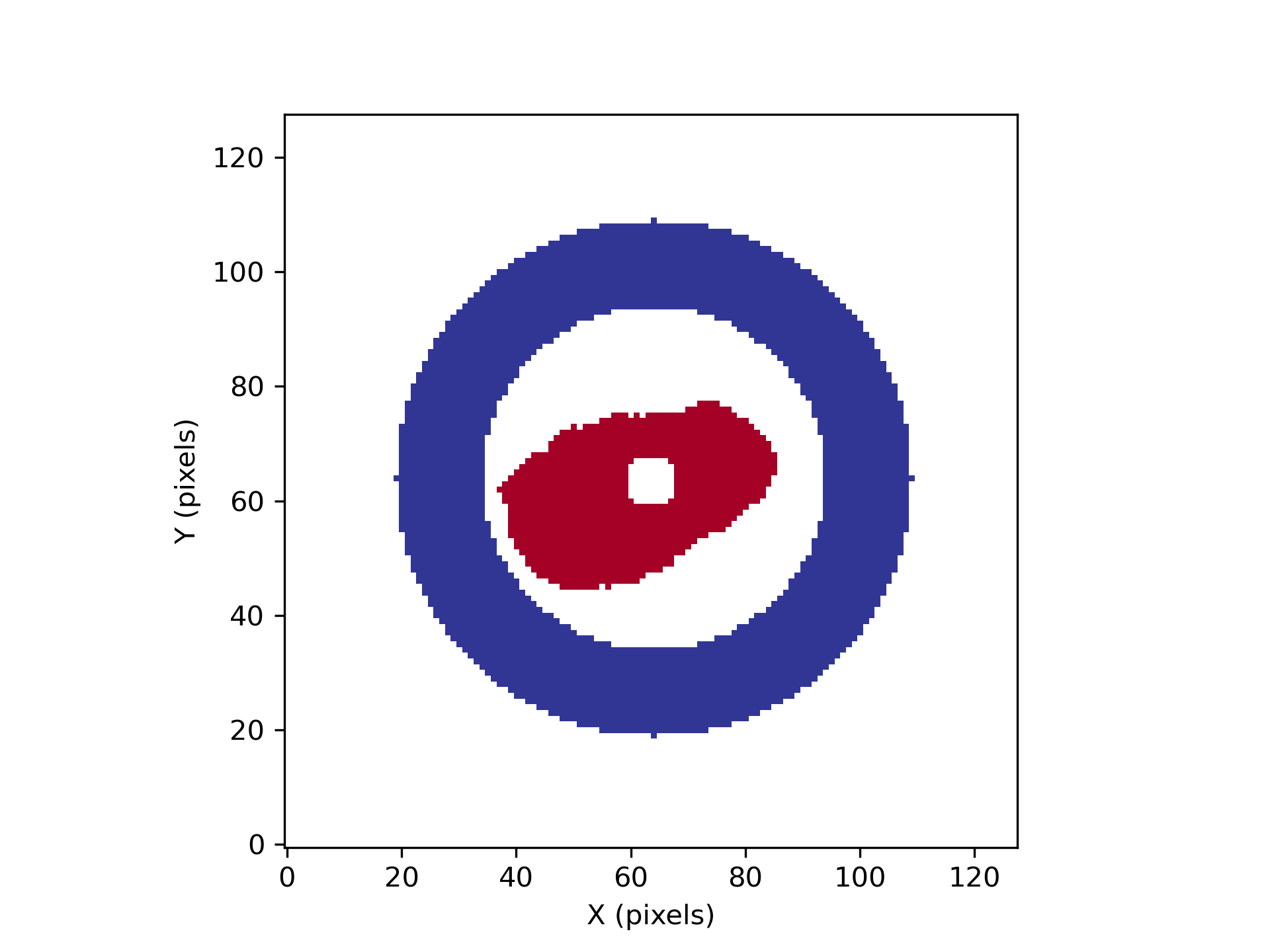}
\end{overpic}

\caption{(Continuation of Fig.\,\ref{fig:FNFP_test_Cave_1}) \SAUNAS\ processing test using the double jet model as a function of the equivalent exposure time (see Sec.\,\ref{subsubsec:quality_FPFN} and Table \ref{tab:test_model_properties}). Top row shows the simulated event images. Middle row shows the final recovered surface brightness maps after processing with \SAUNAS. Dashed contours represent the $3\sigma$ and dotted contours the $2\sigma$ detection level of X-ray emission. Bottom row represents the false positive (red) and false negative (blue) detection maps for each simulation (see Sec.\,\ref{subsubsec:quality_FPFN}). From left to right columns, the equivalent exposure times are $\tau_{\rm exp}=5\times10^5$, $10^5$, and $5\times10^4$\scm. See the labels in the panels. Colorbars represent the number of events per pixel (event images) and the surface brightness flux (final mosaics).} 
\label{fig:FNFP_test_Cave_2}
\end{center}
\end{figure*}

\section{NGC\,3079 and UGC\,5101 Point Spread Function}
\label{Appendix:Observed_PSFs}

\begin{figure*}[t!]
\begin{center}

\begin{overpic}[trim={80 0 80 0}, clip, width=0.32\textwidth]{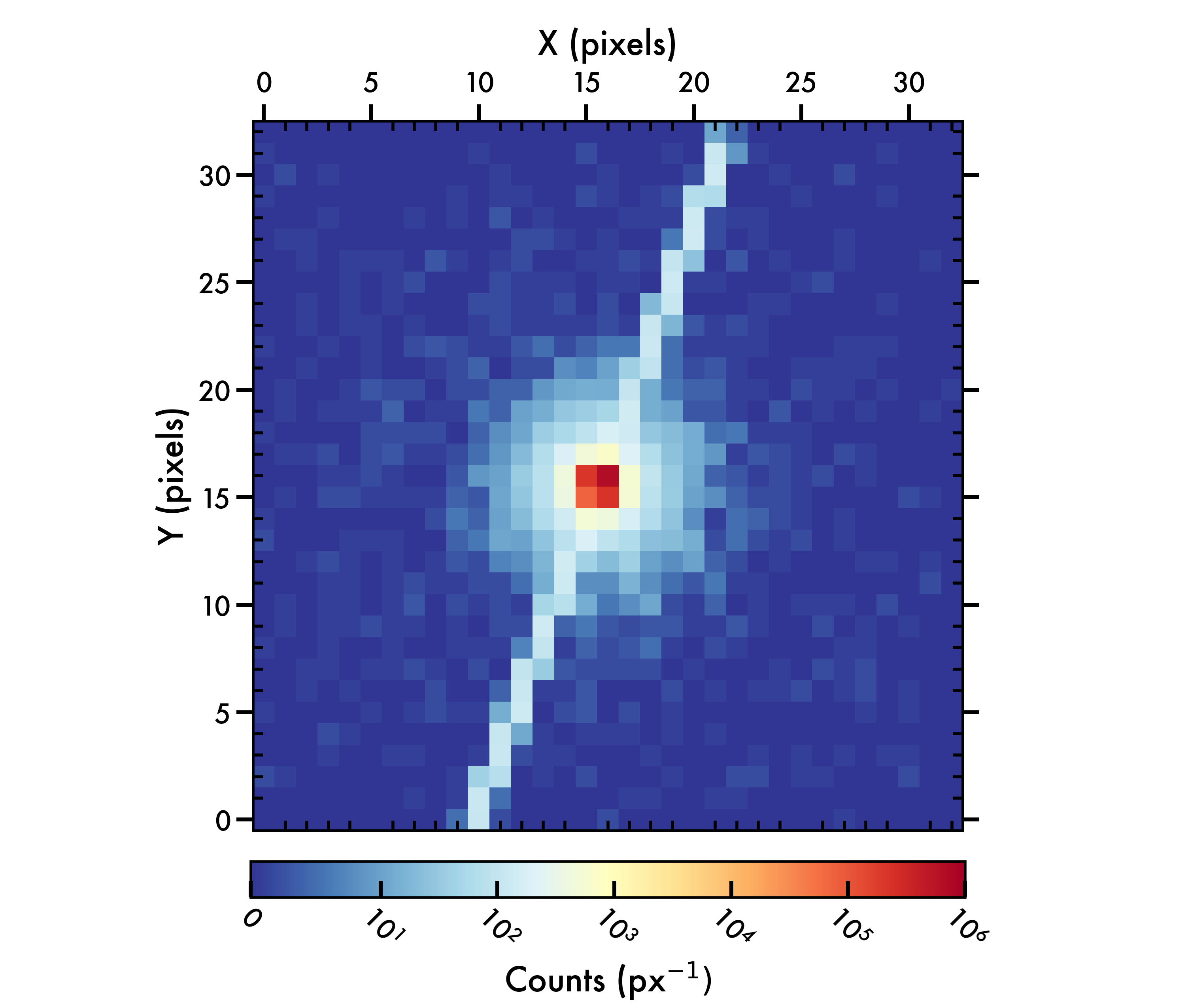}
\put(22,148){\color{black} \colorbox{white}{\textsf{ObsID: 2038}}}
\end{overpic}
\begin{overpic}[trim={80 0 80 0}, clip, width=0.32\textwidth]{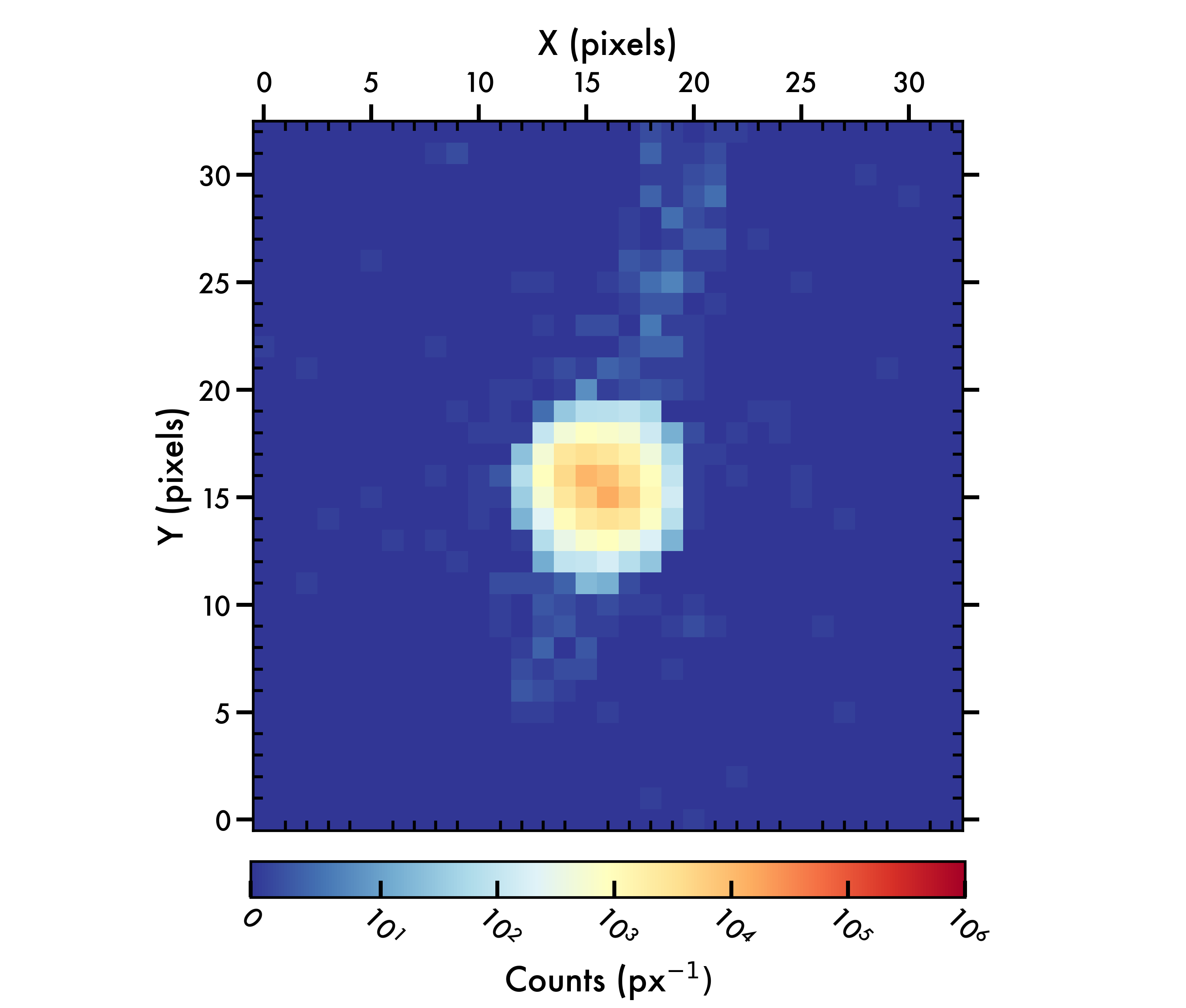}
\put(22,148){\color{black} \colorbox{white}{\textsf{ObsID: 7851}}}
\end{overpic}

\caption{Point Spread Functions (PSF) of NGC\,3079 images processed with \SAUNAS. \emph{Left panel:} 1) PSF of the visit 2038 to NGC\,3079 on the broad 0.3-2.0 keV band. 2) PSF of the visit 7851 to NGC\,3079 on the same broad 0.3-2.0 keV band. The binning (pixelscale) for the NGC\,3079 images and the PSFs is $8\times8$ (3.936 arcsec px$^{-1}$). Notice the logarithmic color scale in the bottom of the panels.} 
\label{fig:NGC3079_psf}
\end{center}
\end{figure*}

\begin{figure*}[t!]
\begin{center}
\begin{overpic}[trim={80 0 80 0}, clip, width=0.32\textwidth]{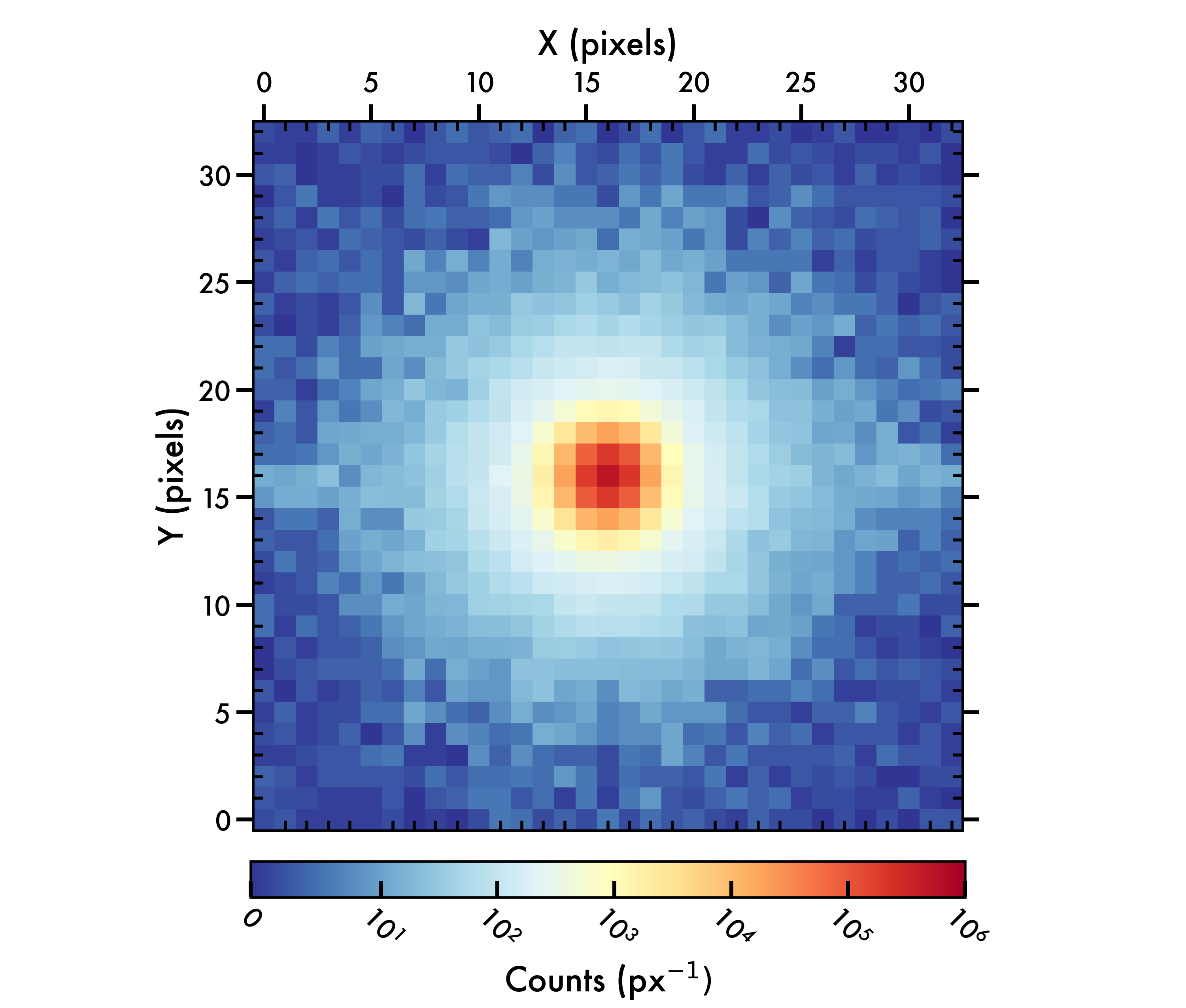}
\put(22,148){\color{black} \colorbox{white}{\textsf{0.3 -- 1.0 keV}}}
\end{overpic}
\begin{overpic}[trim={80 0 80 0}, clip, width=0.32\textwidth]{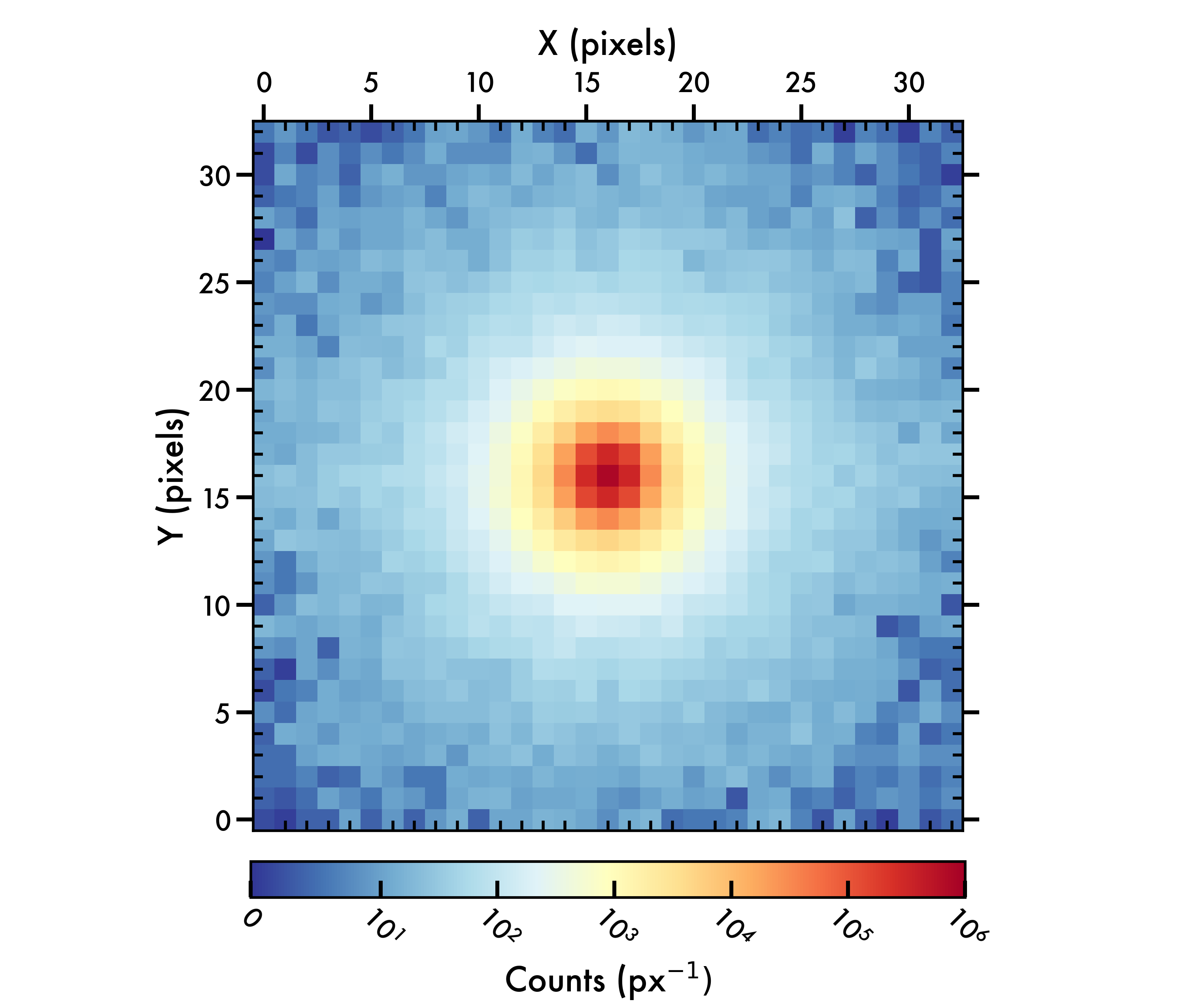}
\put(22,148){\color{black} \colorbox{white}{\textsf{1.0 -- 2.0 keV}}}
\end{overpic}
\begin{overpic}[trim={80 0 80 0}, clip, width=0.32\textwidth]{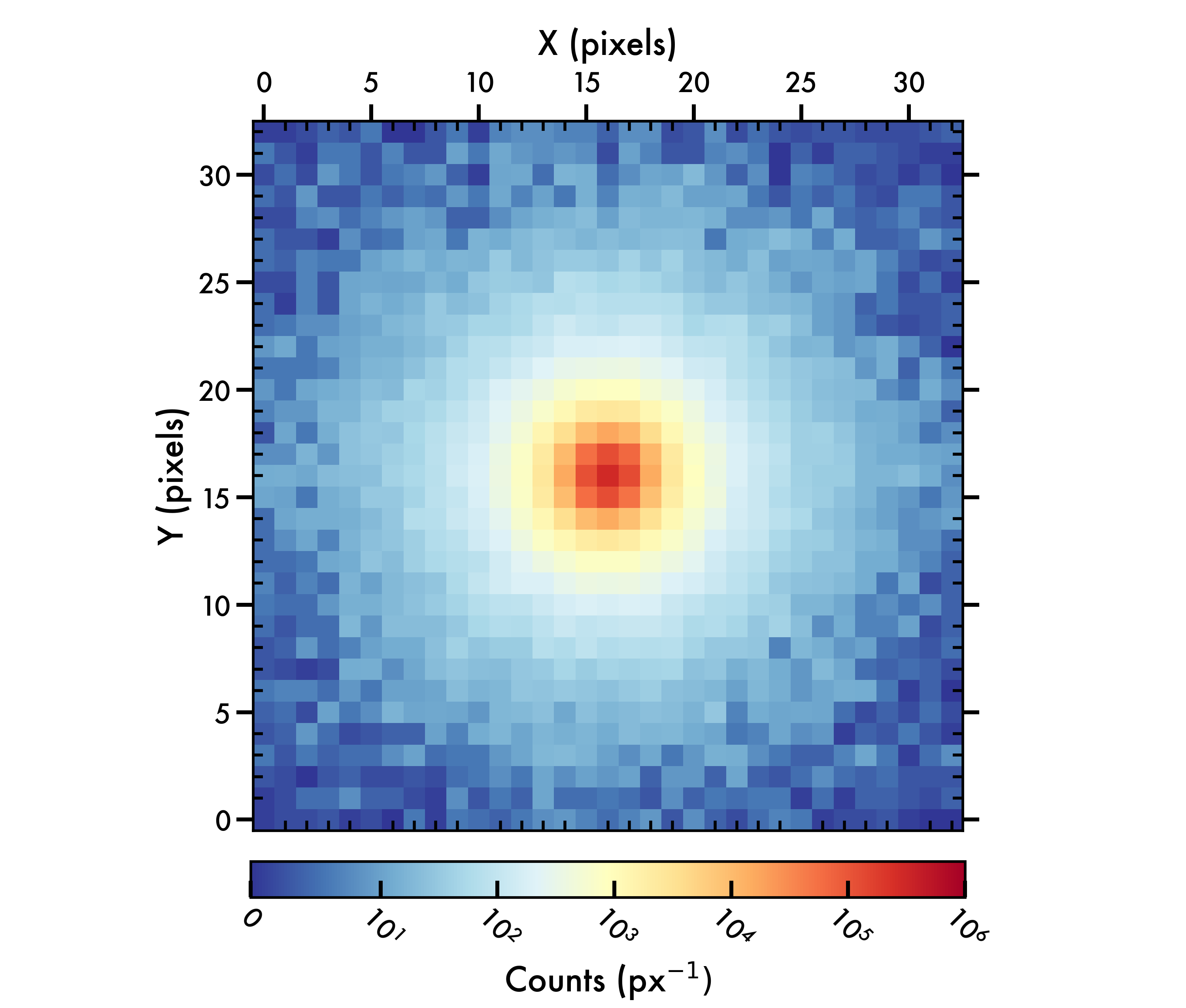}
\put(22,148){\color{black} \colorbox{white}{\textsf{2.0 -- 8.0 keV}}}
\end{overpic}

\caption{Point Spread Functions (PSF) of UGC\,5101 images processed with \SAUNAS. \emph{From left to right:} 1) PSF of the 0.3--1.0 keV (soft) band. 2) PSF of the 1.0--2.0 keV (medium) band. 3) PSF of the 2.0--8.0 keV (hard) band. The binning (pixelscale) for the UGC\,5101 images and the PSFs is $1\times1$ (0.492 arcsec px$^{-1}$). Notice the logarithmic color scale in the bottom of the panels.} 
\label{fig:UGC5101_psf}
\end{center}
\end{figure*}

This section presents the PSFs generated for the NGC\,3079 (see Sec.\,\ref{subsec:NGC3079}) and UGC\,5101 (see Sec.\,\ref{subsec:UGC5101}), \Chandra/ACIS observations. The PSFs were generated using \texttt{MARX} as described in Sec.\,\ref{subsec:methods_saunas}. The panels in Figs. \ref{fig:NGC3079_psf} and Fig.\,\ref{fig:UGC5101_psf} show the different PSFs obtained for the three bands (0.3--1.0 keV, 1.0--2.0 kev, and 2.0--8.0 kev) in UGC\,5101, and for the two datasets analyzed in the broadband (0.3--2.0 keV) for NGC\,3079.

\section{NGC\,3079 and UGC\,5101 event maps}
\label{Appendix:Observed_events}

\begin{figure*}[t!]
\begin{center}
\begin{overpic}[trim={20 0 80 0}, clip, width=0.49\textwidth]{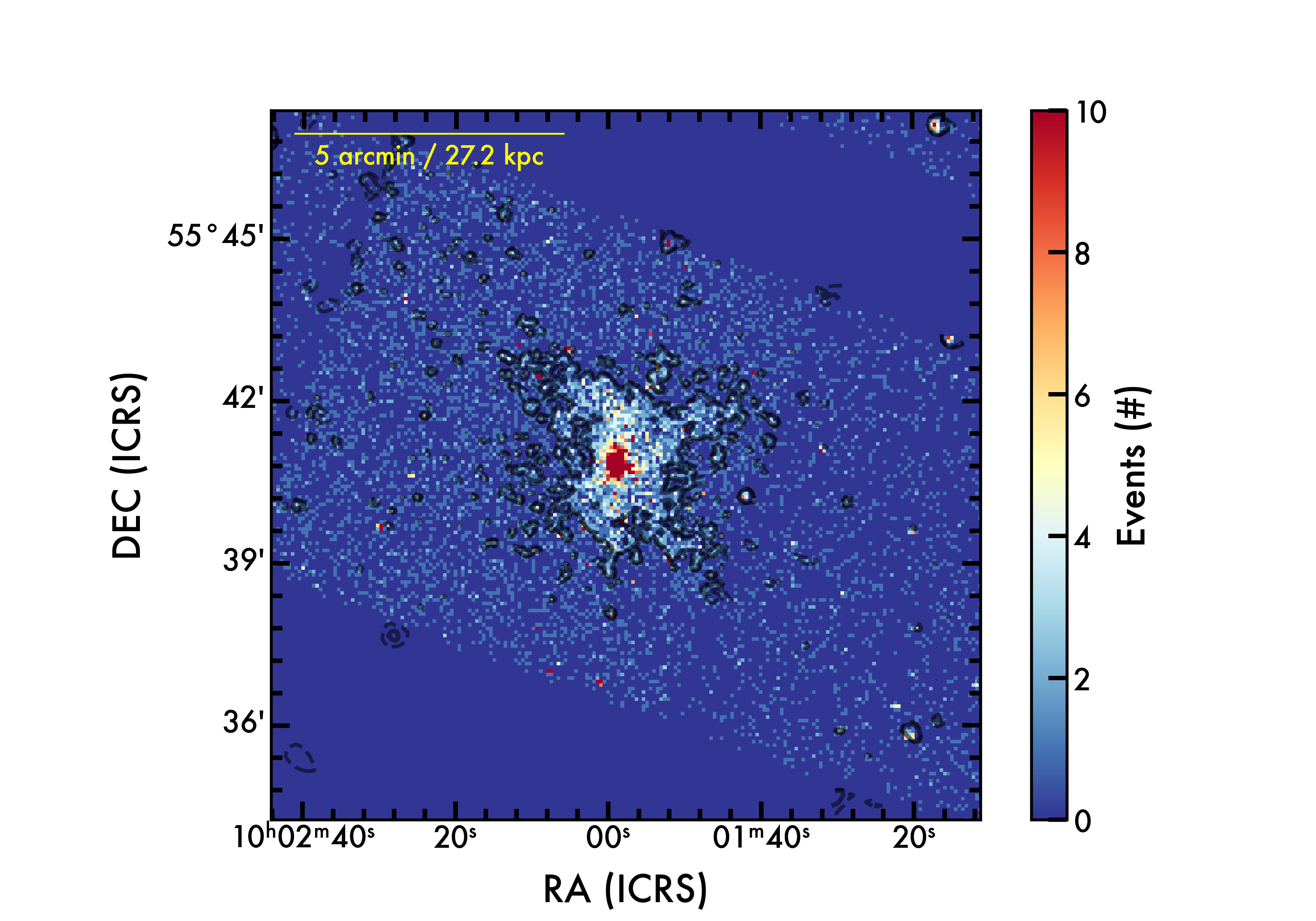}
\put(52,188){\color{black} \colorbox{white}{\textsf{NGC\,3079 - ObsID: 2038 - 0.3--2.0 keV}}}
\end{overpic}
\begin{overpic}[trim={20 0 80 0}, clip, width=0.49\textwidth]{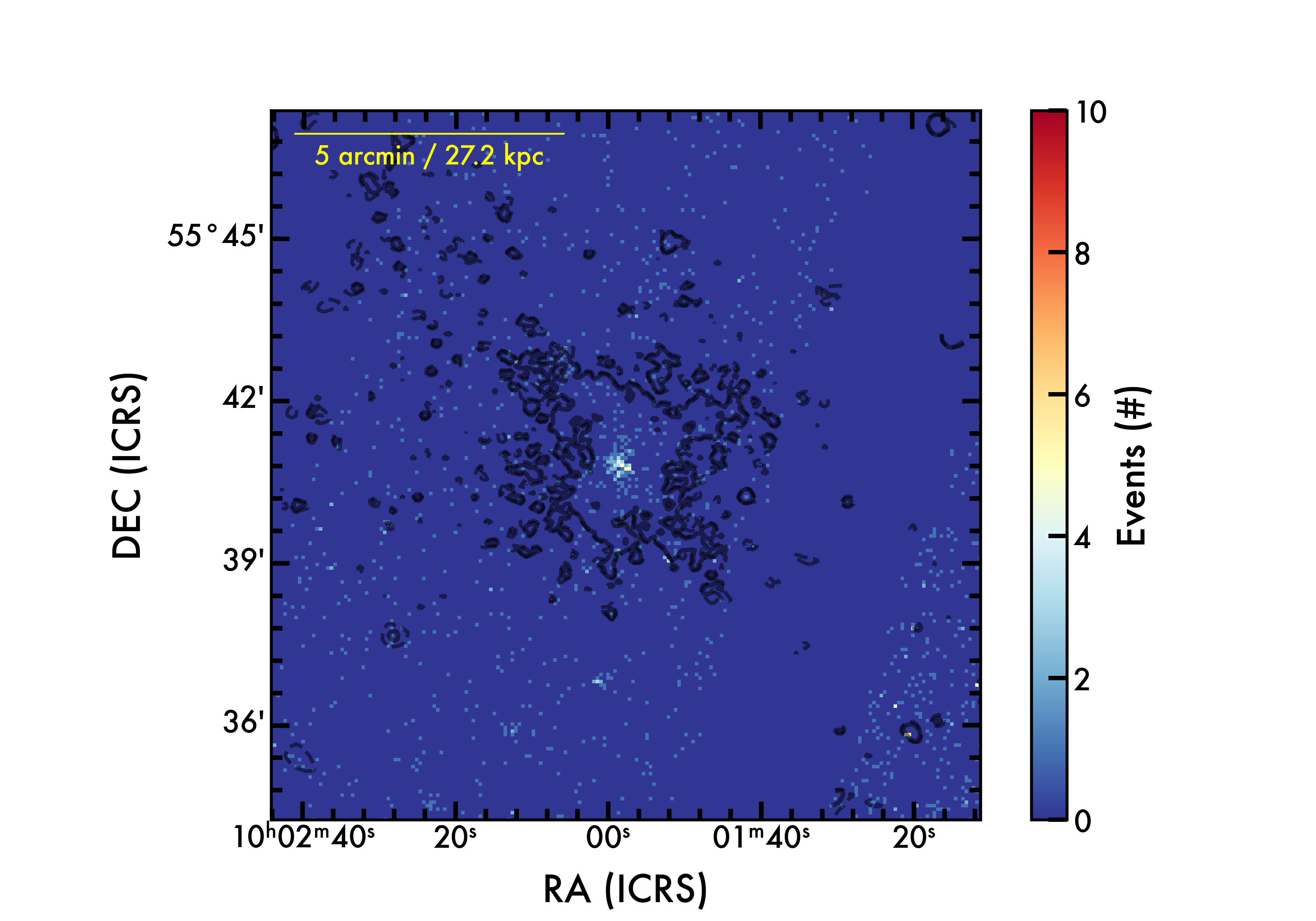}
\put(52,188){\color{black} \colorbox{white}{\textsf{NGC\,3079 - ObsID: 7851 - 0.3--2.0 keV}}}
\end{overpic}

\caption{Event maps of the NGC\,3079 observations, before processing with \SAUNAS. \emph{Left:} Observation ID 2038, 0.3--2.0 keV. \emph{Right:} Observation ID 7851, 0.3--2.0 keV. The binning (pixelscale) for the NGC\,3079 images is $8\times8$ (3.936 arcsec px$^{-1}$). Solid black contours represent the $3\sigma$ and dashed contours the $2\sigma$ detection level of X-ray emission.} 
\label{fig:NGC3079_events}
\end{center}
\end{figure*}

\begin{figure*}[t!]
\begin{center}
\begin{overpic}[trim={20 0 80 0}, clip, width=0.32\textwidth]{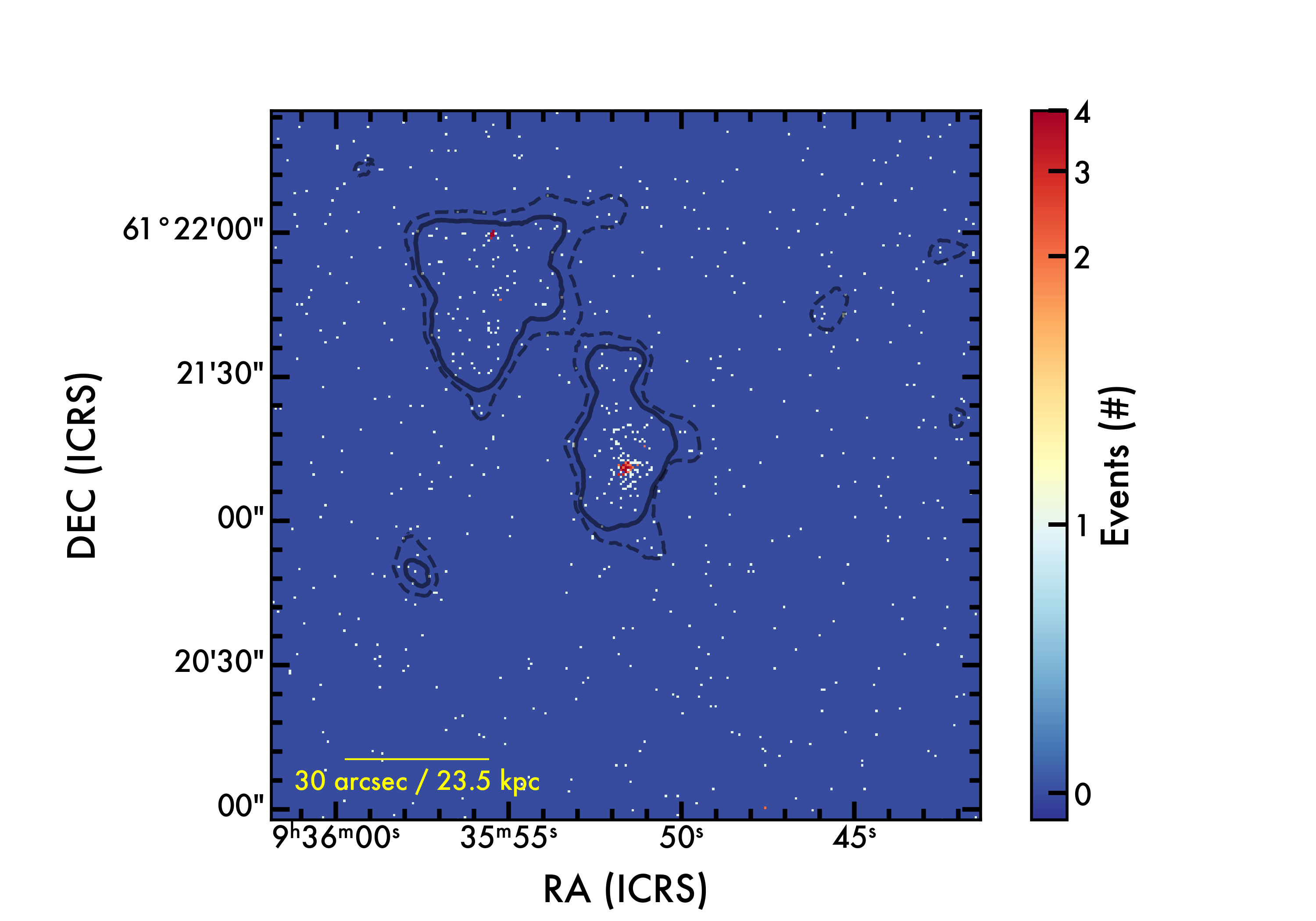}
\put(30,125){\color{black} \colorbox{white}{\textsf{UGC\,5101 - 0.3--1.0 keV}}}
\end{overpic}
\begin{overpic}[trim={20 0 80 0}, clip, width=0.32\textwidth]{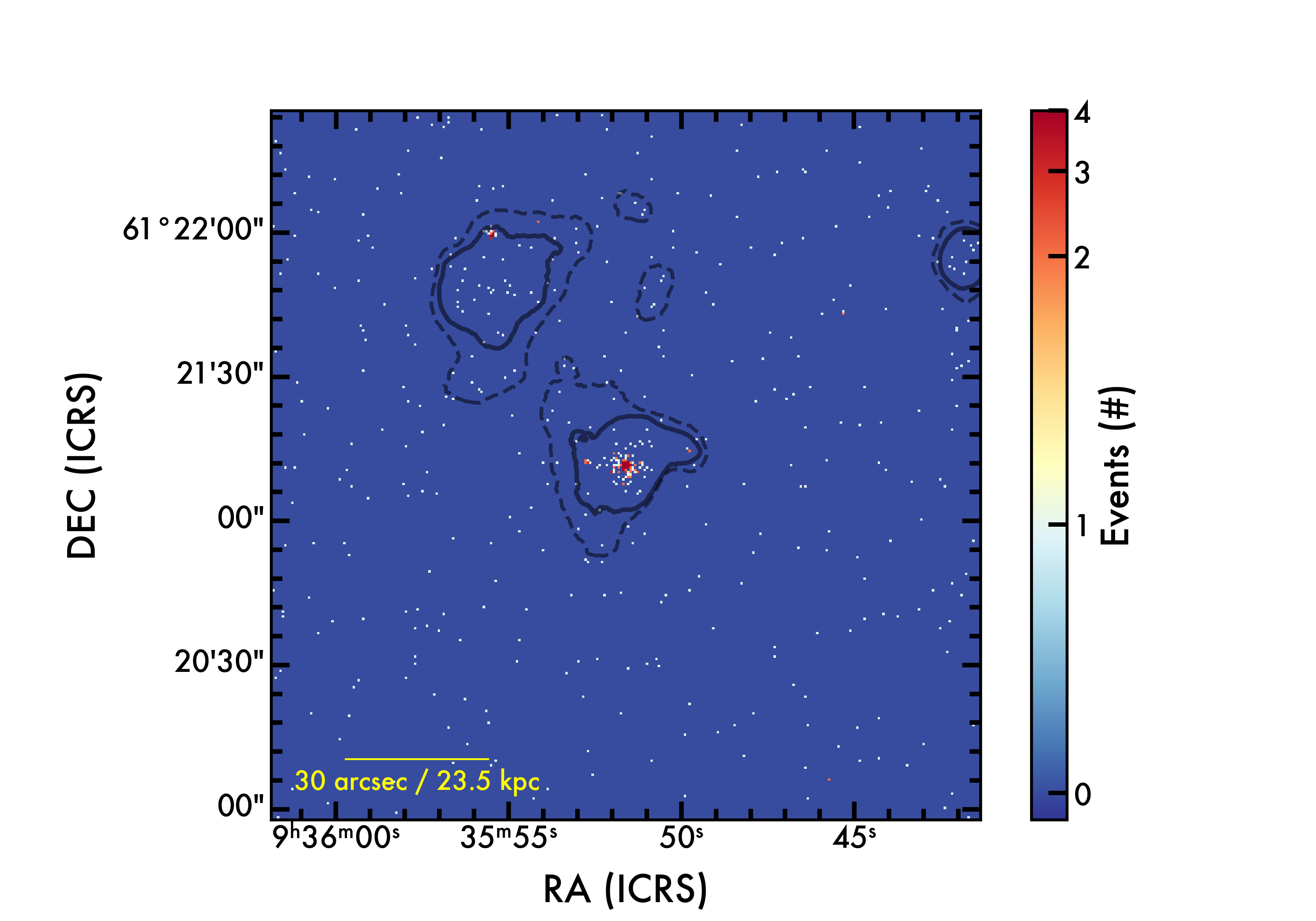}
\put(30,125){\color{black} \colorbox{white}{\textsf{UGC\,5101 - 1.0--2.0 keV}}}
\end{overpic}
\begin{overpic}[trim={20 0 80 0}, clip, width=0.32\textwidth]{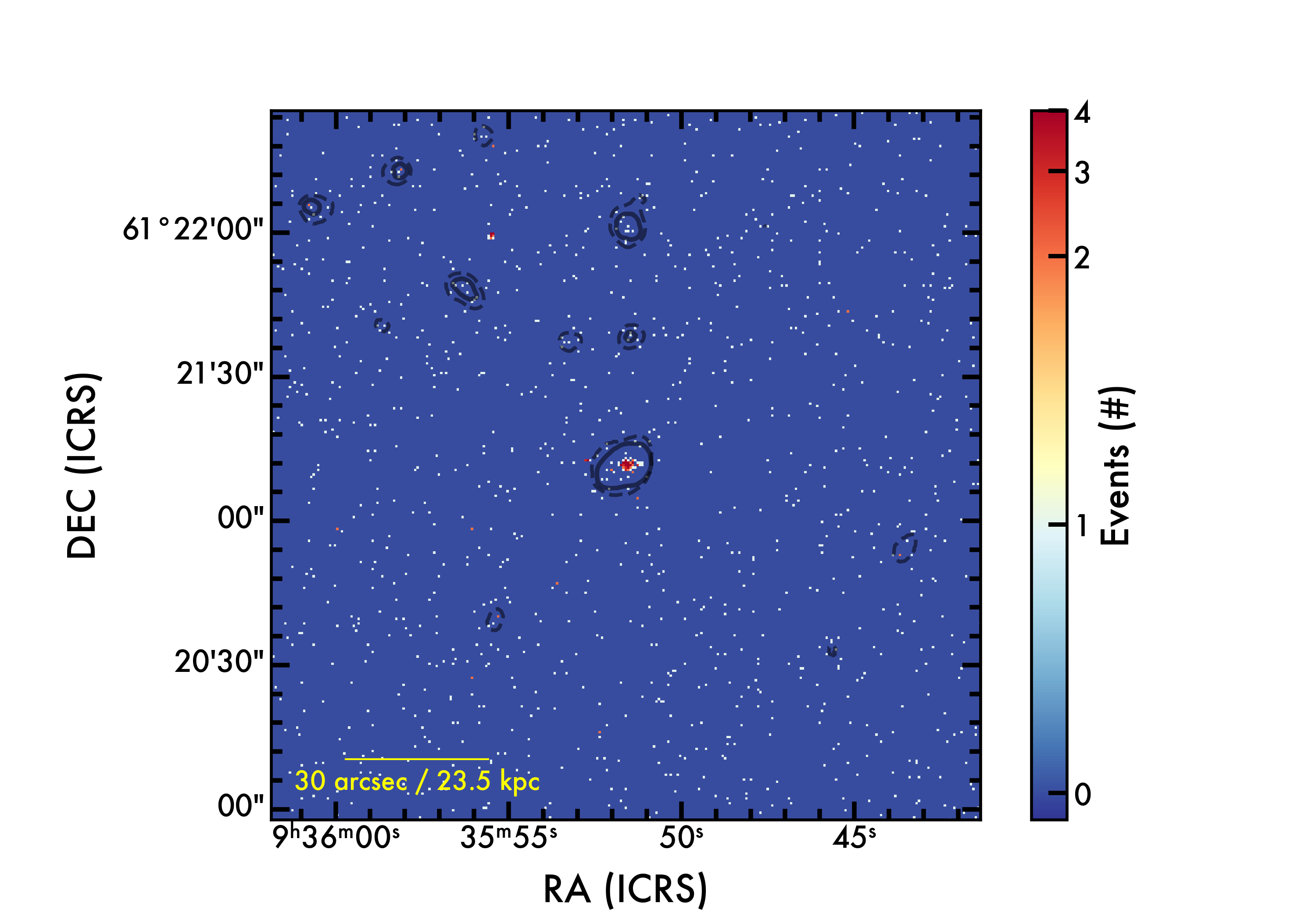}
\put(30,125){\color{black} \colorbox{white}{\textsf{UGC\,5101 - 2.0--8.0 keV}}}
\end{overpic}

\caption{Event maps of the UGC\,5101 observations, before processing with \SAUNAS. \emph{Left to right:} 1) UGC\,3079 on the 0.3-1.0 keV band. 2)UGC\,3079 on the 1.0-2.0 keV band. 3) UGC\,3079 on the 2.0-8.0 keV band. The binning (pixelscale) for the UGC\,5101 images is $1\times1$ (0.492 arcsec px$^{-1}$). Solid black contours represent the $3\sigma$ and dashed contours the $2\sigma$ detection level of X-ray emission.} 
\label{fig:UGC5101_events}
\end{center}
\end{figure*}

This section presents the event maps as observed by \Chandra/ACIS and processed by \texttt{CIAO} for the NGC\,3079 (see Sec.\,\ref{subsec:NGC3079}) and UGC\,5101 (see Sec.\,\ref{subsec:UGC5101}) observations. Note that the events in the panels represent the raw event counts without any \SAUNAS\ processing, and thus they include contamination by sky background, gradients generated by the different equivalent exposure time across the field of view, and point source contamination. The panels in Fig.\,\ref{fig:NGC3079_events} show the events obtained in the \Chandra\ 2038 and 7851 visits to NGC\,3079 in the 0.3--2.0 keV broadband, and Fig.\,\ref{fig:UGC5101_events} show the events obtained for the three bands (0.3--1.0 keV, 1.0--2.0 kev, and 2.0--8.0 kev) in UGC\,5101.

\bibliography{bibFile} 
\bibliographystyle{aasjournal}



\end{document}